\documentclass[longauth]{aa} 
\usepackage{graphicx}
\usepackage{natbib} 
\bibpunct{(}{)}{;}{a}{}{,} 

\usepackage{txfonts}
\usepackage{amssymb}
\def\gapprox{\;\rlap{\lower 2.5pt            
 \hbox{$\sim$}}\raise 1.5pt\hbox{$>$}\;}
\def\lapprox{\;\rlap{\lower 2.5pt            
 \hbox{$\sim$}}\raise 1.5pt\hbox{$<$}\;}

\begin{document} 
   \title{SPHERE / ZIMPOL high resolution polarimetric imager
\thanks{Based on observations collected at La Silla and Paranal 
          Observatory, ESO (Chile), Program ID: 60.A-9249 and 60.A-9255}.}

   \subtitle{I. System overview, PSF parameters, coronagraphy,
               and polarimetry}

    \author{H.M.~Schmid\inst{\ref{instch1}}    
   \and A.~Bazzon\inst{\ref{instch1}}
   \and R.~Roelfsema\inst{\ref{instnl1}}   
   \and D.~Mouillet\inst{\ref{instf1},\ref{instf2}} 
   \and J.~Milli\inst{\ref{insteso2}} 
   \and F.~Menard\inst{\ref{instf1},\ref{instf2}} 
   \and D.~Gisler\inst{\ref{instch3},\ref{instd2}}  
   \and S.~Hunziker\inst{\ref{instch1}}    
   \and J.~Pragt\inst{\ref{instnl1}}    
   \and C.~Dominik\inst{\ref{instnl2}}
   \and A.~Boccaletti\inst{\ref{instf4}}         
   \and C.~Ginski\inst{\ref{instnl3}}      
   \and L.~Abe\inst{\ref{instf5}}
   \and S.~Antoniucci\inst{\ref{insti3}}
   \and H.~Avenhaus\inst{\ref{instd1},\ref{instch1}}
   \and A.~Baruffolo\inst{\ref{insti1}}
   \and P.~Baudoz\inst{\ref{instf4}}      
   \and J.L.~Beuzit\inst{\ref{instf3}}
   \and M.~Carbillet\inst{\ref{instf5}}
   \and G.~Chauvin\inst{\ref{instf1},\ref{instcl1}} 
   \and R.~Claudi\inst{\ref{insti1}} 
   \and A.~Costille\inst{\ref{instf3}} 
   \and J.-B.~Daban\inst{\ref{instf5}}
   \and M.~de Haan\inst{\ref{instnl1}}      
   \and S.~Desidera\inst{\ref{insti1}}
   \and K.~Dohlen\inst{\ref{instf3}}     
   \and M.~Downing\inst{\ref{insteso1}}
   \and E.~Elswijk\inst{\ref{instnl1}}  
   \and N.~Engler\inst{\ref{instch1}}     
   \and M.~Feldt\inst{\ref{instd1}}      
   \and T.~Fusco\inst{\ref{instf3},\ref{instf6}}     
   \and J.H.~Girard\inst{\ref{insteso2}}      
   \and R.~Gratton\inst{\ref{insti1}}
   \and H.~Hanenburg\inst{\ref{instnl1}}  
   \and Th.~Henning\inst{\ref{instd1}}      
   \and N.~Hubin\inst{\ref{insteso1}}  
   \and F.~Joos\inst{\ref{instch1}}      
   \and M.~Kasper\inst{\ref{insteso1}}   
   \and C.U.~Keller\inst{\ref{instnl3}}      
   \and M.~Langlois\inst{\ref{instf7},\ref{instf3}}      
   \and E.~Lagadec\inst{\ref{instf5}}
   \and P.~Martinez\inst{\ref{instf5}}
   \and E.~Mulder\inst{\ref{instnl1}}  
   \and A.~Pavlov\inst{\ref{instd1}}
   \and L.~Podio\inst{\ref{insti4}}       
   \and P.~Puget\inst{\ref{instf1}} 
   \and S.~Quanz\inst{\ref{instch1}}            
   \and F.~Rigal\inst{\ref{instnl1}}  
   \and B.~Salasnich\inst{\ref{insti1}}
   \and J.-F.~Sauvage\inst{\ref{instf3},\ref{instf6}}        
   \and M.~Schuil\inst{\ref{instnl1}}   
   \and R.~Siebenmorgen\inst{\ref{insteso1}}
   \and E.~Sissa\inst{\ref{insti1}}
   \and F.~Snik\inst{\ref{instnl3}}      
   \and M.~Suarez\inst{\ref{insteso1}}
   \and Ch.~Thalmann\inst{\ref{instch1}}  
   \and M.~Turatto\inst{\ref{insti1}}   
   \and S.~Udry\inst{\ref{instch2}}
   \and A.~van Duin\inst{\ref{instnl1}}  
   \and R.~van Holstein\inst{\ref{instnl3}}
   \and A.~Vigan\inst{\ref{instf3}}  
   \and F.~Wildi\inst{\ref{instch2}} 
          }

\institute{
ETH Zurich, Institute for Particle Physics and Astrophysics, 
Wolfgang-Pauli-Strasse 27, 
CH-8093 Zurich, Switzerland\label{instch1}
\and
NOVA Optical Infrared Instrumentation Group at ASTRON, Oude
Hoogeveensedijk 4, 7991 PD Dwingeloo, The Netherlands\label{instnl1}
\and
Universit\'{e} Grenoble Alpes, IPAG, 38000 Grenoble, France\label{instf1}
\and
CNRS, IPAG, 38000 Grenoble, France\label{instf2}
\and
European Southern Observatory, Alonso de Cordova 3107, Casilla
19001 Vitacura, Santiago 19, Chile\label{insteso2}
\and
Istituto Ricerche Solari Locarno, Via Patocchi 57,
6605 Locarno Monti, Switzerland\label{instch3}
\and
Kiepenheuer-Institut f\"{u}r Sonnenphysik, Schneckstr. 6, D-79104
Freiburg, Germany\label{instd2}
\and
Anton Pannekoek Astronomical Institute, University of Amsterdam,
PO Box 94249, 1090 GE Amsterdam, The Netherlands\label{instnl2}
\and
LESIA, CNRS, Observatoire de Paris, Universit\'{e} Paris Diderot,
UPMC, 5 place J. Janssen, 92190 Meudon, France\label{instf4}
\and
Leiden Observatory, Leiden University, P.O. Box 9513, 2300 RA
Leiden, The Netherlands\label{instnl3}
\and
Laboratoire Lagrange, UMR7293, Universit\'{e} de Nice Sophia-Antipolis, 
CNRS, Observatoire de la C\^{o}te d'Azur, Boulevard de l'Observatoire, 
06304 Nice, Cedex 4, France\label{instf5}
\and
INAF - Osservatorio Astronomico di Roma, via Frascati 33, 
I-00087 Monte Porzio Catone, Italy\label{insti3}
\and
Max-Planck-Institut f\"{u}r Astronomie, K\"{o}nigstuhl 17, 69117
Heidelberg, Germany\label{instd1}
\and
INAF – Osservatorio Astronomico di Padova, Vicolo
dell’Osservatorio 5, 35122 Padova, Italy\label{insti1}
\and
Aix Marseille Universit\'{e}, CNRS, CNES, LAM (Laboratoire
d’Astrophysique de Marseille) UMR 7326, 13388, Marseille,
France\label{instf3}
\and
Unidad Mixta International Franco-Chilena de Astronomia, CNRS/INSU
UMI 3386 and Departemento de Astronomia, Universidad de Chile, Casilla 36-D,
Santiago, Chile\label{instcl1}
\and
European Southern Observatory, Karl Schwarzschild St, 2, 85748
Garching, Germany\label{insteso1}
\and
ONERA, The French Aerospace Lab BP72, 29 avenue de la
Division Leclerc, 92322 Ch\^{a}tillon Cedex, France\label{instf6}
\and
Centre de Recherche Astrophysique de Lyon, CNRS/ENSL
Universit\'{e} Lyon 1, 9 av. Ch. Andr\'{e}, 69561 Saint-Genis-Laval,
France\label{instf7}
\and
INAF - Osservatorio Astrofisico di Arcetri, Largo E. Fermi 5, 
I-50125 Firenze, Italy\label{insti4}
\and
Geneva Observatory, University of Geneva, Chemin des Mailettes
51, 1290 Versoix, Switzerland\label{instch2}
            }

   \date{Received ...; accepted ...}

 
\abstract
{The SPHERE ``planet finder'' is an extreme adaptive
optics (AO) instrument for high resolution and high contrast 
observations at the Very Large Telescope (VLT). We describe the Zurich 
Imaging Polarimeter (ZIMPOL), the visual focal plane subsystem of SPHERE,
which pushes the limits of current AO systems 
to shorter wavelengths, higher spatial resolution, and much 
improved polarimetric performance.}
{We present a detailed characterization of SPHERE/ZIMPOL which 
should be useful for an optimal planning of observations and 
for improving the data reduction and calibration. 
We aim to provide new benchmarks for the performance of
high contrast instruments, in particular for polarimetric differential 
imaging.}  
{We have analyzed SPHERE/ZIMPOL point spread functions 
(PSFs) and measure the normalized peak surface brightness, 
the encircled energy, and the full width half maximum (FWHM) 
for different wavelengths, atmospheric conditions, star brightness, 
and instrument modes. 
Coronagraphic images
are described and the peak flux 
attenuation and the off-axis flux transmission are determined.
Simultaneous images of the coronagraphic focal plane and the pupil plane 
are analyzed and the suppression of the diffraction rings by the pupil stop is
investigated. We compared the performance at 
small separation for different coronagraphs with tests for the binary 
\object{$\alpha$ Hyi} with a separation of 92~mas and a contrast of 
$\Delta m \approx 6^m$. 
For the polarimetric mode we made the instrument calibrations 
using zero polarization and high polarization standard stars 
and here we give a recipe for the absolute calibration of 
polarimetric data. The data show a small ($<1$~mas) but disturbing 
differential polarimetric beam shifts, which can be explained as 
Goos-H\"ahnchen shifts from the inclined mirrors, and we discuss
how to correct this effect. The polarimetric sensitivity 
is investigated with non-coronagraphic and deep, 
coronagraphic observations of the dust scattering around 
the symbiotic Mira variable \object{R Aqr}.} 
{SPHERE/ZIMPOL reaches 
routinely an angular resolution (FWHM) of $22-28$~mas,
and a normalized peak surface brightness 
of ${\rm SB}_0-m_{\rm star}\approx -6.5^m$~arcsec$^{-2}$ for the V-, R- and
I-band. The AO performance is worse for mediocre $\gapprox 1.0''$ 
seeing conditions, faint stars $m_R\gapprox 9^m$, or in the presence of the 
''low wind'' effect (telescope seeing). The coronagraphs are effective
in attenuating the PSF peak by factors of $>100$, and the
suppression of the diffracted light improves the contrast performance
by a factor of approximately two in the separation range $0.06''-0.20''$. 
The polarimetric sensitivity is $\Delta p<0.01$~\% and the 
polarization zero point can be calibrated to better than  
$\Delta p\approx 0.1$~\%. The contrast limits 
for differential polarimetric imaging for the 400~s I-band data of
R Aqr at a separation of $\rho=0.86''$ are for the surface brightness 
contrast 
${\rm SB}_{\rm pol}({\rho})-m_{\rm star}\approx 8^m\,{\rm arcsec}^{-2}$ 
and for the point source contrast
$m_{\rm pol}({\rho})-m_{\rm star}\approx 15^m$ 
and much lower limits are achievable with deeper observations.} 
{SPHERE/ZIMPOL achieves imaging performances in the visual range 
with unprecedented characteristics, in particular very high 
spatial resolution and very high polarimetric contrast. 
This instrument opens up many new research opportunities for
the detailed investigation of circumstellar dust, in scattered 
and therefore polarized light, for the investigation of faint companions, 
and for the mapping of circumstellar H$\alpha$ emission.}

\authorrunning{H.M. Schmid}

\titlerunning{SPHERE / ZIMPOL high resolution polarimetric imager}

   \maketitle

\section{Introduction} \label{Introduction}

The SPHERE ``Planet Finder'' instrument has been successfully 
installed and commissioned in 2014 at the VLT. 
The main task of this instrument is the
search and investigation of extra-solar planets around
bright stars $m_{\rm R}\lapprox 10^m$. Therefore SPHERE
is optimized for high contrast and diffraction limited resolution observation
in the near-IR and the visual spectral region using an extreme adaptive
optics (AO) system, stellar coronagraphs, and three focal plane 
instruments for differential imaging. General technical descriptions
of the instrument are given in \citet{Beuzit08,Kasper12}, and
the SPHERE user manual and related technical 
websites\footnote{www.eso.org/sci/facilities/paranal/instruments/sphere} 
of the European Southern Observatory (ESO). SPHERE is a very powerful
facility instrument which provides a broad suite of sophisticated
instrument modes for the very demanding investigation of 
extra-solar planetary systems. Essentially all of these 
modes also provide unique observing opportunities for the 
study of the immediate circumstellar environment of 
bright stars. Technical results about the on-sky performance 
of the SPHERE instrument are given in \citet{Dohlen16}, and on-sky results 
for the AO-system are described in \citet{Fusco16} and \citet{Milli17}. 
A series of first SPHERE science papers demonstrates
the performance of various observing modes of this instrument 
\citep[e.g.,][]{Vigan16, Maire16a, Zurlo16, Bonnefoy16}. However, 
the SPHERE instrument is complex and therefore
it is appropriate to give more specific descriptions on individual
subsystems and this is the first of a few technical papers  
for the visual focal plane instrument ZIMPOL.  

ZIMPOL, the Zurich
Imaging Polarimeter, works in the spectral range from 500 nm to 900 nm 
and provides, thanks to the SPHERE AO system and visual coronagraph, high
resolution ($\approx$ 20-30 mas) and high contrast imaging 
and imaging polarimetry for the immediate surroundings ($\rho<4$ arcsec)
of bright stars. SPHERE/ZIMPOL includes a very innovative concept
for high performance imaging polarimetry using a fast modulation -
demodulation technique and it is tuned for very high contrast
polarimetry of reflected light from planetary system. Beside this it
can also be used as a high contrast imager offering angular 
differential imaging and simultaneous spectral differential imaging. 

Previous publications on ZIMPOL describe the science goal
\citep{Schmid06a}, the expected performance \citep{Thalmann08}, 
and give reports about the concept of ZIMPOL \citep{Gisler04,
Joos07,deJuanOvelar12}, the instrument design and component tests 
\citep{Roelfsema10,Roelfsema11,Pragt12,Bazzon12,
Schmid12} and system testing \citep{Roelfsema14,Roelfsema16}. Some early
science results based on SPHERE/ZIMPOL observations are given 
in \citet{Thalmann15,Garufi16,Kervella16,Stolker16,Khouri16,Avenhaus17,
Ohnaka17a,Engler17}. \citet{Schmid17} also gives technical information about 
H$\alpha$ imaging and the flux calibration of ZIMPOL data.

Many technical aspects must be considered for carrying out well
optimized observations and calibrations with an 
instrument like ZIMPOL, which combines diffraction-limited imaging 
using extreme adaptive optics, coronagraphy, and differential 
techniques like polarimetry, or angular and spectral 
differential imaging. It is not possible to cover 
all these topics in detail in one paper and therefore we focus
on a basic technical description and on
aspects which are special to SPHERE/ZIMPOL when compared to 
other high contrast instruments. This should serve as a starting point for
potential SPHERE/ZIMPOL users to carry out well optimized
observations and data analyzes for exploiting the full potential
of this instrument. We plan that subsequent papers will address 
other aspects of SPHERE/ZIMPOL, such as
astrometry, precision photometry, a detailed technical
assessment of the high performance polarimetry mode and more. 

This paper is organized as follows. The next section gives
a brief overview on the SPHERE common path and a detailed description
of the ZIMPOL subsystem, imaging properties, the ZIMPOL polarimetry,
the detectors and detector calibrations, and the filters. 
Section 3 characterizes the ``typical'' point spread functions (PSFs)
and describes special cases, like faint stars, poor atmospheric
conditions, or particular instrumental effects. The topic of
Sect.~4 is the SPHERE visual coronagraph and
the comparison of coronagraphic test measurements
taken with different
focal plane masks. SPHERE/ZIMPOL polarimetry is described
in detail in Sect.~5 including the concept for the control
of the polarimetric signal and the correction of the
measurements based on calibrations of the telescope, the instrument,
and the detectors. Further we discuss the polarimetric differential 
beamshift, a disturbing effect which is new for astronomical
optics and which was not anticipated in the design of this instrument.
Then, we illustrate the very good polarimetric performance of 
ZIMPOL with test observations of the system R Aqr. We conclude 
in Sect.~6 with a summary of the most 
outstanding technical properties of SPHERE/ZIMPOL and an 
outline of the new research opportunities offered by 
this instrument. 

\section{The visual channel of SPHERE}

The SPHERE visual channel covers the wavelength range 
from $500-900$~nm and provides observational modes for
imaging, spectral differential imaging, angular differential
imaging and polarimetric differential imaging. The next subsection
gives a brief overview of the SPHERE common path while the
visual focal plane instrument ZIMPOL is described in
detail in the following subsections.

\subsection{Common path and infrastructure - CPI}

\begin{figure}
\includegraphics[trim=3.5cm 13cm 4cm 3.5cm,clip,width=8.8cm]{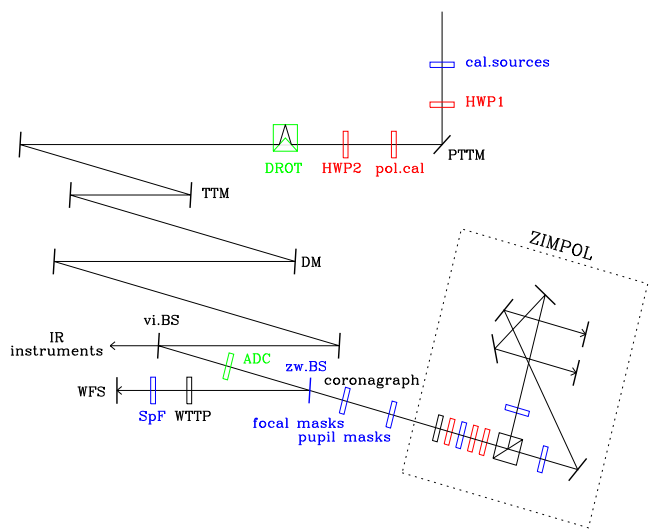}
\caption{Block diagram of the SPHERE common path (CPI) up to
the beam splitter vi.BS and the SPHERE visual channel. The blue color indicates
exchangeable components, green are rotating components, and red components are
only inserted for polarimetry. The ZIMPOL box
is shown in detail in Fig.~\ref{ZIMPOLblock}.}
\label{SPHEREblock}
\end{figure}

Figure~\ref{SPHEREblock} gives a simplified block diagram of 
those parts of the SPHERE main bench or ``common path and 
infrastructure'' (CPI) system which are relevant for visual 
observations. Table~\ref{CPIcomp} lists the components along
the beam indicating the rotating, insertable, and exchangeable components 
and those which are only in the beam for polarimetry (see also the
colors in Fig.~\ref{SPHEREblock}).

The heart of the SPHERE instrument is the extreme adaptive 
optics (AO) system, which corrects for the variable wave-front 
distortions introduced by the rapidly changing Earth's atmosphere.
At the same time the AO corrects also for aberrations introduced 
by the telescope and the SPHERE instrument \citep{Fusco14,
Sauvage16a}. The AO system needs a bright natural guide 
star in the center of the science field as wave 
front probe, preferentially with a brightness $m_{\rm R}\lapprox 10^m$.
The AO performance depends strongly on the atmospheric 
conditions and the guide star brightness \citep{Sauvage16a} as
described in Section~\ref{SectAO}. Essential components of
the AO system are the 
Shack-Hartmann wave front sensor (WFS), which measures the
wave front distortions, the fast high-order deformable 
mirror (DM), the fast tip-tilt mirror (TTM), and the pupil 
tip-tilt mirror (PTTM) which correct for the measured distortions. 

The CPI includes in addition an image derotator (DROT) which can 
be used in three different rotation modes: (i) to stabilize the 
sky image on the detector, (ii) to fix the orientation
of the telescope pupil, or (iii) to keep the instrument
polarization stabilized. The visual-infrared beam splitter (vi.BS) transmits 
long wavelengths $\lambda>950$~nm to the IR science channel 
and reflects the short wavelengths $\lambda<950$~nm
to the wave front sensor arm and ZIMPOL. The IR channel
includes the IR-coronagraph \citep{Boccaletti08} and two 
focal plane instruments, the infrared double beam imager and spectrograph
IRDIS  \citep{Dohlen08,Vigan14} and the integral field spectrograph 
IFS \citep{Claudi08}. 

The visual beam is further split after 
the visual atmospheric dispersion corrector (ADC) by one of two 
exchangeable beam splitters (zw.BS) which reflect part of the 
light to the wave front sensor arm and transmits the other part 
to the visual coronagraph and ZIMPOL. There is a gray beam-splitter
transmitting about 79~\% of the light to ZIMPOL and 21~\% to the
WFS, and a dichroic beam-splitter transmitting
the wavelengths $600 - 680$~nm to ZIMPOL and reflecting the other
wavelengths within the $500 - 950$~nm range, or about 80~\% of
the light depending on the color of the central star, to the WFS.  

The wave front sensor arm includes a tip-tilt plate (WTTP) for the
fine centering of the central AO guide star on a coronagraphic 
focal plane mask, or another position in the field of view 
within about $0.6''$ from the optical axis. In front of the
WFS one can also select between a large, medium or small field mask as 
spatial filter to optimize the AO performance 
\citep{Fusco16}.  

Furthermore, the CPI includes two insertable and
rotatable half-wave plates (HWP1 and HWP2) 
and polarimetric calibration components (pol.cal) for 
polarimetric imaging with ZIMPOL as will be described later in 
Sect.~\ref{Sectpolcontrol}.    

Different calibration light sources and components can be inserted
inside SPHERE at the VLT-Nasmyth focus \citep{Wildi09,Wildi10}.
For the visual science channel there is a flat field source with
a continuous spectrum for detector flat-fielding. This source
can be combined with a mask with a grid of holes
for measurements of the SPHERE/ZIMPOL image scale and distortions.
In addition, there is a point source with a 
continuous spectrum for measurements and checks of the instrument alignment. 
The brightness of the sources can be adjusted with neutral density
filters also located in the calibration unit.  

\begin{table}
\caption{Optical components of the VLT and SPHERE CPI visual path.}
\label{CPIcomp}
\begin{tabular}{p{0.85cm}p{2.5cm}p{1.0cm}p{3.0cm}}
\noalign{\smallskip\hrule\hrule\smallskip}
\hspace{-0.3cm}
Abbr.  & Name                & ~r~~i~~e~p~ & Comment \\
\noalign{\smallskip\hrule\smallskip}
\noalign{\smallskip\noindent {\sl VLT telescope} \smallskip}
\hspace{-0.3cm}
M1,M2 
       & telescope mirrors   & ~r~~~~~~~~~ & alt.-az. orientation  \\
\noalign{\smallskip}
\hspace{-0.3cm}
M3     & telescope mirror    & ~r~~~~~~~~~ & strong pol. effects \\
\noalign{\medskip\noindent {\sl CPI common path} \smallskip}
\hspace{-0.3cm}
  --       & calibration sources & ~~~~i~~e~~~ & point and flat field sources, hole grid \\
\noalign{\smallskip}
\hspace{-0.3cm}
  HWP1   & half-wave plate 1   & ~r~~i~~~~p~ & rotates M3 pol. \\ 
\noalign{\smallskip}
\hspace{-0.3cm}
PTTM   & pupil tip-tilt $\phantom{ww}$ 
                mirror   &             & AO, compensates M3 polarization \\
\noalign{\smallskip}
\hspace{-0.3cm}
--     & pol. calibration $\phantom{w}$ components
                             & ~~~~~i~~e~p~ & pol. characterization \\
\noalign{\smallskip}
\hspace{-0.3cm}
HWP2   & half-wave plate 2   & ~r~~i~~~~p~ & polarization control and switch \\
\noalign{\smallskip}
\hspace{-0.3cm}
DROT   & image derotator     & ~r~~~~~~~~~ &
                                       rotates image, strong pol. effects \\
\noalign{\smallskip}
\hspace{-0.3cm}
TTM    & tip tilt mirror &                 & AO  \\
\noalign{\smallskip}
\hspace{-0.3cm}
DM     & deformable mirror &                 & AO  \\
\noalign{\smallskip}
\hspace{-0.3cm}
vi.BS     & beam splitter       &                 & visual-infrared \\
\noalign{\medskip\noindent {\sl CPI visual common path} \smallskip}
\hspace{-0.3cm}
ADC    & atmospheric dispersion corrector 
                             & ~r~~~~~~~~~  & for visual \\
\noalign{\smallskip}
\hspace{-0.3cm}
zw.BS     & ZIMPOL-WFS beam splitter
                             & ~~~~~~~e~~~  & gray or dichroic BS \\
\noalign{\medskip\noindent {\sl wave front sensor (WFS) arm} \smallskip}
\hspace{-0.3cm}
WTTP   & WFS tip-tilt plate  &                 & source positioning  \\
\noalign{\smallskip}
\hspace{-0.3cm}
SpF    & WFS spatial filter  & ~~~~~~~e~~~ & AO tuning  \\
\noalign{\smallskip}
\hspace{-0.3cm}
WFS    & WFS-detector        &              & AO tuning \\
\noalign{\medskip\noindent {\sl CPI visual coronagraph} \smallskip}
\hspace{-0.3cm}
--     & focal plane masks   & ~~~~~~~e~~~ & exchange wheel \\
\noalign{\smallskip}
\hspace{-0.3cm}
--     & pupil masks         & ~~~~~~~e~~~ & exchange wheel \\
\noalign{\smallskip\hrule\smallskip}
\end{tabular}
\tablefoot{This incomplete list gives all important components 
for visual science observations. ``Short'' are
abbreviations as used in this paper and columns 3 to 7
indicate whether the components are rotating (r), insertable (i), 
exchangeable (e), or/and only inserted for visual polarimetry (p).}
\end{table}

\subsection{The Zurich Imaging Polarimeter}
\label{SectZIMPOL}
\begin{figure}
\includegraphics[trim=3.5cm 13cm 4cm 3cm,clip, width=10.0cm]{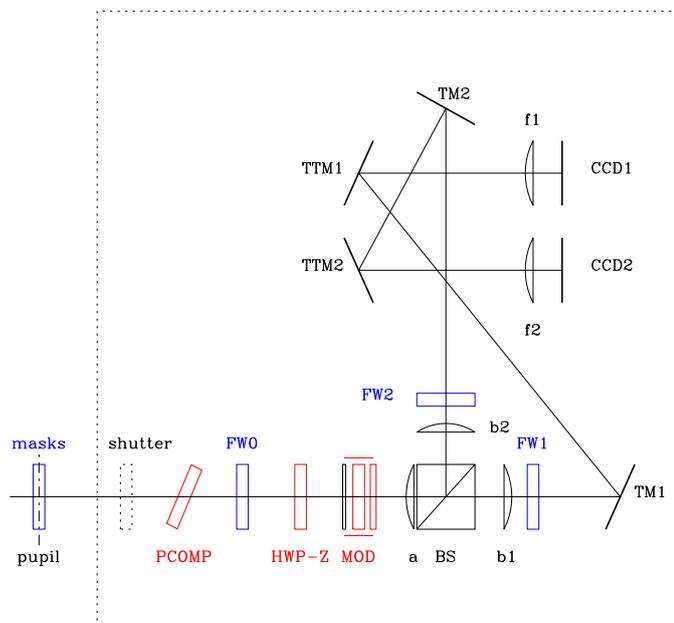}
\caption{Block diagram for ZIMPOL with exchangeable components plotted
in blue, while red components are only inserted for polarimetry. }
\label{ZIMPOLblock}
\end{figure}

A block diagram for the Zurich IMaging POLarimeter (ZIMPOL)
is shown in Fig.~\ref{ZIMPOLblock} and a list of all components
is given in Table~\ref{ZIMPOLcomp}. ZIMPOL is a two arm imager
with a polarization beam splitter (BS). The ZIMPOL common
path consists of a collimated beam section with an intermediate 
pupil just before ZIMPOL, at the position of the 
coronagraphic pupil mask wheel. This pupil has a diameter
of 6~mm and it defines the 
interface between CPI and ZIMPOL. Polarimetric components can be inserted
and removed in the ZIMPOL common path without changing the  
image focus. The following subsections describe the imaging 
properties of this setup, the ZIMPOL 
polarimetric principle, the polarimetric components, the special
detector properties, and the ZIMPOL filters. 
Previous publications on ZIMPOL give more information about the 
opto-mechanical design \citep{Roelfsema10}, optical alignment
procedures \citep{Pragt12}, test results at various phases of the project 
\citep{Roelfsema11,Roelfsema14,Roelfsema16},
the detectors \citep{Schmid12}, and the polarimetric calibration 
concept \citep{Bazzon12}.

\subsubsection{ZIMPOL imaging properties}

The ZIMPOL subsystem is optimized for polarimetric imaging but
provides at the same time also very good imaging
capabilities. In this section we describe the imaging properties 
of ZIMPOL, which apply for imaging
and polarimetric imaging. For polarimetry, additional 
components are inserted in CPI and ZIMPOL and different detector 
modes are used, which
reduce the overall instrument throughput by about
a factor 0.85 with respect to non-polarimetric imaging.  
     
For imaging the three red polarimetric
components in the ZIMPOL block diagram (Fig.~\ref{ZIMPOLblock})  
are removed from the beam and there remains 
only the shutter and filter wheel FW0 in the common path.
The polarization beam splitter creates then the
two camera arms 1 and 2, each equipped with its own 
filter wheel FW1 or FW2, own detectors CCD1 or CCD2, 
own imaging optics (lenses b1,f1 or b2,f2), and own
movable folding mirrors TM1,TTM1 or TM2,TTM2 for 
dithering and the selection of the detector field of view.

\begin{table}
\caption{List of all components in ZIMPOL.}
\label{ZIMPOLcomp}
\begin{tabular}{p{1.0cm}p{2.5cm}p{1.0cm}p{2.7cm}}
\noalign{\smallskip\hrule\smallskip}
\hspace{-0.3cm}
Abbr.  & Name               & ~~r~~i~~e~p~ & Comment \\
\noalign{\smallskip\hrule\smallskip}
\noalign{\smallskip\noindent {\sl ZIMPOL common path} \smallskip}
\hspace{-0.3cm}
--     & shutter             &            & cut transfer smear  \\
\noalign{\smallskip}
\hspace{-0.3cm}
{PCOMP\hspace{-0.2cm}}    
        & polarization$\qquad$ compensator
                             & ~r~~i~~~~p~ & 
                                compensate DROT $\quad$ polarization   \\
\noalign{\smallskip}
\hspace{-0.3cm}
FW0    & filter wheel 0
                             & ~~~~~~~e~~~ & 
                               color filters, NDs, $\quad$ pol. calibration \\
\noalign{\smallskip}
\hspace{-0.3cm}
HWPZ   & half wave plate $\qquad$ ZIMPOL
                             & ~r~~i~~~~p~ & 
                                rotate polarization, $\quad$ calibrations \\
\noalign{\smallskip}
\hspace{-0.3cm}
MOD    & modulator (FLC) $\qquad$
         and 0-HWP           & ~~~~i~~~~p~ & 
                              pol. modulation, $\phantom{+}$ 
                               blocking filter (BF) \\
\noalign{\smallskip}
\hspace{-0.3cm}
BS     & polarization beam splitter
                             &        & with camera lens ``a'' \\
\noalign{\medskip\noindent {\sl ZIMPOL arms 1 and 2} \medskip}
\hspace{-0.3cm}
FW1/2  & filter wheels 1,2 
                             & ~~~~~~~e~~~  &   
                               color filters,~~~ $\qquad$ pupil lens in FW2  \\
\noalign{\smallskip}
\hspace{-0.3cm}
b1/2     & camera lenses ``b''  &        &   \\
\noalign{\smallskip}
\hspace{-0.3cm}
TM1/2    & tip-mirrors 1,2      &         & mask illumination \\
\noalign{\smallskip}
\hspace{-0.3cm}
{TTM1/2\hspace{-0.2cm}} & tip-tilt mirrors 1,2 &         & dithering, offsets \\
\noalign{\smallskip}
\hspace{-0.3cm}
f1/2     & field lenses 1,2     &       & mask illumination \\
\noalign{\smallskip}
\hspace{-0.3cm}
CCD1/2  & detectors 1,2   &        & pol. demodulation \\ 
\noalign{\smallskip\hrule\smallskip}
\end{tabular}
\tablefoot{``Short'' gives abbreviations as used 
in this paper. Columns 3 to 6
indicate whether the components are rotating (r), insertable (i), 
exchangeable (e), or/and only inserted for visual polarimetry (p).}  
\end{table}
 
The beam-splitter includes on the entrance side lens ``a'', a first
common component of the camera optics. It is glued onto the
beam splitter in order to avoid back-reflections
(and ghost images) from the surfaces of the beam-splitter. Lens ``a''
produces together with ``b1'' or ``b2'' the converging beams 
with an $f$-ratio of 221 producing an image scale
of 0.12 arcsec/mm or a pixel scale of about 3.6~mas/pixel on the
CCD detectors. The detectors have an active area of about 
$3.0\times 3.0$~cm$^2$ or about $1000\times 1000$ pixels 
covering a detector field of view of
about $3.6''\times 3.6''$. There are optical image distortions and
the dominant effect is an anamorphism which originates from the 
SPHERE common path optics located after the image derotator. 
This stretches for ZIMPOL the image scale expressed as mas/pix  
by a factor of about 1.006 in the detector row or x-direction, which
is perpendicular to the CCD charge shifting direction. In field 
stabilized observations without field angle offset a square pixel
covers an area, which is slightly elongated in east-west direction when
projected onto the sky, like for the IRDIS and IFS 
near-IR instruments \citep[see also][]{Maire16b}. Detailed 
measurements of these distortions and the derivation of
an accurate astrometric calibration for ZIMPOL will be described in 
Paper~II (Ginski et al., in preparation).
 
ZIMPOL allows imaging in two channels using either a filter 
in the common path wheel FW0, or combining filters
from the wheels FW1 in arm 1 and FW2 in arm 2 (see Sect~\ref{SectFilters}). 
Differential spectral imaging can be achieved by using different filters in
FW1 and FW2. The filters in FW1 and FW2 can be combined
with a neutral density filter located in FW0 in order to avoid
detector saturation of bright sources. 

\begin{figure}
\includegraphics[trim=0.5cm 0.5cm 2.2cm 1.5cm,clip,width=8.8cm]{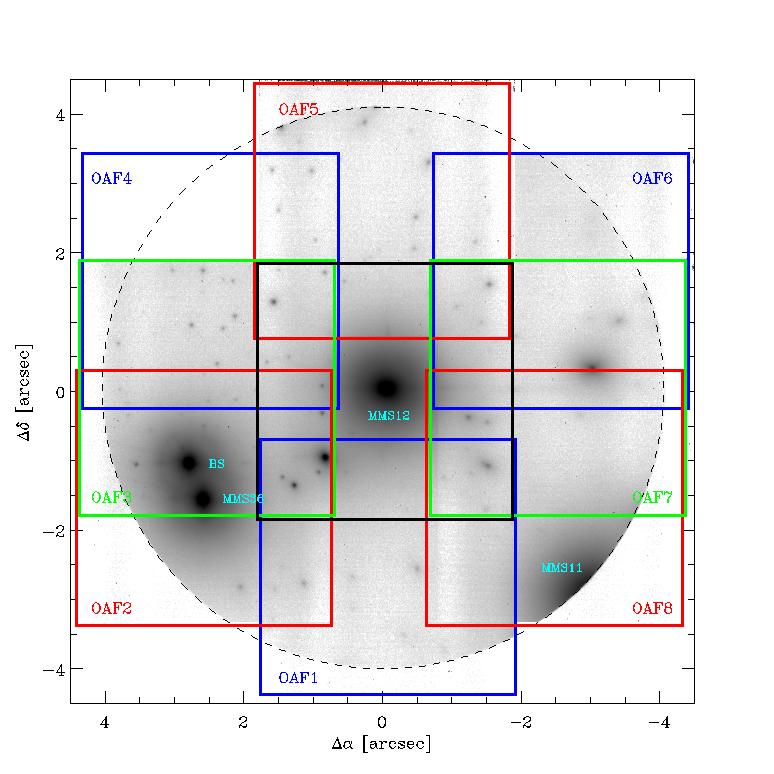}
\caption{Full ZIMPOL instrument field for the 
astrometric field 47 Tuc. The central detector field is
plotted in black, while colors are used for the 8 pre-defined 
off-axis fields OAF1-OAF8. The off-axis fields have been 
shifted slightly (red to the left, blue to the right, 
and green up) for better visibility. No data are available for OAF4 because
of an observational error.} 
\label{Figfields}
\end{figure}

\paragraph{Off-axis fields.}
The SPHERE/ZIMPOL optical field of view has a diameter of $8''$ 
(6.67 cm) and is about four times larger than the detector field of view. 
This $8''$-field is defined by the wide field ``WF''
focal plane masks in the coronagraph. 
To access the whole field of view offered by SPHERE/ZIMPOL, 
it is possible to move the image
on the detector with the tip-tilt mirrors TTM1 or TTM2 and tip mirrors 
TM1 or TM2 in the two arms. Field lenses f1 or f2 and tip-mirrors 
are required to achieve,
also for off-axis fields, a perpendicular illumination of the 
cylindrical micro-lenses and stripe masks in front of the 
special ZIMPOL detectors (see Sect.~\ref{Sectdetectors}).
To simplify the operation of ZIMPOL the instrument software allows
only identical dithering and field offsets in arm 1 and arm 2. 
Optimized mirror settings have been pre-defined and tested for
observations of the field center and eight off-axis fields 
(OAF1 - OAF8). The selectable off-axis fields are shown in 
Fig.~\ref{Figfields} and approximate values for the offsets are
given in Table~\ref{OAFtab}. The off-axis fields avoid the central
star and the surrounding $\rho\lapprox 0.8''$ strong light halo.
This enables long exposures with broad-band filters in the off-axis
fields without saturation by light from the typically
much brighter central star. 

The sky region shown in Fig.~\ref{Figfields} is the 
SPHERE astrometric reference field 47 Tuc = NGC 104 \citep{Maire16b}.  
This field is centered on the bright asymptotic giant 
branch (AGB) star \object{2MASS J$00235767-7205296$}, which is also
star ``MMS12'' in the mid-IR study of \citet{Momany12}. Also ``MMS36'',
the southern star of the pair 
in the SE and the star ``MMS11'' (also \object{2MASS J$00235692-7205325$}) 
just outside the field in the SW are bright AGB stars. 
The northern star of the SE pair is the well studied post-AGB star 
\object{Cl* NGC 104 BS} (BS for bright star) with spectral type B8 III,
the brightest star in 47 Tuc in the V-band $V=10.7^m$ and even
more prominent in the near and far UV \citep[e.g.,][]{Dixon95,Schiavon12}. 
The catalog of \citet{McLaughlin06} and a study of
A. Bellini {\citep[private communication, see also][]{Bellini14}}
provide accurate HST astrometry
of most stars visible in Fig.~\ref{Figfields} and they can be used
for the astrometric calibrations for the central field, but also the
off-axis fields of ZIMPOL. 

\begin{table}
\caption{Approximate offsets for the centers of the off-axis fields (OAF).}
\label{OAFtab}
\begin{tabular}{lccccc}
\noalign{\smallskip\hrule\smallskip}
Field    & \multispan{2}{\hfil Offset in pixels \hfil} 
                  & \multispan{2}{\hfil Offset in arcsec \hfil} & on sky \\ 
         & $\Delta x$    & $\Delta y$               
                  & $\Delta \alpha$   & $\Delta \delta$  &  \\
\noalign{\smallskip\hrule\smallskip}
center   &   0           &    0       &    0         &   0        \\
OAF1     &   0           & $-715$     & $0.00''$     & $-2.57''$  & S  \\   
OAF2     &  $-715$       & $-428$     & $+2.57''$    & $-1.54''$  & E-SE \\    
OAF3     &  $-715$       & 0          & $+2.57''$    & $0.00''$   & E   \\
OAF4     &  $-715$       & $+428$     & $+2.57''$    & $+1.54''$  & E-NE \\
OAF5     &   0           & $+715$     & $0.00''$     & $+2.57''$  & N  \\   
OAF6     &  $+715$       & $+428$     & $-2.57''$    & $+1.54''$  & W-NW \\    
OAF7     &  $+715$       & 0          & $-2.57''$    & $0.00''$   & W   \\  
OAF8     &  $+715$       & $-428$     & $-2.57''$    & $-1.54''$  & W-SW \\     
\noalign{\smallskip\hrule\smallskip}
\end{tabular}
\tablefoot{Pixel offsets $\Delta x$ and $\Delta y$ refer to
preprocessed images with image flips applied to cam1 and cam2 images.  
Columns~4, 5 and 6 give the on-sky offsets 
for the right ascension $\Delta\alpha$ and declination $\Delta\delta$
for field stabilized observations taken without field position angle offset 
as shown in Fig.~\ref{Figfields}.
The actual offsets deviate by about $\pm 3$~pixels or $\pm 0.01''$
from the given design values.}  
\end{table}

\paragraph{The ZIMPOL image rotation modes.}
The SPHERE bench is fixed to the VLT Nasmyth platform A of UT3 and 
the sky image rotates in the telescope focus (``tf'') or 
the entrance focus of the instrument like 
\begin{equation}
\theta_{\rm tf}(t) = \theta_{\rm para}(t)+a(t)
\end{equation}
where $\theta_{\rm para}$ is the parallactic angle and $a$ the telescope 
altitude. The field orientation on the detector is defined
by the image rotation introduced by DROT. 

For the ZIMPOL imaging mode one can choose between field stabilized
and pupil stabilized observations. In the first case the sky image is 
fixed and N$_{\rm CCD}$, the north direction on the CCD-detector 
after applying image flips in the data preprocessing, is given by
\begin{equation}
{\rm N}_{\rm CCD} = \theta_0 - \delta\theta\,,
\label{Eqthoffset1}
\end{equation}
where the angle ${\rm N}_{\rm CCD}$ is measured from the vertical 
or y-direction in counter-clockwise direction. There is a small 
rotational offset $\theta_0\approx 2^\circ$,
which is not exactly identical for CCD1 and CCD2, and which can 
be accurately determined with astrometric calibrations (Ginsky et al.,
in prep.).  
The term $\delta\theta$ stands for the user defined 
field orientation angle offset (see Fig.~\ref{CoroRotation}). 

In pupil stabilized mode the telescope pupil 
is stabilized on the detector and the field rotates 
in step with the parallactic angle. 
This may introduce image smearing for long integrations, 
if the rotation during $t_{\rm DIT}$ is too large.

For polarimetric imaging one can choose between static 
derotator mode, called P1, and field stabilized mode P2. 
In P1 mode the derotator is fixed and the field rotates on the CCD as
\begin{equation} 
{\rm N}_{\rm CCD}(t) = \theta_{\rm para}(t)+a(t) +\theta_0
\end{equation}
but also the telescope pupil moves with $a(t)$. 
The advantage of this mode is an accurate
calibration of the telescope polarization, because derotator 
and all following components (except for the atmospheric dispersion corrector) 
are in a fixed orientation. The rotation law for P2 is identical to
the field stabilized mode in imaging. 

\subsubsection{The ZIMPOL principle}
\label{SectZIMprinciple}
\begin{figure}
\includegraphics[trim=3.2cm 14.8cm 2.5cm 3.5cm,clip, width=8.8cm]{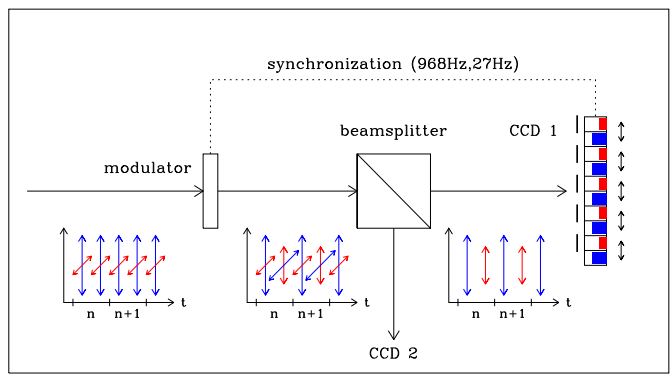}
\caption{ZIMPOL principle: The modulator switches in one cycle $n$ the
polarization direction between $I_\perp$ and $I_\parallel$. 
The polarization beam splitter selects for each channel only one 
polarization mode so that a polarization signal is converted 
into an intensity modulation. The masked CCDs demodulate the signal
with charge-shifting, which is synchronized
with the modulator.}
\label{ZIMPOLprinciple}
\end{figure}

Strong, variable speckles from the bright star are the main problem 
for high contrast imaging from the ground. 
ZIMPOL is optimized for high precision imaging
polarimetry under such conditions because the speckle noise
can be strongly reduced with an imaging polarimeter based on a fast 
modulation-demodulation technique. A polarization modulator and
a polarizer (or a polarization beam splitter) convert the fractional
polarization signal into a fractional modulation of the intensity 
signal, which is then measured by a masked, demodulating imaging detector as
shown schematically in Fig.~\ref{ZIMPOLprinciple}.

A polarimetric modulation with a frequency of about 1~kHz 
is sufficient to ``freeze'' the speckle variations introduced 
by the atmospheric turbulence in the differential polarimetric measurement. 
This requirement is realized in SPHERE/ZIMPOL with a 
modulation using a ferro-electric liquid crystal (FLC) retarder and CCD 
array detectors for the demodulation. 
On the CCD ``every second row'' is masked so that photo-charges 
created in the open rows during
one half of the modulation cycle are shifted for the second half of the
cycle to the next masked row, which is used as temporary buffer.
The charge shifting is synchronized with the modulator switching,
so that two images, the ``even-row'' and the ``odd-row'' subframes, 
with opposite linear polarization modes 
$I_\perp$ and $I_\parallel$ are built up. Photo-electrons can be 
collected during hundreds or thousands of 
modulation cycles before the detector is read out.
The difference of the two images is proportional to the polarization
flux and the sum proportional to the intensity
\begin{equation}
P^Z=I_\perp-I_\parallel \quad {\rm and} \quad I=I_\perp+I_\parallel\,.
\label{EqPZ}
\end{equation}

\noindent
Important advantages of the ZIMPOL technique are:
\begin{itemize}
\vspace{-0.2cm}
\item{} the images for opposite polarization states are created 
essentially simultaneously because the modulation is faster 
than the speckle variations and instrument drifts,
\item{} allows a fast modulation without high frame rates so that
the read-out noise is low, 
\item{} both images are recorded with the same pixels, reducing significantly
flatfielding requirements and alignment issues for the 
difference image, 
\item{} differential aberrations between the two images with opposite 
polarization are very small.
\end{itemize}

ZIMPOL was initially developed for solar applications \citep{Povel90,Povel95}
and it proved to be an extremely sensitive
technique for polarimetric imaging of the Sun 
\citep[e.g.,][]{Stenflo96,Stenflo97,Gandorfer97}. 
For SPHERE, the ZIMPOL concept was adapted for a
higher transmission, longer integration times and broad band capabilities
with the development of a new type of achromatic polarization modulator 
\citep{Gisler03,Gisler04}, a CCD detector with more pixels, and a stripe 
mask equipped with a cylindrical micro-lens array to reduce the photon 
loss on the stripe mask \citep{Schmid12}.   

ZIMPOL is a single beam technique which could also be used with a
polarizer, but then 50~\% of the light is lost. With a 
polarization beam splitter all light is used and the same
polarimetric information is encoded in both channels. The two channels
can be combined but they can also be used as separate but simultaneous
measurements, for example by using different filters in the two arms.

\subsubsection{ZIMPOL polarimetric setup}
\label{Sectpolsetup}
This section describes the properties of the polarimetric components
in ZIMPOL while Sect.~\ref{Sectpolcontrol} explains how they are 
used to obtain well calibrated polarization data. ZIMPOL
has three polarimetric components, the ferro-electric liquid crystal (FLC)
modulator assembly, the polarization 
compensator (PCOMP) and a half-wave plate (HWPZ) 
which are only inserted for polarimetric observations (indicated in red
in Fig.~\ref{ZIMPOLblock}). Further polarimetric components are 
the polarization beam splitter and the polarimetric calibration components
in FW0. The other key elements for polarimetry are the demodulating CCD 
detectors described in the next section. 

\paragraph{PCOMP and HWPZ.}
The polarization compensator plate (PCOMP) is required to reduce 
the instrument polarization of about $2-3~\%$ introduced by 
the DROT (derotator) mirrors in CPI. A low instrument polarization
is important for a good charge trap compensation and
for reducing the impact of non-linearity of the detectors on
the achievable polarization sensitivity. 
PCOMP is an uncoated, inclined glass plate (fused silica $n=1.45$ to 1.46), 
where the two surfaces deflect more $I_\perp$ than $I_\parallel$ 
so that a linear polarization of the incoming beam 
$p^i=(I^i_\perp-I^i_\parallel)/(I^i_\perp+I^i_\parallel)$ can be 
reduced for the transmitted beam $p^t=p^i-\Delta p$, 
if the inclined plate has a perpendicular 
orientation $\theta_{\rm PCOMP}=\theta_{\rm pol}+90^\circ$.
The polarimetric compensation 
depends on the inclination angle of the plate, according to the
Fresnel formulae \citep[e.g][]{Born99,Collett92}. 
The PCOMP orientation rotates in step with DROT
and the inclination can be adjusted. In the commissioning a 
good compensation for the entire wavelength range and all DROT 
orientations was found for $i_{\rm PCOMP}=25^\circ$. For this 
angle the two surfaces deflect together 
about $I_\perp^r=0.089 I_\perp^i$ and $I_\parallel^r=0.051 I_\parallel^i$.
For small polarization $p^i<5~\%$ the total transmission 
is $I^t\approx 0.93 I^i$ and the polarization in the transmitted
beam is reduced according to $p^t\approx p^i-2.0~\%$

ZIMPOL has two DROT-modes for polarimetry, P1-mode with fixed DROT 
and a rotating sky field on the detector, and P2-mode with rotating
DROT and fixed field on the detector. In 
P1-mode, DROT and PCOMP are both fixed and in P2 they
rotate synchronously. In P2-mode an additional
achromatic half-wave plate (HWPZ) must be 
introduced, to rotate the polarization to be measured, 
which passes the DROT as $I_\perp$ and $I_\parallel$, into 
the $I_\perp$ and $I_\parallel$ orientation of the polarimeter. 
For this, also HWPZ must be on a rotational stage and its 
orientation is $\theta_{\rm HWPZ}=\theta_{\rm DROT}/2$.  

ZIMPOL measures the linear polarization 
$P^{\rm Z}=I_\perp-I_\parallel$ perpendicular and parallel 
to the SPHERE bench only. HWP2 and DROT 
in the common path and HWPZ within ZIMPOL are responsible for
the correct rotation of the sky polarization 
into the ZIMPOL system. The achromatic half-wave plates 
are made of quartz and MgF$_2$ retarders in optical
contact\footnote{from Bernhard Halle Nachfl.}.

\paragraph{The FLC modulator.}
The ferro-electric liquid crystal (FLC) polarization modulator is a key
component in ZIMPOL. An FLC retarder is a zero order half wave plate
where the orientation of the optical axis can be switched by about
$45^\circ$ by changing the sign of the applied voltage 
through the FLC layer.
A fast switch time in the range of 50~$\mu$s is 
required to achieve a good efficiency for a modulation cycle 
frequency on the order 1~kHz. The selected FLC retarder achieves 
this fast switching only when the operation temperature is in the 
range $20^\circ-30^\circ$ Celsius. Also the switching angle depends 
slightly on temperature \citep{Gandorfer99,Gisler03}. 
Because the FLC retarder has 
to be warmer than the rest of the SPHERE instrument ($0^\circ-15^\circ$~C)
it is thermally insulated in a vacuum housing to avoid
air turbulence in the instrument. 

The FLC retarder is a half wave plate only at the 
nominal wavelength $\lambda_0$ because the retardance varies 
roughly like $1/\lambda$ and therefore the modulation efficiency
depends strongly on wavelength. In order to cover the broad wavelength 
range of ZIMPOL, a combined design with a static zero order half wave plate 
(0-HWP) is used which reduces significantly 
the chromatic dependencies of the modulator
\citep{Gisler04,Bazzon12}. The zero-order half wave plate is placed 
on the exit window of the FLC modulator housing.
The entrance window includes an out-of-band blocking filter as described
in Sect.~\ref{SectFilters}. 

\paragraph{Polarization beam splitter.} The polarization beam splitter
is a cube made of two $90^\circ$-prisms of Flint-glass in optical 
contact. The transmitted beam consists of more than 
99.9~\% of $I_\parallel$, while
the reflected beam consists of about $97~\%$ of $I_\perp$ and
$3~\%$ of $I_\parallel$ light. The polarimetric efficiency of the ZIMPOL
arm 2 is therefore about 
$\epsilon_{\rm arm2}=(I_\perp-I_\parallel)/(I_\perp+I_\parallel)\approx 94~\%$ 
or about 6~\% lower than for arm 1 while the total intensity 
throughput is 6~\% higher than for arm 1. 

The polarization beam splitter has a lens ``a'' glued to the first 
surface. Together with lenses ``b1'' and ``b2'' in the two arms 
they form the camera lenses. Component ``a'' is
added to the beam-splitter to reduce the impact of back-reflections 
from the flat beam-splitter surfaces into the collimated beam where
subsequent reflections could produce disturbing ghost images.

\paragraph{Polarimetric calibration components.} The
filter wheel FW0 includes three
polarization calibration components: a linear polarizer,
a quarter wave plate, and a circular polarizer which are 
described in \citet{Bazzon12}.  
The linear polarizer produces essentially a 100~\%
polarized illumination of ZIMPOL which is
used to determine and calibrate the modulation efficiency 
$\epsilon_{\rm mod}$ as described in Sect.~\ref{Sectmodeff}. 

The quarter wave plate and the circular polarizer are
used for polarization cross-talk measurements and other 
instrument tests. Identical polarization calibration components 
like for ZIMPOL are also available in the SPHERE common path 
(cal.pol in Fig.~\ref{SPHEREblock})
and together they allow a detailed characterization of the
instrument polarization of the entire SPHERE/ZIMPOL visual channel
as outlined in \citet{Bazzon12}. 
   
\subsection{ZIMPOL detectors}
\label{Sectdetectors}
The ZIMPOL CCD detector properties are quite special because of the
demodulation functionality required for the polarimetric mode. This
section gives a brief description of detector properties and the
resulting observational data, while more details are given 
in \citet{Schmid12}. 

The detectors are two back-illuminated,
frame transfer CCDs with an imaging area of 2k~$\times$~2k 
and 15~$\mu$m $\times$ 15~$\mu$m pixels. The CCDs
are operated like 1k~$\times$~1k pixel frame transfer CCDs with 2 x 2 pixel 
binning providing an effective pixel size of 30 $\mu$m $\times$ 30 $\mu$m. 
In the following a ``pixel'' always means this
30~$\mu$m binned pixel. The quantum efficiencies of the (bare) CCDs 
are about 0.95, 0.90 and 0.65 at $\lambda=600$~nm, 700~nm and 800~nm
respectively, while the photo-response non-uniformity is $\leq 2~\%$ 
up to 800~nm as measured with 5~nm band widths. For longer
wavelengths fringing is visible, which becomes
dominant for $\lambda > 750$~nm, but remains for 
$\Delta\lambda=5$~nm at a level $< 4~\%$ up to 900~nm.             

\begin{figure}
\includegraphics[trim=2.5cm 15.2cm 2.5cm 3cm,clip, width=9.2cm]{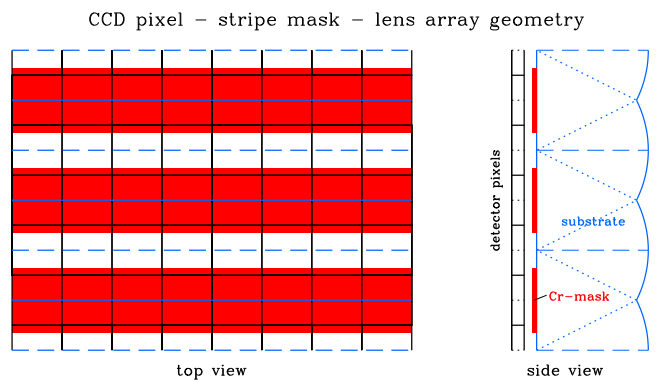}
\caption{Schematic setup of a small section of the ZIMPOL 
detector assembly with the stripe mask (red), the cylindrical 
micro lens array with the dashed focus line (blue) and 
the detector pixels (black).}
\label{detector}
\end{figure}

In front of the CCD imaging area is 
a stripe mask and a cylindrical micro-lens array
as illustrated in Fig.~\ref{detector}.  
There is a substrate with a photo-lithographic mask on the backside 
with 512 stripes and a width of 40 $\mu$m which are separated by 
gaps of 20 $\mu$m. On the
front side are an equal number of cylindrical micro-lenses with a width
of 60 $\mu$m which focus the light through the gaps onto the CCD. The
stripes and micro-lenses assembly are fixed about 10 $\mu$m above the CCD and
they are aligned with the pixel rows of the detector.   

On one detector there are 512 open rows with 1024 pixel each 
separated by one masked row. In polarimetric mode the
photo-charges created in the illuminated rows are shifted up and 
down in synchronization with the polarimetric modulation and 
the final frame consists of an ``even rows'' subframe $I_e$ for one 
polarization state $I_\perp$, and 
an ``odd rows'' subframe $I_o$ for the opposite polarization state $I_\parallel$
with 1024~$\times$~512 pixels each.

In imaging mode there is no charge shifting and the final frame
consists of an ``illuminated rows'' subframe with 1024 x 512 pixels 
with the scientifically relevant data and a ``covered rows'' subframe 
with some residual signal from light or photo-charges which diffused 
from the open rows into the covered rows.

\begin{figure}
\includegraphics[trim=0.1cm 0.1cm 1.0cm 0.1cm,clip,width=8.8cm]{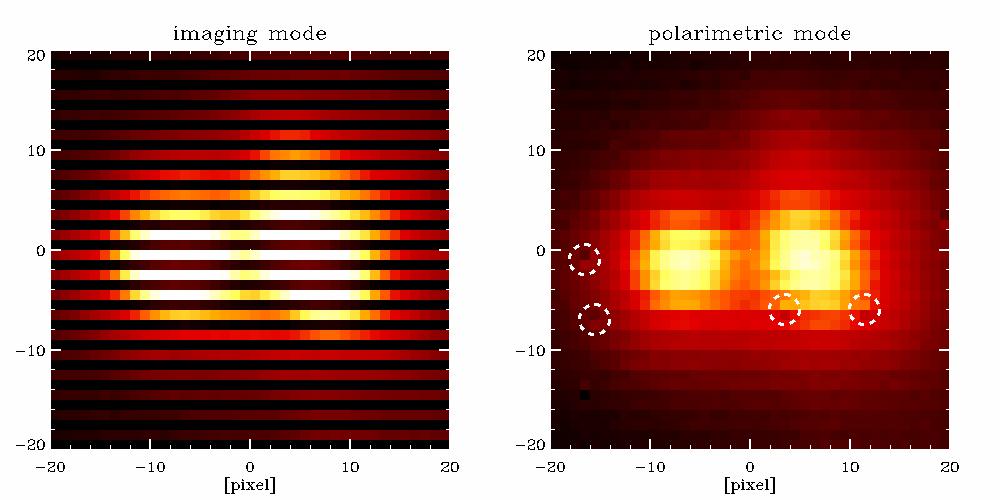}
\caption{Raw counts for the central $40\times 40$ pixel regions of the
H$\alpha$ images for the central \object{R Aqr} binary, separation 45~mas, taken
in imaging mode (left) and in polarimetric mode (right). The dotted
circles in the polarimetric data indicate spurious charge 
shifting effects.}
\label{rawcounts}
\end{figure}

Figure~\ref{rawcounts} shows small sections of two raw H$\alpha$ frames 
of the central R Aqr binary taken in polarimetric and imaging mode
\citep[from][]{Schmid17}. 
In imaging mode, there is a large illumination difference between
the open rows and covered rows. The entire image is illuminated
in polarimetry despite the stripe mask because of the charge shifting 

The extracted subframes
(even/odd or illuminated/covered rows) have an unequal dimension
with twice the number of pixels in the row direction. The raw frames can be
converted into a 512~$\times$~512 pixel square format by a two pixel binning in
row direction or into a 1024~$\times$~1024 pixel square format by 
inserting everywhere between two adjacent rows one additional row 
with a flux conserving linear interpolation in column direction.

The pixel scale for even/odd or the illuminated/covered subframes 
are about 3.6 mas in row direction and 7.2 mas in column
direction. The two pixel or Nyquist sampling in column direction 
is therefore $14.4$~mas and this corresponds to the
diffraction limit of the VLT telescope $\lambda/D$ at a wavelength
of 570~nm. However, the spatial resolution achieved 
with ZIMPOL is typically above 20~mas because of telescope and
instrument vibrations, optical aberrations and other effects,
so that the pixel sampling in column direction is also adequate
for the shortest wavelength filter V\_S (532~nm).
Nonetheless, it can be beneficial for the data analysis 
to arrange interesting structures of the target along the
better sampled row direction, like the close binary in Fig.~\ref{rawcounts}.

\subsubsection{Detector modes} 
Three detector modes are fully characterized
and tested for ZIMPOL, one for imaging and two for imaging polarimetry.
Important parameters for these modes are given in 
Table~\ref{Detmodes}. The imaging and the fast polarimetry modes are
conceived for high contrast observations of bright stars and the slow
polarimetry mode for longer integrations of images with lower
illumination levels.

The two ZIMPOL detectors are controlled for all detector modes
strictly in parallel. The CCDs are operated as frame transfer 
devices where read-out of a frame in the shielded read-out area occurs during
the integration (and demodulation in polarimetric mode) of the next
frame in the imaging area \citep{Schmid12}.
Therefore, the shortest possible integration $t_{\rm DIT}^{\rm min}$ is essentially
the read-out time for one frame. After the integration a frame 
is shifted with a fast frame transfer from the image area
to the read-out area and a new cycle of integration and read-out 
starts. The fast frame transfer introduces a small detector overhead
of $t_{\rm ft}$ per frame and a smearing of the image which can be easily
recognized as trails in column direction in short integrations of 
bright or saturated point sources (see Fig.~\ref{PSFVI}).
ZIMPOL includes a fast shutter to suppress the frame transfer smearing
but it has been disabled because of technical problems. 
The ZIMPOL design does not require a shutter for the basic
instrument modes. 

Table~\ref{Detmodes} lists the electronic parameters for the different
CCD modes, like read-out noise (RON), dark current (DC), and pixel gain.
RON and DC need to be measured regularly for the
calibration of the science data. The two CCDs are read out 
each by two read-out registers,
one for the left half and the other for the right half of one detector.
Each read-out register has a slightly different bias 
level which can vary by a few counts from frame to frame
if the device is run with high frame rates. 
For this reason, the read-out registers provide for each half row 
also 25 pre-scan and 41 overscan pixel readings for each frame 
to correct properly the bias level in the data reduction. The raw frame format
resulting from this read-out scheme is described in \citet{Schmid12}. 

\begin{table}
\caption{Parameters for the ZIMPOL imaging and 
polarimetric detector modes.}
\label{Detmodes}
\begin{tabular}{lcccc}
\noalign{\smallskip\hrule\smallskip}
Parameter   & Unit &  Imaging      & \multispan{2}{\hfill Polarimetry\hfill}  \\
            &      &               & fast       & slow           \\
\noalign{\smallskip\hrule\smallskip}
min. DIT $t_{\rm DIT}^{\rm min}$   
            & s          & 1.1       & 1.1        & 10            \\
transfer time $t_{\rm ft}$
            & ms         & 56         & 56        & 74         \\
max. frame rate
            & Hz         & 0.86       & 0.86       & 0.1            \\
read out freq.
            & kpix/s     & 625       & 625        & 100            \\
pixel gain 
            & e$^-$/ADU  & 10.5      & 10.5       & 1.5            \\
read out noise 
            & e$^-$/pix  & 20        & 20         & 3              \\ 
well depth  & ke$^-$/pix & 670       & 670        & 100            \\    
dark current  
            & e$^-$/(s~pix)  & 0.2   & 0.2        & 0.015          \\
bias level  & ct/pix     & $\approx$1000 
                             & $\approx$1000   &
                                             $\approx$1000        \\
\noalign{\smallskip polarimetry \smallskip}
mod. freq. $\nu_{\rm mod}$
            & Hz         &           & 967.5     & 26.97          \\
row shift time
            & $\mu$s     &           & 54.7      & 72.3           \\
\noalign{\smallskip\hrule\smallskip}
\end{tabular}
\end{table}

The {\it standard imaging} detector mode provides fast read-out 
and a high pixel gain of 10.5 e$^-$/ADU. The read-out noise 
is about 2~ADU/pix, but because of the high gain 
this corresponds to a rather large ${\rm RON}\approx 20\,{\rm e}^-$. 
Therefore, the standard imaging mode is not ideal for 
low flux observations and a low RON imaging mode for faint targets should be  
considered as a possible future detector upgrade. At the moment
one can use the slow polarimetry mode as low-RON imaging
mode.  
\smallskip

\noindent
A key feature of the {\it polarimetric detector modes} is the
charge up and down shifting with a cycle period of
$P_{\rm mod}=1/\nu_{\rm mod}=1.03$~ms for fast or 
$P_{\rm mod}=37$~ms for slow modulation. 

{\it Fast modulation} is designed to search for polarized sources, 
for example planets or disks, near ($<0.3''$) bright stars 
with short integration times ($\approx 1-5$ sec), high gain,
and a deep full well capacity for collecting
many photo-electrons per pixel ($>10^5$) for high precision polarimetry.
The modulation between $I_\perp$ and $I_\parallel$ is faster than the typical
atmospheric coherence time $\tau_0\approx 2-5$~ms for medium and good 
seeing conditions $(\lapprox 1'')$ at the VLT and therefore it is possible
to suppress the speckle noise and reach a polarimetric sensitivity level at 
the photon noise limit of up to $10^{-5}$. Such a 
performance can only be reached if $\approx 10^{10}$ photo-electrons
can be collected per spatial resolution element with a diameter of
about 28~mas, or a synthetic aperture with 
$\approx 50$~pix$^2$ per detector what requires $n > 1000$ 
well exposed frames. Therefore, 
the fast modulation mode is tuned for short detector integration times $t_{\rm DIT}<10$~s and
high exposure levels $>1000~{\rm e}^-$/pix for which the high read-out noise 
${\rm RON}\approx 20~{\rm e}^-$/pix is not limiting the performance. 
The fast modulation mode produces a fixed bias pattern consisting
of two pixel columns with special bias level values because 
the read-out of a pixel row must be interrupted to avoid interference
with the simultaneous charge shifting in the image area. The pattern can be
removed with a standard bias subtraction procedure \citep[see][]{Schmid12}. 

The {\it slow modulation mode} is conceived and useful for
polarimetry of sources around fainter stars or with narrow filters. 
This mode provides a slow
modulation, but allows long integration times
$\ge 10$~s and delivers data with low ${\rm RON}\approx 2-3\, {\rm e}^-$/pix
appropriate for low flux levels.

\subsubsection{Charge traps in polarimetric imaging} 
The up and down shifting of charges during 
the on-chip demodulation causes single pixel effects 
due to charge traps. This anticipated problem was minimized 
with the selection of CCDs with a charge transfer 
efficency of better than 99.9995~\%.  
A charge trap can hold
for example one electron during a down shift and then release it
in the following up-shift. In this way a trap can dig after 1000 shifts
a hole of 1000 e$^-$ in the subframe of one modulation state, 
e.g. $I_\perp$, and produce a corresponding spike in the $I_\parallel$
subframe. A few examples are marked in the polarimetric frame of 
Fig.~\ref{rawcounts}.
This problem is solved with a phase switching in
which the charge shifting is reversed in every second frame with
respect to the polarization modulation. With such a double phase
measurements one can construct a double difference  
\begin{equation}
P^Z= {(I_{\rm o}(z)-I_{\rm e}(z)) 
         - (I_{\rm o}(\pi)-I_{\rm e}(\pi))\over 2}\,,
\end{equation}
for which the charge trap effects are cancelled or at least strongly 
reduced in the polarization signal \citep{Gisler04,Schmid12}. 
$I_{\rm o}$ and $I_{\rm e}$ are the counts registered in the 
odd and even detector rows for either the zero (z) or $\pi$ phase 
shifts between modulation and demodulation. In ZIMPOL, the alternating phase
shifts are implemented automatically for polarimetric detector
integrations and therefore only even number of frames can be taken
per exposure.
The charge trap effects increase with the number of modulation cycle
and they are therefore small for short integrations $t_{\rm DIT}$ and slow 
modulation mode. Unfortunately, the charge trap effects 
are not corrected for the intensity
signal $I=I_\perp+I_\parallel$ obtained with polarimetric imaging,
and the affected pixels must be treated and cleaned like ``bad''
detector pixels.

\subsubsection{The polarimetric modulation-demodulation efficiency} 
\label{Sectmodeff}

The ZIMPOL modulation-demodulation efficiency $\epsilon_{\rm mod}$
is regularly measured with a standard calibration 
procedure (p\_cal\_modeff). This calibration takes fully polarized 
$P=I_\perp=I^{100}$ ($I_\parallel=0$) flat field 
illumination using the calibration lamp in front of SPHERE and 
the polarizer in the ZIMPOL filter wheel FW0. 
The calibration measures either the mean fractional
polarization $\epsilon=\langle P^Z/I\rangle$ or a 2-dimensional 
efficiency frame   
\begin{displaymath}
\epsilon(x,y)=P^Z(x,y)/I^{100}(x,y)\,. 
\end{displaymath}
The efficiency $\epsilon_{\rm mod}$ is less
than one because of 
several static and temporal effects which 
depend on many factors, mainly on 
the modulation frequency and the detector arm, 
but also on wavelengths (or filter) and 
on the location on the detector. 

On the masked CCD, there is a leakage of photons and newly created 
photo-charges from the e.g. $I_\perp$-subframe in the illuminated pixel rows 
to the adjacent covered pixel rows of the $I_\parallel$ subframe,
of about $\delta_{\rm static}\approx 5$~\%. This reduces the efficiency
by the factor $\epsilon_{\rm static}=1-2\delta_{\rm static}\approx 0.9$.
The effect is slightly field dependent because the alignment of the
stripe mask with the pixel rows, and therefore the leakage to the
covered rows, is not exactly equal everywhere on the detector. 

The polarization beam splitter
is essentially perfect for the transmitted light in arm1, while about 
$\delta_{\rm arm2}\approx 3~\%$ of the ``wrong'' $I_\parallel$ intensity 
is deflected together with $I_\perp$ into arm2. This reduces
the relative efficiency of arm2 by 
$\epsilon_{\rm arm2}=(1-\delta_{\rm arm2})/(1+\delta_{\rm arm2})\approx 0.94$.

For fast modulation a temporal efficiency loss occurs because 
a finite time of about $75\,\mu$s 
is required for the FLC modulation switch and the CCD line shift. 
This reduces the modulation-demodulation efficiency of ZIMPOL by about 10~\% 
or $\epsilon_{\rm temp}\approx 0.9$ for the fast modulation mode. For slow 
modulation the temporal effect can be neglected. 

It results a ZIMPOL overall polarimetric efficiency of 
about $\epsilon_{\rm mod}\approx\epsilon_{\rm temp}\epsilon_{\rm static}\approx 0.8$ 
for fast modulation polarimetry 
and $\epsilon_{\rm mod}\approx \epsilon_{\rm static}\approx 0.9$ 
for slow modulation polarimetry with arm1. 
For arm2 an additional factor of $\epsilon_{\rm arm2}\approx 0.95$ 
needs to be included \citep[see][for further
details]{Bazzon12,Schmid12}. 

The fast frame transfer (with open shutter) causes also
a reduction of the measured efficiency for the measured
fractional polarization $P^Z/I$. During the frame transfer
the detector is still illuminated and the intensity $I_\perp/2+I_\parallel/2$
is added during the frame transfer time $t_{\rm ft}$
to both, the $I_\perp$- and $I_\parallel$-subframes.
This reduces for a polarized flat-field illumination, or 
a full frame aperture measurement like for a standard star, 
the fractional polarization $P^Z/I$ by 
\begin{equation}
\epsilon_{\rm ft}=t_{\rm DIT}/(t_{\rm DIT}+t_{\rm ft})\,. 
\label{Eqft}
\end{equation}
The frame transfer effect is stronger for short integration times 
in fast modulation. The factor is 
$\epsilon_{\rm ft}=0.973$, 0.986, and 0.995 for integration times 
of $t_{\rm DIT}=2$~s, 4~s, and 8~s (and $t_{\rm ft}=56$~ms).
For slow polarimetry with $t_{\rm ft}=74$~ms
there is $\epsilon_{\rm ft}=0.993$ for 
$t_{\rm DIT}^{\rm min}=10$~s and higher for longer integrations.  

Table~\ref{Modeff} gives the mean values from 
many calibrations taken throughout the year 2015 for 
$\epsilon_{\rm ft}\epsilon_{\rm mod}=P^Z/I$, while $\epsilon_{\rm mod}$ are
the efficiencies corrected for the transfer smearing according to 
Eq.~\ref{Eqft}. The calibrations show clearly the 
$\epsilon_{\rm mod}$-differences
between cam1 and cam2 and between fast and slow
modulation in particular with the ratios given in the last column and
the bottom row. The N\_R data taken with $t_{\rm DIT}=4$~s, 2~s, and 1.1~s
illustrate well the $\epsilon_{\rm ft}$ dependence. The efficiency
$\epsilon_{\rm mod}$ shows also a small dependence on wavelength.
The efficiencies $\epsilon_{\rm mod}$ given in boldface are
recommended values for a particular camera, modulation mode, and
filter. The statistical uncertainties in the 
$\epsilon_{\rm mod}$ calibrations are 
less than $\sigma < 0.005$ as derived from sets of four or
more measurements taken during 2015 with the same instrument configuration.  
Thus, the $\epsilon_{\rm mod}$ calibrations are very stable 
for a given configuration and the obtained values
or calibration frames are valid for a month typically and
possibly even longer. 
 
The degradation of the polarization flux because of the
non-perfect ($<1$) modulation-demodulation
efficiency of ZIMPOL is corrected with a calibration
frame $\epsilon_{\rm mod}(x,y)$ according to
\begin{equation}
P^Z_{\rm c1}(x,y) = P^Z(x,y)/\epsilon_{\rm mod}(x,y)\,,
\label{EqModeff}
\end{equation}
which we call polarmetric correction ``c1''.
This type of correction considers also the field dependent detector
effects but might also introduce pixel noise from the calibration 
frame if not corrected. For not very high signal to noise 
data $P/\Delta P<50$, it is
good enough to correct the $P^Z(x,y)$ just with a mean value
$\langle\epsilon_{\rm mod}\rangle$ as given in Table~\ref{Modeff}. 
For the derivation of the fractional polarization $P^Z/I$ in
a large aperture one needs also to account for the frame transfer
illumination by using $\epsilon_{\rm mod}\epsilon_{\rm ft}$.   

\begin{table}
\caption{Calibration measurements taken during the year 2015
for the modulation - demodulation 
efficiencies $\epsilon_{\rm ft}\epsilon_{\rm mod}$
and $\epsilon_{\rm mod}$
in various filters.}
\label{Modeff}
\begin{tabular}{lccccccc}
\noalign{\smallskip\hrule\smallskip}
Filter & {\hspace{-2mm}$t_{\rm DIT}$ \hspace{-2mm}} 
            & n & \multispan{2}
              {\hfil $\epsilon_{\rm ft}\epsilon_{\rm mod}$\hfil}
                     &  \multispan{2}{\hfil $\epsilon_{\rm mod}$ \hfil} 
                                                          & Ratio \\
       & [s]         &   & cam1   & cam2  & cam1   & cam2 & 
                   ${{\rm cam1}\over {\rm cam2}}$      \\
\noalign{\smallskip\hrule\smallskip}
\noalign{\smallskip fast modulation \smallskip}
V      & 8          & 9 & 0.772  
                           & 0.748  & {\bf 0.778}  & {\bf 0.753} & 1.032 \\
\noalign{\smallskip}
{R\_PRIM\hspace{-4mm}}
                   & 2          & 3 & 0.805   & 0.747 & 0.828  & 0.768 &    \\
N\_R   & 4          & 8 & 0.818  
                           & 0.751 & {\bf 0.829}  & {\bf 0.761} & 1.089 \\
N\_R   & 2          & 4 & 0.808  & 0.744 & 0.831  & 0.765 \\
N\_R   & 1.1        & 1 & 0.794  & 0.730 & 0.834  & 0.767 \\
B\_Ha  & 10         & 1 & 0.824  & 0.778  & 0.829  & 0.782 &      \\
\noalign{\smallskip}
VBB    & 1.1        & 2 & 0.783  & 0.744  & 0.823  & 0.782 &      \\
TiO717 & 8          & 1 & 0.827  &        & 0.833  \\
Cnt847 & 8          & 1 &        & 0.783  &        & 0.788 \\
\noalign{\smallskip}
Cnt820 & 8          & 2 & 0.807  & 0.767  & 0.813  & 0.772 \\
N\_I   & 4          & 8 & 0.802  & 0.774  
                               & {\bf 0.813}  & {\bf 0.785} & 1.036 \\
{I\_PRIM \hspace{-4mm}} 
                    & 2          & 2 & 0.802  & 0.769  & 0.824  & 0.791 &       \\
\noalign{\smallskip slow modulation \smallskip}
V      & 10         & 6 & 0.848  
                           & 0.821  & {\bf 0.854}  & {\bf 0.827} & 1.033 \\
N\_R   & 10         & 7 &  0.904  
                           & 0.831  & {\bf 0.911}  & {\bf 0.837} & 1.088 \\
N\_I   & 10         & 1 & 0.894  & 0.862  
                                     & {\bf 0.900}  & {\bf 0.868} & 1.037 \\
{Cnt820 \hspace{-3mm}} & 10         & 4 & 0.890  & 0.844  & 0.897  & 0.850 &     \\
\noalign{\smallskip}
       &            
 \multispan{4}{\hfil $\epsilon_{\rm mod}({\rm fast})/\epsilon_{\rm mod}({\rm slow})$\hfil}    & 0.908 & 0.908 &  \\
\noalign{\smallskip\hrule\smallskip}
\end{tabular}
\tablefoot{The third column indicates the number of calibrations 
used for the mean value given in the following four columns. 
Recommended values for the V, R, and I bands are 
highlighted in boldface, while the last column gives the
corresponding cam1/cam2 ratios and the bottom row the fast to slow 
modulation for these filters. 
The scatter for multiple calibrations taken with 
the same setup is 
$\sigma_\epsilon\approx 0.003$ to $0.005$.} 
\end{table}

\subsection{ZIMPOL filters}
\label{SectFilters}
\begin{figure}
\includegraphics[trim=2.8cm 13cm 0.5cm 3.2cm,clip, width=16cm]{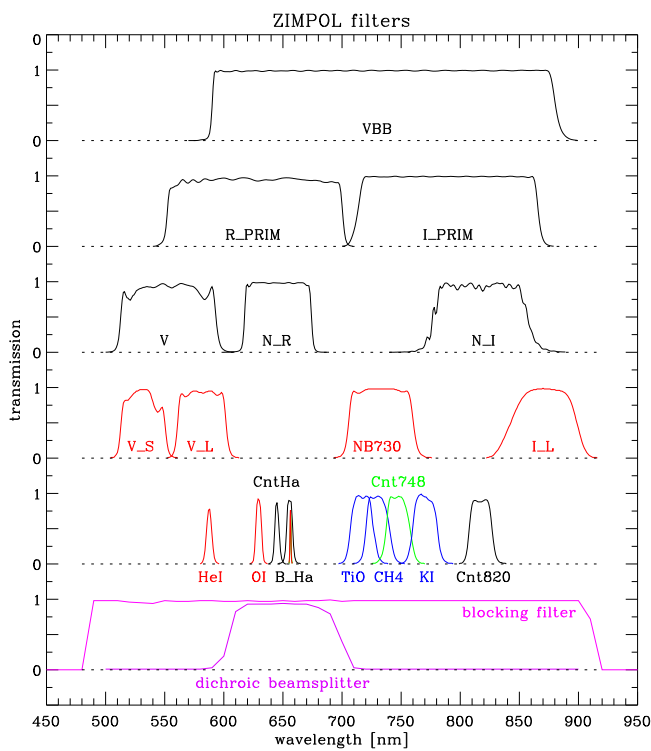}
\caption{Transmission curves for the ZIMPOL color filters,
the blocking filter and the dichroic beamsplitter.  
Black curves are used for filters located in FW1 and FW2, 
red for filters in FW0, blue are filters in FW1 only, and green
filters in FW2 only.}
\label{FigFilters}
\end{figure}

The pass bands of the available ZIMPOL color filters are shown
in Fig.~\ref{FigFilters} and listed in Table~\ref{FWcomps}
with central wavelengths $\lambda_c$, filter widths $\Delta\lambda$ 
(FWHM), their location in the filter wheels, and whether they can
be combined with the dichroic beam splitter or with one
of the two coronagraphic four-quadrant phase masks 
4QPM1 or 4QPM2 (see Sect.~\ref{Sectcoro}). Table~\ref{FWcomps}
includes also the calibration and test components in the filter
wheels. 

\begin{table}
\caption{Pass-band filters, neutral density filters, and
calibration components available in the filter wheels FW0,
FW1 and FW2 of ZIMPOL.}
\label{FWcomps}
\begin{tabular}{lccccccc}
\noalign{\smallskip\hrule\hrule\smallskip}
Component & $\lambda_c$ & $\Delta\lambda$ & 
          \multispan{3}{\hfil Wheel FWx\hfil} & Dicr. & 
                                           4QPM\tablefootmark{b} \\ 
          &  [nm]       & [nm]            &  0  &  1  & 2    
                                           & BS\tablefootmark{a}    & 1/2  \\ 
\noalign{\smallskip\hrule\smallskip}
open      &             &                 & +   & +   & +   & y   \\
\noalign{\smallskip\noindent broadband filters \smallskip}
VBB      & 735          & 290             &     & +   & +   &    \\
R\_PRIM  & 626          & 149             &     & +   & +   &    \\
I\_PRIM  & 790          & 153             &     & +   & +   &    \\
\noalign{\smallskip}
V        & 554          & 81              &     & +   & +   &    \\
N\_R     & 646          & 57              &     & +   & +   & y   &  1  \\
N\_I     & 817          & 81              &     & +   & +   &     &  2  \\
\noalign{\smallskip\noindent narrowband filters \smallskip}
V\_S     & 532          & 37              & +   &     &     &    \\
V\_L     & 582          & 41              & +   &     &     &    \\
NB730    & 733          & 55              & +   &     &     &    \\
I\_L     & 871          & 56              & +   &     &     &    \\
\noalign{\smallskip}
TiO717   & 716.8        & 19.7            &     & +   &     &    \\
CH4\_727 & 730.3        & 20.5            &     & +   &     &    \\
Cnt748   & 747.4        & 20.6            &     &     & +   &    \\
KI       & 770.2        & 21.1            &     & +   &     &    \\
Cnt820   & 817.3        & 19.8            &     & +   & +   &    & 2  \\
\noalign{\smallskip\noindent line filters \smallskip}
HeI      & 588.0        & 5.4             & +   &     &     &    \\
OI\_630  & 629.5        & 5.4             & +   &     &     & y  & 1 \\
CntHa    & 644.9        & 4.1             &     & +   & +   & y  & 1 \\
B\_Ha    & 655.6        & 5.5             &     & +   & +   & y  & 1 \\
\noalign{\smallskip}
N\_Ha    & 656.9        & 1.0             & +   &     &     & y  & 1 \\
Ha\_NB   & 656.9        & 1.0             &     &     & +   & y  & 1 \\
\noalign{\medskip\noindent neutral density filters\tablefootmark{c} \smallskip}
ND1      & \multispan{2}{500-900}         & +   &     &     & y   \\
ND2      & \multispan{2}{500-900}         & +   &     &     & y   \\
ND4      & \multispan{2}{500-900}         & +   &     &     & y   \\
\noalign{\medskip\noindent calibration components \smallskip}
\multispan{3}{\hspace{0.15cm} polarizer \hfil}   
                                 & +   &      &    & +  \\
\noalign{\vspace{0.05cm}}
\multispan{3}{\hspace{0.15cm} quarter wave plate \hfil}          
                                 & +   &     &    & +  \\
\noalign{\vspace{0.05cm}}
\multispan{3}{\hspace{0.15cm} circular polarizer \hfil}         
                                 & +   &     &    & +  \\
\noalign{\vspace{0.05cm}}
\multispan{3}{\hspace{0.15cm} pupil imaging lens \hfil}         
                                 &     &     & +   &  + \\
\noalign{\smallskip\hrule\smallskip}
\end{tabular}
\tablefoot{
\tablefoottext{a}{This column indicates components (y = yes), which can
be used together with the dichroic beam splitter} 
\tablefoottext{b}{this column marks the the recommended color 
filters to be used with 4QPM1 (indicated by 1) or 
4QPM2 (indicated by 2).}
\tablefoottext{c}{The neutral density (ND) filters attenuate the
throughput by about a factor $\approx 10^{-1}$ for ND1, $\approx 10^{-2}$ 
for ND2, and $\approx 10^{-4}$ for ND4.}}
\end{table}

The ZIMPOL filters were selected based on several   
technical and scientific requirements: 

For {\it imaging}, all pass band filters can be used. The filters in
FW1 and FW2 can be combined with one of the three neutral density 
filters located in FW0 to avoid image saturation. One can use the same
filter type in FW1 and FW2 to optimize the sensitivity in that
pass band or one can also use simultaneously two different 
filters in FW1 and FW2. The latter mode provides
spectral differential imaging, e.g. combining an H$\alpha$ filter
with the continuum filter CntHa. Alternatively, using different filters
provides an efficient way to get angular differential
imaging with large field rotation in two pass bands simultaneously, 
e.g. R\_PRIM and I\_PRIM, during one single meridian passage of a target. 

When combining different filters in FW1 and FW2 it needs to
be considered that the detector integration times are equal
in both channels. Broad-band filters combined with narrow
band filters can therefore cause strongly different illumination
levels on the two detectors. 

The filters in FW1 and FW2 are recommended for {\it polarimetry}, because
they are located after the polarization beam-splitter and the 
polarization is encoded as intensity modulation. Therefore,
polarization dependent pass bands and other polarization effects
of the filters do not affect the polarization measurements.  
Only for these filters a modulation-demodulation efficiency $\epsilon_{\rm mod}$
can be determined, because this calibration requires
the polarizer located in FW0. Polarimetry using the neutral density
filters or the color filters in FW0 is also possible but with an increased
polarimetric calibration uncertainty. More work is required to characterize 
accurately such non-optimal polarimetric modes.  

Different filters can be used in FW1 and FW2 
simultaneously for ZIMPOL polarimetry, because each arm provides 
a full polarization measurement. Thus
one can perform a combination of simultaneous spectral and 
polarimetric differential imaging with ZIMPOL. 

There are some instrument configurations which can only be
combined with certain filters. Observations with the dichroic
beam-splitter between ZIMPOL and WFS provide
more photons for the WFS and is therefore particularly beneficial for 
faint stars. This mode must use the filters 
N\_R, B\_Ha, N\_Ha, NB\_Ha, CntHa, and OI\_630, which have their 
pass bands in the  
transmission window of the dichroic BS plate (Tab.~\ref{FWcomps}, col.~7). 
The R-band four quadrant phase mask coronagraph 4QPM1 has its 
central working wavelength at 650~nm and is also designed for these
filters. The I-band 4QPM2 with central wavelength 820~nm requires
the filters Cnt820 or N\_I for good results (Tab.~\ref{FWcomps}, col.~8). 
\smallskip

The following list gives important scientific criteria which were
considered for the selection of the different filters:
\begin{itemize}
\item{} V, N\_R and N\_I broad band filters for imaging and 
        polarimetry of stars and circumstellar dust, 
\item{} RI (or VBB for very broad band), R\_PRIM, and I\_PRIM broad band 
        filters 
        for demanding high contrast observation where a high photon rate
        is beneficial for the detection,
\item{} TiO\_717, CH4\_727, KI, Cnt748 and Cnt820 for narrow band imaging, 
        spectral differential imaging and polarimetry of cool stars, 
        substellar objects and solar system objects, 
\item{} B\_Ha, N\_Ha and CntHa for imaging, spectral differential imaging,
        and polarimetry of circumstellar H$\alpha$ emission,
\item{} V\_S, V\_L, 730\_NB, I\_L for better coverage of the ZIMPOL 
        spectral range with intermediate band filters,
\item{} OI\_630 and HeI as additional line filters for the imaging of 
        circumstellar emission, 
\item{} N\_R, the H$\alpha$ filters B\_Ha, N\_Ha (NB\_Ha), CntHa, 
        and OI\_630 in combination with the dichroic beam splitter for imaging,
        differential imaging and polarimetry of fainter targets 
        $R\gapprox 9$ mag, where the AO performance profits from the
        higher photon throughput to the wave front sensor. 
\end{itemize}

\paragraph{Photometric instrument throughput.}
The total instrument throughput for each filter, or the photometric
zero points, should be determined for flux measurements. 
This value depends on the used system 
configuration (polarimetry, imaging, ND-filters, etc.), 
and some preliminary values
are given in \citet{Schmid17}. However, for many applications 
one can use an unsaturated image of the central bright star, 
for example taken with a ND-filter, as photometric reference 
for the flux calibration of a faint circumstellar source. 

\paragraph{Blocking filter.} A problem of the pass band filters
is their insufficient blocking of the transmission
(often not less than 0.001) for wavelengths outside the ZIMPOL 
range $\lambda<500$~nm and $\lambda>900$~nm. This can be 
particularly harmful for coronagraphic observations, where radiation 
from the central, bright source is displaced for wavelengths outside the 
ZIMPOL spectral range, because the atmospheric dispersion corrector 
is not designed to correct for these wavelengths. For coronagraphic 
images some of the central star radiation, e.g. the blue light 
($420-480$~nm), might not fall onto the focal plane mask. This
would create a point like signal slightly outside the mask despite an
out-of-band attenuation of ${\rm ND}\approx 3$ of the color filter.
For this reason a blocking filter is added on the ``free'' position of
the modulator slider for imaging, and one is added to the 
entrance window of the modulator vacuum housing for polarimetry.

\section{The AO corrected point spread function}
\label{SectAO}

The PSF obtained with SPHERE/ZIMPOL are complex
and depend on wavelengths, atmospheric conditions,
star brightness and observing modes. This section describes 
characteristic PSF parameters for many different cases.

\begin{figure}
\includegraphics[width=8.8cm]{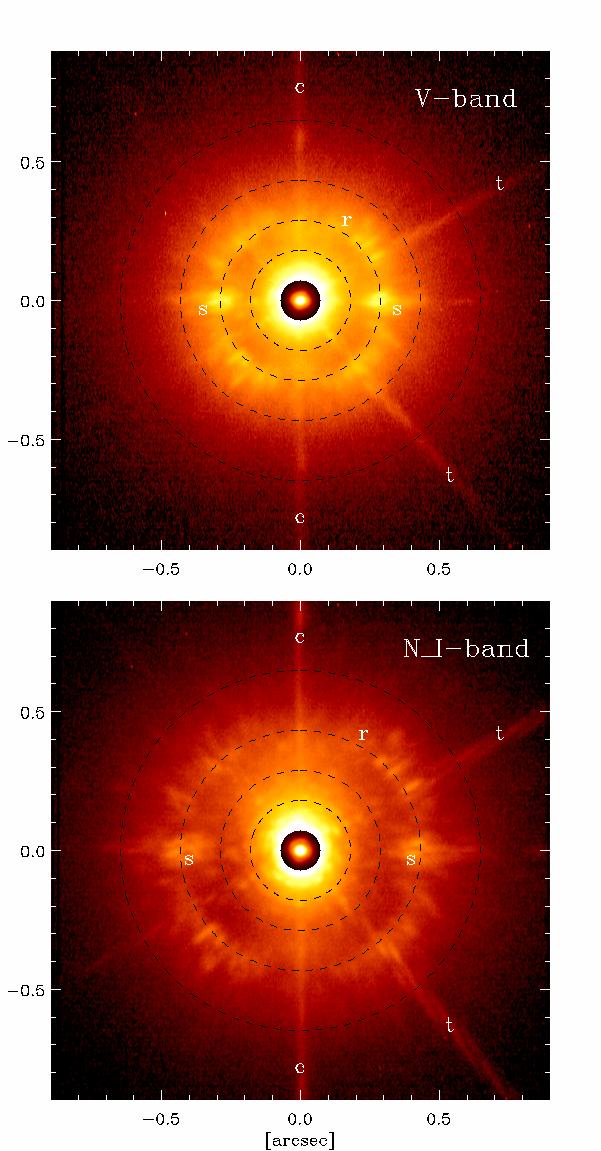}
\caption{Normalized PSFs of \object{HD 183143} for the V-band (top) and
the N\_I-band (bottom) with the color scale reduced by a factor of 100
for the central peak within $r<20$~pixels. Marked PSF features are
the speckle ring near the AO control radius (r), strong fixed 
speckles from the AO system (s), two telescope M2 spider features 
(t), and the CCD frame transfer trail 
of the PSF peak (c). The dashed rings illustrate the location of
the azimuthal cuts shown in Fig.~\ref{VIprofiles}.}
\label{PSFVI}
\end{figure}

\subsection{Two-dimensional PSF structure}
\label{SectPSFAO}

Figure~\ref{PSFVI} shows the averaged PSFs of HD 183143
in the V-band and the N\_I band for 48s of integration each
taken on 2015-09-18, the N\_I-frames about 2 minutes after the V-frames.
These are ``typical'' PSFs for a bright star and for 
stable and good atmospheric conditions.
  
The polarimetric mode P1 with derotator fixed was used,
and the PSF is displayed in the orientation of the 
detectors with $x$ and $y$ in row and column directions, respectively.   
The corresponding polarization images for the V-band PSF are
discussed in Sect.~\ref{SectPSFpol}.

\begin{figure}
\includegraphics[trim=3cm 13cm 2.5cm 3cm,clip,width=8.8cm]{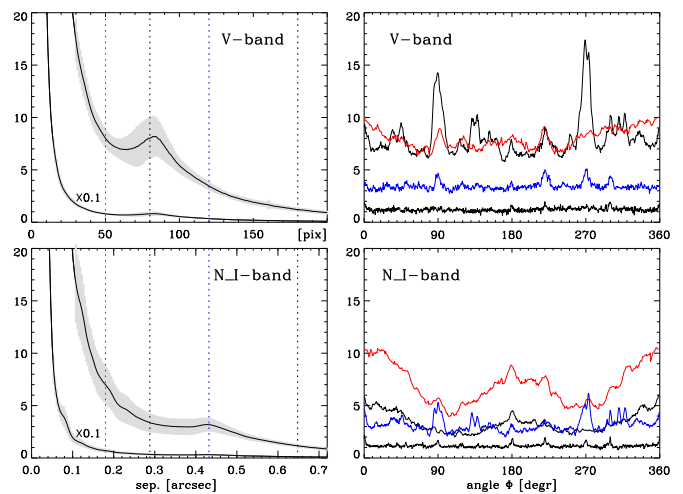}
\caption{Azimuthally averaged radial profiles ct$_{\rm n6}(\rho)$ (left)
and azimuthal profiles ct$_{\rm n6}(r,\phi)$ for $r=50$, 80, 120 and 
180~pix (right) corresponding to $\rho=0.18'',\,0.29'',\,0.43''$, 
and $0.65''$, respectively for the PSFs of HD 183143 shown in
Fig.~\ref{PSFVI}. Also indicated as gray shading are the standard 
deviations of the azimuthal points $\sigma({\rm ct}_{\rm n6}(r,\phi)$. 
The location of the azimuthal 
cuts are indicated in the ${\rm ct}_{\rm 6n}(r)$-panels and in 
Fig.~\ref{PSFVI}.} 
\label{VIprofiles}
\end{figure}

Prominent PSF features in Fig.~\ref{PSFVI} are the strong speckle ring at
a spatial separation corresponding roughly to the AO control radius of
$r_{\rm AO}=20\,\lambda/D$ up to which
the AO corrects the ``seeing'' speckles. 
This corresponds to $\rho_{\rm AO}\approx 0.3''$
for the V-band and $\rho_{\rm AO}\approx 0.45''$ for the N\_I-band. 
Dashed rings indicate azimuthal cuts through the PSF,
which are shown in Fig.~\ref{VIprofiles}. Strong quasi-static 
speckles from the AO system, marked with ``s'', 
are always present left and right from the PSF peak 
on the speckle ring near $r=80$~pix for V and $r=120$~pix for N\_I. 
Another feature of the PSF are the diffraction pattern of the 
four spiders holding the M2 telescope mirror and two of the
four appear particularly bright in Fig.~\ref{VIprofiles}. The vertical line
through the PSF peak is the frame transfer trail, which
is caused by the illumination of the detector during the 
fast frame transfer. The trail is particularly strong for non-coronagraphic
observations of bright stars with short integrations (see Eq.~\ref{Eqft}).   

Figure~\ref{VIprofiles} gives averaged radial profiles and azimuthal cuts
of the PSF for a more quantitative description. 
The PSFs in Figs.~\ref{PSFVI} and \ref{VIprofiles} are given
in units of ct$_{\rm n6}$ where the counts are   
normalized to $10^6$ counts within a round aperture with a diameter of 3$''$.  
The normalization of the surface brightness SB [mag/arcsec$^2$] 
of a PSF to the total stellar flux $m_{\rm star}$ [mag] defines a 
normalized surface brightness or a
surface brightness contrast $C_{\rm SB}$ [mag/arcsec$^2$] according to:
\begin{equation}
C_{\rm SB}(x,y) = {\rm SB}(x,y) - m_{\rm star}\,.
\end{equation} 
This translates, for the applied PSF normalization of $10^6$~ct 
in the $3''$-aperture, to the following normalized surface brightness 
conversion between  
ct$_{\rm n6}$ [pix$^{-1}$] and mag\,arcsec$^{-2}$ 
\begin{equation}
C_{\rm SB}(x,y) 
     = -2.5^m\,{\rm log\,(ct}_{\rm n6}(x,y)) + 2.78^m \,,
\label{eqContrast}
\end{equation}  
using $-2.5^m\, {\rm log}(77160/10^6)=+2.78^m$ because one 
arcsec$^2$ contains 77160 pixels on one ZIMPOL detector.
Thus, a PSF peak surface brightness of ${\rm ct}_{\rm n6}(r$=0$)=5000$~ct/pix 
corresponds to a normalized surface brightness of 
$C_{\rm SB}(0)= -6.47$ mag/arcsec$^2$. 
A signal of 1~ct$_{\rm n6}$ for example from 
a faint companion with a point source contrast of about 
$\Delta m \approx 9.25^m$, is at the level to be visible 
for $r\gapprox 80$~pix in the normalized 
images and cuts shown in Figs.~\ref{PSFVI} and \ref{VIprofiles} except
for the regions of the speckle ring. 

The radial PSF profiles in Figure~\ref{VIprofiles} show
the azimuthal mean with the standard deviation $\pm \sigma$ 
indicated in gray. Four radii were selected to show the
azimuthal profiles ${\rm ct}_{\rm n6}(r,\phi)$
of the PSF as function of position angle $\phi$
measured counterclockwise from top. The innermost ring profiles $r=50$~pix
show a sine-like pattern, especially for the N\_I band, with two maxima 
and minima within 360$^\circ$. They originate from the slightly
elongated base of the very strong 
${\rm ct}_{\rm n6}(0)\approx 5000$~ct/pix
PSF peak (see also Fig.~\ref{FigPeak}). Such PSF extensions are often
present and they can be caused by a dominant 
wind direction. 

The speckle rings at $r\approx 80$~pix in the V-band and at 
$r\approx 120$~pix for the N\_R band produce very strong 
noise features on small angular scales. 
The strongest speckles are at 90$^\circ$ and 270$^\circ$ degrees from
the AO system as already discussed above (Fig.~\ref{PSFVI}). 
Speckles
are weaker inside the speckle ring and outside they are hardly
recognizable besides the faint traces from the spiders at 
$\phi=220^\circ$ and
$300^\circ$ and the frame-transfer trail at $0^\circ$ and 180$^\circ$.
Because of the location of the speckle rings the N\_I-band
observations would be much more sensitive in the radial range
0.2$''$ to 0.4$''$ than V-band observations. On the other side
a faint target at a separation 
between $\rho=0.4''-0.5''$ might be easier to
detect (assuming gray color distributions) just outside the 
speckle ring in the V-band than on the speckle ring in I-band.

The dominant noise sources are the distortions close to the center,
the strong speckles further outside, and the read-out noise outside
of the control ring. 
The read-out noise regime can be pushed outwards easily 
with a stronger frame illumination, 
where the PSF peak is saturated or close to saturation, or with
coronagraphic observations. 

\subsection{Parameters for the radial PSF}

To simplify our discussion we focus on the radial profiles of the PSF 
keeping in mind the significant deviations from rotational symmetry
discussed above (Figs.~\ref{PSFVI} and \ref{VIprofiles}). We use
as basis for this brief overview the polarimetric
standard star data from the ESO archive taken in polarimetric 
mode in the V, N\_R, N\_I band filter, which are regularly 
obtained by ESO staff as part of the SPHERE instrument calibration plan. 
In addition we include a few special PSF cases. 
As a starting point, we have selected the data of HD 183143 
from Sept. 18, 2015 (Figs.~\ref{PSFVI} and \ref{VIprofiles})
as typical examples of SPHERE/ZIMPOL PSFs for the
V, N\_R and N\_I-band taken under good atmospheric conditions.  

\begin{figure}
\includegraphics[trim=2cm 12.5cm 3cm 3cm,clip,width=12cm]{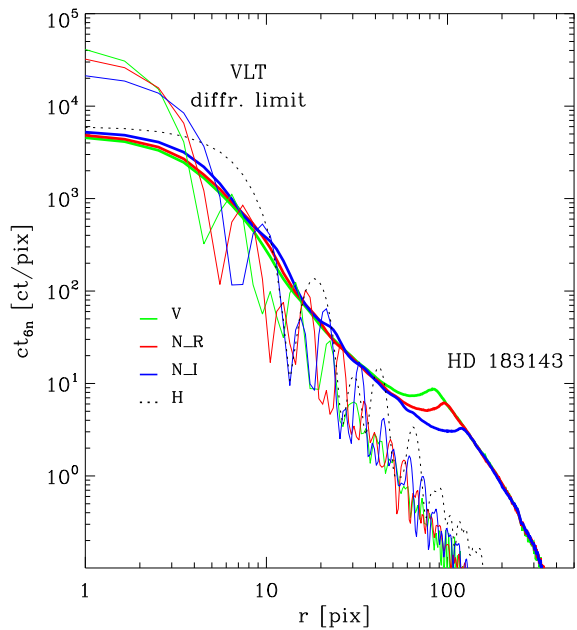}
\caption{Flux normalized PSFs ${\rm ct}_{\rm n6}(r)$
for HD 183143 (thick profiles) and the calculated 
diffraction limited PSF of the VLT (thin profile) for the 
filters V ($\lambda=545$~nm), N\_R (645~nm), 
N\_I (817~nm), and including the H-band (1.62~$\mu$m) 
for the VLT. One pixel is 3.6~mas.} 
\label{FigAOoo}
\end{figure}

\paragraph{PSF wavelength dependence.} 
Figure~\ref{FigAOoo} compares the flux normalized 
radial profiles ${\rm ct}_{\rm n6}(r)$, with equally normalized 
diffraction limited profiles $f(r)/f_{\rm 3dia}$ calculated for the VLT 
with an 8.2~m primary mirror and a 1.1~m central
obscuration by the secondary mirror and where $f_{\rm 3dia}$ is the
total flux within an aperture with a diameter of 3$''$. Also shown 
is the H-band (1.62~$\mu$m) diffraction profile.
For each filter $f(r)$ is
the mean PSF for several wavelengths equally distributed over
the filter range to account for the filter widths. 

The normalized peak flux ${\rm ct}_{\rm n6}(0)$
for HD 183143 (Fig.~\ref{FigAOoo}) are roughly at the same 
level for all filters. The
most prominent difference between the PSFs is
the wavelength dependence for the radius of the local maximum 
associated with the speckle ring. 
Contrary to the observed profiles the peak flux of the normalized 
VLT diffraction PSFs 
depend strongly on wavelength. For the pixel size of 
$3.6 \times 3.6~{\rm mas}^2$ there is $f(0)/f_{\rm 3dia}=5.13~\%$, 3.80~\%, 
and 2.36~\% for the V, N\_R, and N\_I filters, respectively 
(see bottom lines in Table~\ref{TabPSFAO}). 

We use the normalized peak flux ${\rm ct}_{\rm n6}(0)$ 
as key parameter to compare the PSF quality of different
SPHERE/ZIMPOL observations. This value can be
determined easily and under good atmospheric conditions
similar values in the range $0.4 - 1.0$~\% are obtained 
for different wavelengths. 

For a comparison with the performance of other instruments one
should translate the peak flux into a Strehl ratio. An approximate
Strehl ratio $S_0$ can be calculated with the relation 
\begin{displaymath}
S_0 = {{\rm ct}_{n6}(0)\over f(0)/f_{\rm 3dia}}\,.
\end{displaymath}  
One should note, that a lower Strehl ratio at shorter wavelengths
is not equivalent with a lower normalized peak flux because
the diffraction peak depends strongly on wavelength, like
$f(0)/f_{\rm 3dia}\propto 1/\lambda^2$.

The approximate Strehl ratio $S_0$ provides
a useful parameter for a simple assessment of the 
SPHERE/ZIMPOL system PSF. However, the $S_0$ value does not describe 
well the SPHERE AO performance, 
because there exist instrumental effects which degrade the PSF peak
which are not related to the adaptive optics 
\citep[as discussed in][]{Schmid17}.
A more sophisticated AO characterization should be based on the analysis 
of the Fourier transform of the aberrated image as described 
in \citet{Sauvage07}. Such an analysis yields for the N\_I-filter 
PSF of HD 183143 an AO Strehl ratio of 33~\% instead of the 23~\% 
resulting from the values given in Table~\ref{TabPSFAO}. 
The difference can be explained by a residual background at low 
spatial frequencies caused for example by 
instrumental stray light, not corrected frame transfer smearing,
and other effects.

\begin{figure}
\includegraphics[trim=1.5cm 13.cm 2.5cm 2.0cm,clip,width=8.8cm]{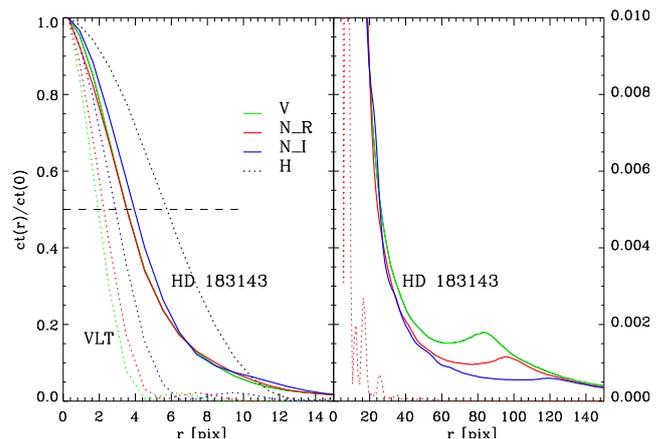}
\caption{Same PSFs for HD 183143 and the VLT diffraction limit 
as in Fig.~\ref{FigAOoo} but normalized to the peak flux. The dashed line
in the left panel marks the half width at 
half maximum of the profiles. The dotted curves are the calculated 
diffraction limited profiles. 
The flux scale in the right panel is
100 times lower.} 
\label{FigPeak}
\end{figure}

Figure~\ref{FigPeak} shows the peak normalized PSF of HD 183143
for the three filter on a linear scale. The PSF full
width at half maximum (FWHM) are
significantly larger, by about $2-3$~pixels or 7-11~mas,
when compared to the diffraction limited profile (Table~\ref{TabPSFAO}). 
Different effects contribute to this degradation like small 
pointing drifts, residual PSF jitter because of
telescope and instrument vibrations, residuals from the
atmospheric dispersion correction and other 
uncorrected optical abberations, cross-talks between detector
pixels, and possibly more. Also shown in this plot are the 
maxima of the speckle rings.
The mean profiles can be misleading when considering the
strong azimuthal dependence discussed in Fig.~\ref{VIprofiles}.

\begin{figure}
\includegraphics[trim=1.5cm 12.8cm 2.5cm 2.2cm,clip,width=8.8cm]{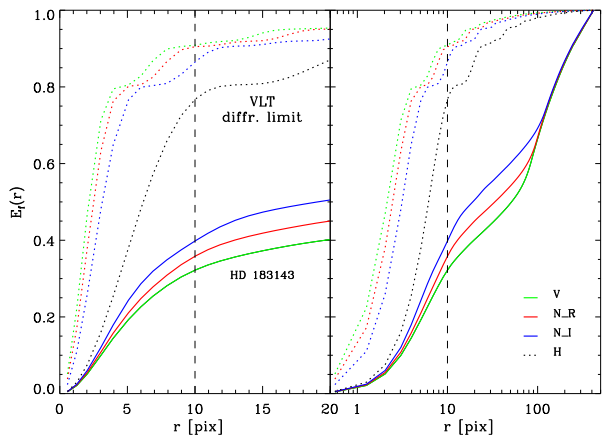}
\caption{Profiles for the normalized encircled flux $E_f(r)$ 
for the same PSFs of HD 183143 and
the VLT diffraction as in Fig.~\ref{FigAOoo}. The dashed
lines illustrate the parameter $E_{f10}$.} 
\label{FigEF}
\end{figure}

The profiles of the normalized encircled flux $E_f(r)$ shown
in Fig.~\ref{FigEF} is another way to characterize the PSFs.
This plot illustrates that only about 20~\% of the total PSF
flux is encircled in an aperture with $r=5$~pix (18 mas), 
or about 30~\% for $r=10$~pix. 
The encircled flux is an 
important parameter for flux determinations using synthetic 
apertures. We select as characteristic
parameter the halo and background corrected encircled flux $E_{f10}$ for an
aperture radius of $r=10$~pix, using the mean flux value 
$\langle {\rm ct}_{\rm n6}(r=11) \rangle$ of the pixel ring with $r=11$~pix 
as background and halo level, which is subtracted from all 
317~pixels $i$ within the aperture $r_i\leq 10$~pix
\begin{equation}
E_{f10} = {1\over 10^6}
\sum_{r_i\leq 10}\bigl[{{\rm ct}_{\rm n6}(x_i,y_i)
        -\langle {\rm ct}_{\rm n6}(r=11) \rangle}\bigr]  
\end{equation}
This type of encircled energy measurement is also applicable to 
the PSF peak of faint companions, for which it is not possible 
to measure the outer part of the PSF. For high contrast measurements
we need $E_{f10}$, or encircled energies for other apertures with
small radii, to derive the flux contrast between central star and
faint companion. The measured $E_{f10}$-values for HD 183143 and 
other test cases are given in Table~\ref{TabPSFAO}.

\begin{table*}
\caption{PSF characteristics for three typical standard stars and 
several special cases. The last four lines give the PSF parameters for 
the calculated, diffraction limited PSFs for the VLT.}
\label{TabPSFAO}
\begin{tabular}{lllccccccccc}
\noalign{\smallskip\hrule\smallskip}
Star &            &          & \multispan{3}{\hfil Atmosphere \hfil}
                               & \multispan{3}{\hfil Instrument mode \hfil}
                                  &\multispan{3}{\hfil PSF parameters \hfil} \\
Name & $m_{\rm R}$  & Date    &  Seeing & $\tau_0$ & $v_{\rm wind}$ 
                                & Mode\tablefootmark{a}
                  & Filter & $n_{\rm DIT}$$\times$$t_{\rm DIT}$ 
                       & ${\rm ct}_{\rm n6}(0)/10^6$ & $E_{f10}$ & FWHM \\
      & [mag]     & Airmass  &  ['']   &  [ms]    & [m/s]       & $\nu({\rm AO})$/SpF  
                  &        & [s]       &  [\%]               & [\%]    & [mas] \\  
\noalign{\smallskip\hrule\smallskip}
\noalign{\smallskip standard stars \smallskip}

\object{HD 183143} 
                 & 5.74  & 15-09-18 &  0.73  & 8.3 &  5.2 &  P1 pol.
                 & V     & $12\times 4$          &   0.486 &  28.1 &  25.4    \\   	
\multispan{2}{~~''good'' \hfil}  
                 & 1.72  &        &              &         & 1.2~kHz / S           
                 & N\_R  & $12\times 4$          &   0.533 &  29.8 &  25.4    \\
                 &       &            &          &         &       &
                 & N\_I  & $12\times 4$          &   0.548 &  32.4 &  28.4    \\
\noalign{\smallskip}

HD 129502 & 3.48  & 15-04-19 & 1.20      & 1.4       &   2.2    & P1 pol.
                 &  V        & $16\times 2$          &   0.148  &  14.7 &  43.8   \\  
\multispan{2}{~~''bad''\hfil}   
                 & 1.06      &       &               &          &  1.2~kHz / M 
                 & N\_R      & $8\times 4$           &   0.389  &  20.2 &  25.3   \\
                 & &         &        &              &          &          
                 & N\_I      & $16\times 2$          &   0.369  &  18.1 &  27.7   \\
\noalign{\smallskip}

HD 161096 & 1.96   & 15-04-28& 0.88      & 4.6       & 5.7      & P1 pol 
                   & V       & $16\times 2$          & 0.733    & 36.5    & 22.6    \\ 
\multispan{2}{~~''excellent'' \hfil}          
                   & 1.20    &           &           &          & 1.2~kHz / S 
                   & N\_R    & $16\times 2$          & 0.919    & 43.5    & 22.5    \\
                   & &       &        &              &          &          
                   & N\_I    & $16\times 2$          & 0.842    & 41.5    & 25.3    \\

\noalign{\smallskip special cases \smallskip}
HD 142527 & 8.3  
                  & 17-06-01 & 0.65          & 11.6     & 1.5   & P2 pol. 
                  & VBB      & $1\times 3$   & 0.196    & 35.1  & 53.8    \\  
\multispan{2}{~~``low wind effect'' \hfil}
                  & 1.05     &               &          &       & 600~Hz / M \\

\noalign{\smallskip}
MMS 12     & 10.5 & 14-10-11 & 1.40       & 2.1      & 3.8     & imaging
                   & I\_PRIM  & $3\times 40$          & 0.039   & 7.0    & 53.1   \\
\multispan{2}{~~''faint'' \hfil} 
                   & 1.49     &            &          &         & 300~Hz / L \\

\noalign{\smallskip}
R Aqr     & 8.8\tablefootmark{b}
                   & 14-10-11 & 0.83       & 2.8      & 13.6    & P2 pol.
                   & V        & $8\times 10$          & 0.091   & 10.5  & 34.1    \\
\multispan{2}{~~``red source''\hfil}
                   & 1.13    &           &           &          & 1.2~kHz / M 
                   & Cnt820  & $20\times 1.2$        & 0.541 & 41.0    & 28.0    \\   

\noalign{\smallskip}
$\alpha$ Eri & 0.49 & 14-10-13 & 1.40      & 1.60     & 3.0   & imaging    
                  & CntHa    & $50\times 0.01$      & 1.02  & 48.2      & 19.3    \\
\multispan{2}{~~``snap shot''\hfil}
                  & 1.25     &           &          &          & 1.2~kHz / S   \\

\noalign{\smallskip}
$\alpha$ Hya & 2.55 & 14-10-10 & 0.83   & 2.8     & 13.6     & ima 
                  & R\_PRIM  & $10\times 1.1$     & 0.798    & 40.5  &  20.5    \\  
\multispan{2}{~~``coro test source'' \hfil}  
                  & 1.36     &        &         &          & 1.2~kHz / M 
                  & I\_PRIM  & $10\times 1.1$     & 0.781    & 46.5  &  23.8    \\     
\noalign{\medskip}

          &  &    &        &   & \multispan{2}{\hfil {\bf VLT diff. limit}} 
          & V     &                       & 5.13    & 91.0  & 14.2  \\
          &       &          &        &         &         &
          & N\_R  &                       & 3.80    & 90.6  & 16.8  \\
          &       &          &        &         &         &
          & N\_I  &                       & 2.36    & 86.7  & 21.0  \\
          &       &          &        &         &         &
          & H     &                       & 0.583   & 76.3  & 41.6  \\
                        
\noalign{\smallskip\hrule\smallskip}
\end{tabular}
\tablefoot{
\tablefoottext{a}{$\nu$(AO) is the AO loop frequency and SpF the used 
spatial filter (S=small, M=medium, L=large) for the WFS}
\tablefoottext{b}{brightness in other bands is 
$m_V = 11.4^m$ and $m_I=4.4^m$ \citep{Schmid17}}
}
\end{table*}

\subsection{PSF variations}
The PSFs obtained with SPHERE/ZIMPOL show a large diversity depending on
atmospheric conditions, central star brightness, AO performance, 
and instrumental mode, and a few typical cases are discussed in
this subsection. Table~\ref{TabPSFAO} lists for the
analyzed profiles the measured PSF values for the normalized peak flux
${\rm ct}_{n6}(0)$, the encircled energy $E_{f10}$, and 
full width at half maximum FWHM. Also given are atmospheric and instrument 
parameters taken from the ESO data file headers. Seeing and coherence 
time $\tau_0$ are measured for the vertical direction with the DIM-MASS
systems. 

\begin{figure}
\includegraphics[trim=2cm 15.5cm 3cm 3cm,clip,width=12cm]{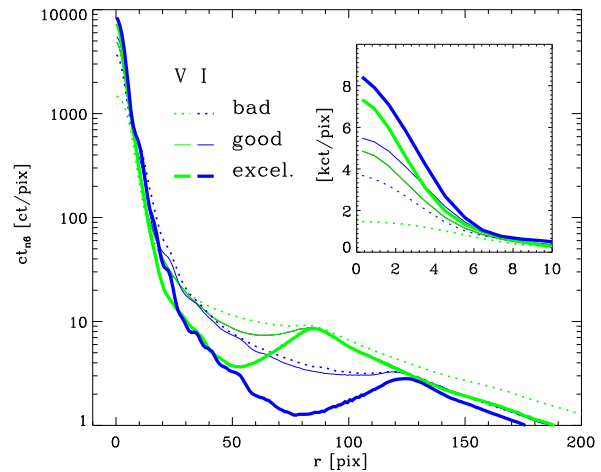}
\caption{Normalized radial profiles ${\rm ct}_{\rm 6n}$
for V- and N\_I-band observations of HD 161096 with ``excellent'' , 
for HD 183143 with ``good'', and HD 129502 with ``bad'' quality PSFs.} 
\label{PSFprofgoodbad}
\end{figure}

\paragraph{Atmospheric conditions.}
The PSF of SPHERE/ZIMPOL changes often strongly within one night and
under instable conditions even from frame to frame. Figure~\ref{PSFprofgoodbad} 
compares for bright standard stars the typical or ``good'' PSFs
of HD 183143 for the V-band and the N\_I band with the
``excellent'' PSFs of observations of HD 161096 
and the ``bad'' PSFs of HD 129502. These three examples represent
quite well the much larger data set of bright standard stars available
in the ESO archive. These archive data
show an overall correlation between good 
PSFs parameters and long atmospheric coherence times scales 
$\tau_0\gapprox 3$~ms or good seeing $\lapprox 0.9''$  and bad PSFs 
for short time scales $\tau_0\lapprox 2$~ms or mediocre 
seeing $\gapprox 1.0''$. This 
is roughly in agreement with the study of \citet{Milli17} on 
SPHERE PSF properties in the near IR. 
Bad atmospheric conditions as for the observations of HD 129501, 
affect much more the short-wavelength V-band profile. 
Other effects, for example the high airmass for the observations 
of HD 183143, play also a role. 

Let us consider in more detail the contrast characteristics of the 
three N\_I profiles bad, good, and excellent in 
Fig.~\ref{PSFprofgoodbad} and Table~\ref{TabPSFAO}.
Normalized peak fluxes and encircled fluxes scale roughly 
like 0.7 : 1.0 : 1.4 between
bad : good : excellent conditions. The normalized mean flux level 
at $r=80$~pix is much lower $\approx 1.5$~ct for the ''excellent''
PSF, and  $\approx 5$~ct for the ``good'' and ``bad'' PSF. 
The speckle noise is measured as standard deviation of fluxes in apertures
at $r=80$~pix ($0.29''$) as illustrated for the case of
\object{$\alpha$ Hyi B} in Fig.~\ref{CoroPSFcenter}(a). This 
yields the 5-$\sigma$ raw contrasts for the
faint point source detection at $\rho=0.29''$ of about 
$8 \cdot 10^{-4}$ : $4\cdot 10^{-4}$ : $2.3\cdot 10^{-4}$ 
for the bad : good : excellent PSFs discussed here. 

This rough estimate does not consider differential
imaging techniques for speckle noise suppression, which may change the 
picture. In any case, there exist dramatic differences of factors $2-4$ 
in the SPHERE/ZIMPOL contrast performance for bad, good or 
excellent atmospheric conditions which are of great importance 
when defining the seeing requirements for observations 
of a particular object.

\begin{figure*}
\includegraphics[width=18.8cm]{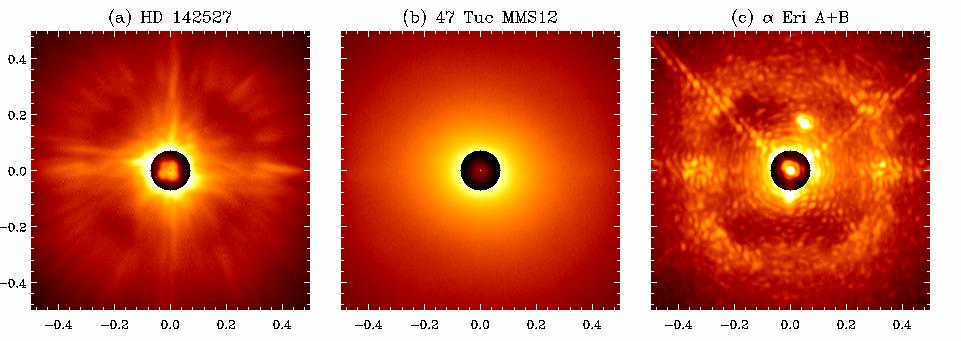}
\caption{Normalized PSFs for special cases: (a) VBB-filter image 
of HD 142527 as example for the low wind effect, (b) the faint star 47 Tuc MMS12 
in I\_PRIM, and (c) a 10~ms snap-shot image of \object{$\alpha$ Eri A} and B 
in the line filter CntHa. The color scale is reduced by a factor of 
100 for the PSF center within $\rho<0.072''$. Axes are in arcsec.}
\label{PSFextreme}
\end{figure*} 

\paragraph{Low wind effect.} 
The PSF of HD 142527 in Fig.~\ref{PSFextreme}(a) is an extreme
example for the so called low wind effect, which leads to
multiple PSF peaks in the center. This is explained 
in \citet{Sauvage16b} by a
discontinuity in the wave front phase in the pupil plane at
the location of the mirror M2 telescope spiders. These spiders 
are cold and cause a temperature difference between the air in upwind and 
downwind direction. Such phase offsets are not easily recognized
by the Shack-Hartmann WFS and therefore different PSF peaks result.  
This effect is only observed when the wind 
is particularly slow, $\lapprox 2$~m/s, so that the heat exchange 
between spider and air induces a substantial temperature
difference. The atmospheric conditions for the observation of HD 142527 
were in principle excellent with a very good seeing of $0.65''$ and a long
coherence time of 11.5~ms, but a wind speed of only 1.5~m/s 
(Table~\ref{TabPSFAO}). The FWHM is 53~mas 
for the multiple peaked PSF and the relative
peak flux ${\rm ct}_{n6}(0)$ is a factor $2.5-4.0$ lower than for 
other PSFs taken under sub-arcsecond seeing conditions.

Still relatively high is the encircled energy $E_{f10}=35$~\%, 
which is comparable to ``good'' atmospheric conditions. Thus the low
wind effect splits the central PSF peak and degrades the resolution,
but at larger separation the $E_f(r)$-profile is not much
affected. This means that the sensitivity for high contrast 
imaging of extended circumstellar scattering regions 
is not strongly 
degraded by the low wind effect, apart from the reduced spatial resolution. 
For example, the ZIMPOL observations
of the proto-planetary disk around TW Hya described by 
\citet{vanBoekel17} suffered from the low wind effect, but  
despite this the quality of the resulting disk images is good and certainly
competitive with near-IR observations from other 
AO instruments \citep{Akiyama15,Rapson15}.    

\paragraph{Central star brightness and color.}
The AO performance degrades for faint stars, because of the lack
of photons for accurate measurements and corrections of the wave front 
distortions. In addition the WFS shares the photons 
in the ``visual'' range $500-900$~nm with the ZIMPOL science channel.
The gray beamsplitter (zw.BS) reflects only 21~\% of the light 
to the WFS and therefore the AO performance degrades significantly for stars
fainter than about $R\approx 8^m$ \citep[see][]{Sauvage16a}. 
This limit is relaxed to about $R=9.2^m$, 
if the dichroic beam-splitter is used instead of the gray beam
splitter, but then the useful spectral range for ZIMPOL is reduced to
the N\_R-filter and the line filters B\_Ha, N\_Ha, CntHa and OI (see
Fig.~\ref{FigFilters}).  
For faint central stars there are means to optimize the AO system with 
longer integrations with the WFS camera, running with reduced 
AO-loop frequencies 
of 600~Hz or 300~Hz instead of 1200 Hz, and the use of a 
large spatial filter in the WFS arm \citep[see][]{Sauvage16a}.
A mirror instead of a beam splitter is used 
for infrared science observations and therefore more light
reaches the WFS and the corresponding limit is about
$R\approx 10.0^m$ . 

Figure~\ref{PSFextreme}(b) shows as example for a faint star
the central regions of 47 Tuc MMS12, the central
star of the astrometric field from Fig.~\ref{Figfields}. This star
has only $R=10.5^m$ and was observed with the gray beam splitter
under mediocre atmospheric conditions
at high airmass and therefore the resulting PSF is strongly 
downgraded. When compared to the N\_I band PSF of the bright 
HD 129502, which was observed 
under similar seeing conditions, then
the faint star in 47 Tuc MMS12 has a $10 \times$ lower normalized peak flux
(or Strehl ratio), $3\times$ lower encircled flux $E_{f10}$, and 
a $2\times$ enhanced FWHM (Table~\ref{TabPSFAO}). Figure~\ref{Figfields} 
demonstrates that the resulting image can still be useful, 
but the broad and extended PSF is strongly reducing the spatial 
resolution and the contrast performance, and produces a much higher 
read-out noise limit for faint stars. 
\smallskip

The AO-correction may also depend on the color of the central star.
A strong wavelength dependence of the PSF parameters is reported
by \citet{Schmid17} for the Mira variable R Aqr
with very red colors $V-I=7^m$. The
normalized peak counts ${\rm ct}_{nb}(0)$ show
a strong wavelength dependence with $0.54$~\% for the I-band but only
$0.09$~\% for the V-band (Table~\ref{TabPSFAO}). The explanation is most 
likely, that the WFS ``sees'' essentially only I-band light, because 
of the very red color of the star, and therefore the
AO-system performs less good in the V-band.      

\paragraph{PSF structure and instrument mode.}
A few special instrumental effects are noticeable in the
PSFs shown in Fig.~\ref{PSFextreme}(a) and (c). 

Figure~\ref{PSFextreme}(a) shows the PSF of the star HD 142527
taken with the VBB filter with a filter widths of $\Delta\lambda=290$~nm
for a central wavelength of $\lambda_c=735$~nm 
or $\lambda_c/\Delta\lambda=2.53$. 
In this case the speckles are strongly extended radially. This has 
the disadvantage that the ring of bright speckles is very broad,
but on the other side the flux of an individual speckle is distributed
over many pixels and therefore the elongated speckles are less prominent,
and they can be distinguished more easily from a faint point source
companion. 

The star $\alpha$ Eri A with its fainter companion \object{$\alpha$ Eri B} 
were taken during the SPHERE commissioning with the narrow 
line filter CntHa with a very short integration of 10~ms 
using the ``snap shot'' engineering mode 
of the ZIMPOL detector, and a single 10~ms exposure is
shown in Fig.~\ref{PSFextreme}(c). For our data we 
measure for the companion a relative separation of $175\pm 4$~mas,
a sky position angle of $-19.5^\circ\pm 2.0^\circ$, and a flux ratio
of about $f_B/F_A=0.014$. 
This position measured by us for 2014-10-13 (2014.78)
coincides with the location for epoch 2007.75
in the partial orbit measured by \citet{Kervella08}, indicating
an orbital period of about $7.0\pm 0.1$~yr for the \object{$\alpha$ Eri} 
binary. 
Because of the narrow filter and the very short
integration the speckles are strong and point like because
there is no or only little radial and temporal smearing, respectively. 
For averaged PSFs 
or PSFs taken with long integration times like in Fig.~\ref{PSFVI}, the 
pattern of the strongly variable speckles will average out and only strong
quasi-static speckles remain clearly visible. We note, that the
PSF parameters given in Table~\ref{TabPSFAO} are derived
from averaged PSFs of $n_{\rm DIT}$-images where the PSF of each image
was re-centered individually before adding it to the others.

The PSF peak for $\alpha$ Eri is very narrow, 
because slow instrumental drifts and vibrations with 
frequencies $<100$ Hz do not broaden the PSF peak in very short 
integration. Further, there is no spectral dispersion because the
line filter CntHa produces essentially monochromatic images. 
For these reasons, the PSF of $\alpha$ Eri has 
a FWHM of $\approx 19$~mas, 
which is about 1 pixel better than for the observations taken
under ``excellent'' conditions or 2 pixels better than for 
``good'' conditions (Table~\ref{TabPSFAO}). 

Not corrected instrumental vibrations can also be recognized in the frame
transfer trails of a bright source. These trails wiggle left
and right with a scatter of $\sigma \approx 1-2$~pix from
the mean position and one such excursion extends over about 
200 pixels in vertical direction
corresponding to a time scale of 10~ms during the frame transfer.
This residual PSF jitter explains partly the 
discrepancy of $2-3$ pix between observed widths of the PSF peak 
and the nominal width of a diffraction limited PSF profile.

Lucky imaging and speckle suppression techniques can be optimized with
a careful selection of instrument and observing parameters, 
which enhance or reduce the speckle variations between frames 
or which modify the geometric appearance of speckles 
\citep[e.g.,][]{Law09,Brandner16}. 
It is beyond the scope of this paper to investigate further such 
techniques apart from pointing out that ZIMPOL offers a broad 
range of instrument modes which can be exploited. 

\section{Coronagraphy}
\label{Sectcoro}
Coronagraphy is a powerful tool to suppress the light of a
bright object for the search of faint sources at small separation
\citep{Malbet96,Sivaramakrishnan01}. 
The basic concept of the SPHERE visual coronagraph is a classical Lyot 
coronagraph where the focal plane mask 
stops the light of the bright star and the pupil mask
stops the diffracted light from the
telescope and the coronagraph in a pupil plane further downstream.  
SPHERE has a focal plane exchange wheel with 16 positions equipped
with 7 round focal plane masks, 2 monochromatic four quadrant 
phase masks (4QPM), two open field stops (see Table~\ref{Coropara}), 
and 4 empty positions. There is also an exchange wheel with 8 
positions for pupil stops (Table~\ref{Pupilpara}).  

The $F$-ratio of the coronagraph is $F\# = 30$ 
and the pupil size in the collimated beam is 6~mm in diameter.
The focal plane masks serve also as field 
stop for ZIMPOL defining either a circular ``wide field'' (WF) with
a radius of $\rho=4''$ or a ``narrow field'' (NF) of $1''\times 1''$
for a very fast, windowed detector read-out mode for polarimetry, 
which was not commissioned yet.

We use the stellar system \object{$\alpha$ Hyi} (HR 591, HIP 9236) 
as a test target for the on-sky characterization of the SPHERE 
visual coronagraphs. $\alpha$ Hyi is a bright $m_{\rm V}=2.9^m$, 
nearby (22~pc), F0 star and was used as test target in the commissioning.
It turned out to have a nearby companion which is ideal 
for the characterization of the SPHERE visual 
coronagraph. \object{$\alpha$ Hyi B} is not in the SIMBAD data base but
the Hipparcos catalog gives for $\alpha$ Hyi a binary orbit for 
the astrometric photo-center motion with an orbital period 
of $P=606$~days and a semi-major axis of 21.7~mas. 
On 2014-10-10 we measure a separation of $91\pm 3$~mas and
a position angle of $-19^\circ\pm 2^\circ$. The companion is about
$5.8^m\pm 0.2^m$ and $6.5^m\pm 0.2^m$ fainter than the primary 
in the I\_PRIM and R\_PRIM bands, respectively 
(see Table~\ref{CoroFlux}).
 
The atmospheric conditions for our coronagraph tests were
photometric, with an atmospheric coherence time of 
about $\tau_0\approx 2.8$~ms and a seeing of roughly $0.83''$
(see Table~\ref{TabPSFAO}).

\subsection{Coronagraphic focal plane masks}

\begin{figure}
\includegraphics[width=8.8cm]{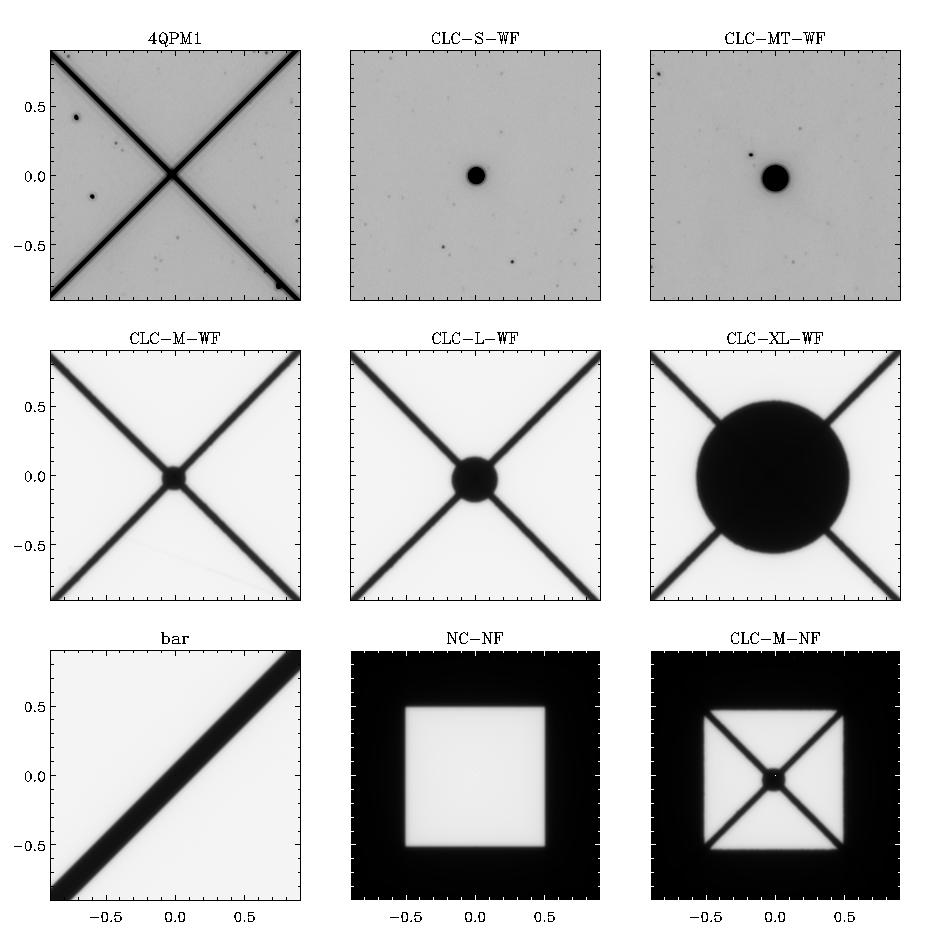}
\caption{Images of the central part ($1.8''\times 1.8''$) of focal plane
masks available in the visual coronagraph of SPHERE. The masks in the 
top row are deposited on a plate and the gray scale is enhanced to 
show the frequency of dust features on the masks.}
\label{CoroMaps}
\end{figure}

Figure~\ref{CoroMaps} shows images of the central region  
($1.8''\times 1.8''$) of nine focal plane masks obtained
with flat-field lamp illuminations. For non-coronagraphic observations, 
the clear field stop NC\_WF (or NC\_NF for window mode) are used. 
Non-coronagraphic observations
can also obtained by offsetting the star from the mask, but then a
shadow from the coronagraph will be present in the image and the pupil
stop is still in place. Such ``offset PSFs'' taken before or after 
coronagraphic observations are very useful for the flux calibration 
and the PSF characterization. 

Several Lyot coronagraphs with different spot sizes are available. 
Small spots, suitable
for small inner working angles, are made of a metallic coating
deposited on transparent plates with the disadvantage that dust 
particles on the plate are visible in the recorded images.  
In Fig.~\ref{CoroMaps} the gray scale for CLC\_S\_WF, CLC\_MT\_WF
(and 4QPM1) is enhanced to illustrate the frequency 
of such dust features as seen for our tests in Oct.~2014.

Larger coronagraphic spots are suspended with thin wires. Dust are no
problem for these masks but the suspension spiders, which have a
full width of 40~$\mu$m = 34~mas, can be an important issue for 
the observing strategy and the data reduction.  

\begin{table}
\caption{Focal plane masks in the 
SPHERE visual coronagraph}
\label{Coropara}
\begin{tabular}{lccclcc}
\noalign{\smallskip\hrule\smallskip}
Name      & p & $\rho$      & Remark     
                        & \multispan{2}{\hfil max(${\rm ct}_{\rm 6n}$)\hfil} \\
          &   & [mas]       &  & {\hspace{-3mm}{\small R\_PRIM}\hspace{-1mm}}
                            & {\hspace{-2mm}{\small I\_PRIM}\hspace{-2mm}}  \\
\noalign{\smallskip\hrule\smallskip}
NC\_WF      &     &         & clear      &  7983         &  7813  \\
NC\_NF      &     &         & clear       \\
\noalign{\smallskip}
CLC\_S\_WF  & $+$ & 46.5    &            &  72            & 52    \\
CLC\_M\_WF  &     & 77.5    &            &  26            & 13    \\
CLC\_MT\_WF & $+$ & 77.5    & astrom. mask \\   
CLC\_L\_WF  &     & 155     &            &  31            & 11    \\
CLC\_XL\_WF &     & 538     &            &  7.5           & 2.7   \\
\noalign{\smallskip}
CLC\_S\_NF  & $+$ & 46.5    & not tested  \\
CLC\_M\_NF  &     & 77.5    & not tested  \\
\noalign{\smallskip}
4QPM1       & $+$ &         & $\lambda_0=666$~nm  \\
4QPM2       & $+$ &         & $\lambda_0=823$~nm  \\
\noalign{\smallskip}
bar         &     &         & $w=155$~mas  \\
\noalign{\smallskip\hrule\smallskip}
\end{tabular}
\tablefoot{The first column gives the mask name where NC stands for 
no coronagraph, CLC for classical Lyot coronagraph, 4QPM 
for four quadrant phase mask, WF for wide field
(radius $\rho=4''$), NF for narrow field ($1''\times 1''$). Masks on
plates $p$ are indicated in the second column and $\rho$ is the mask radius
in $\mu$m and converted to mas using the scale 0.86 mas/$\mu$m.
The last two columns give the maximum normalized counts for 
the $\alpha$ Hyi tests (Fig.~\ref{CoroPSF}) for the R\_PRIM and 
I\_PRIM filters, respectively.} 
\end{table}

\begin{figure}
\includegraphics[trim=0.5cm 0.3cm 2cm 0.5cm,clip,width=8.8cm]{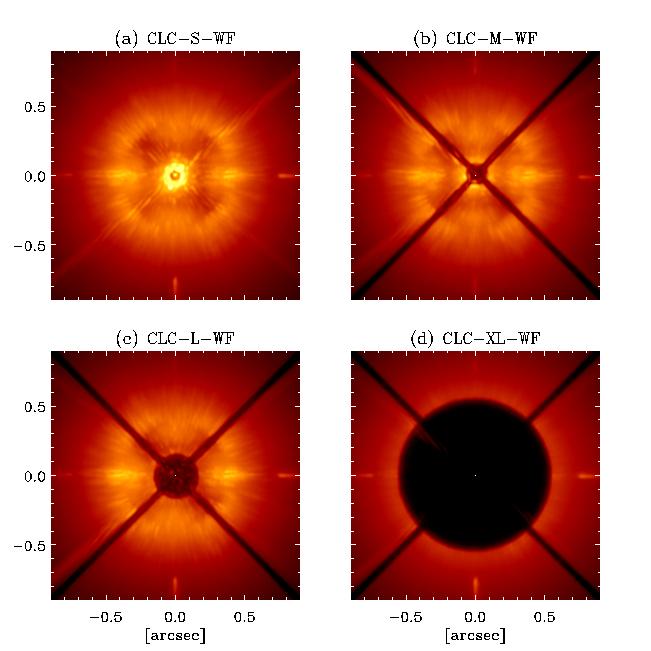}
\caption{Coronagraphic images of $\alpha$ Hyi ($1.8''\times 1.8''$) 
for the I\_PRIM-filter taken with the (a) small S, (b) medium M, (c)
large L, and (d) extra-large XL focal plane masks of SPHERE
All images are normalized ct$_{\rm n6}$ relative to the non-coronagraphic
PSF and displayed with the same color scale.} 
\label{CoroPSF}
\end{figure}

\begin{figure}
\includegraphics[trim=1.5cm 12.8cm 2.5cm 2.0cm,clip,width=8.8cm]{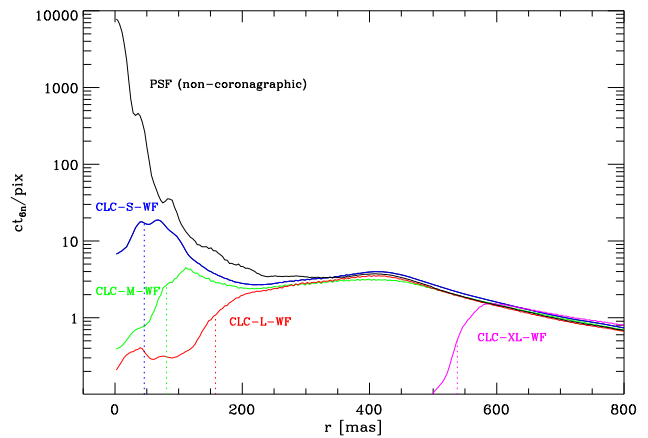}
\caption{Normalized radial profiles ct$_{\rm 6n}(r)$ for the I\_PRIM
coronagraphic images in Fig.~\ref{CoroPSF} of $\alpha$ Hyi 
taken with the small S (blue), medium  M (green), large L (red), and
extra-large XL (magenta) focal plane stops. Also plotted is the
non-coronagraphic PSF profile. The dotted lines indicate the
nominal mask radii.} 
\label{Coroprofiles}
\end{figure}

The coronagraphic attenuation of the PSF of $\alpha$ Hyi for the 
small CLC-S-WF, medium CLC-M-WF, large CLC-L-WF, and extra large 
CLC-XL-WF Lyot spots are shown for the I\_PRIM filter data in 
Fig.~\ref{CoroPSF} as coronagraphic images and in Fig.~\ref{Coroprofiles} as 
azimuthally averaged radial profiles. These coronagraphic data were taken 
simultaneously in the I\_PRIM filter for cam1 and the 
R\_PRIM filter for cam2 using the pupil 
stopB1\_2. Figure~\ref{Coroprofiles} includes a
stellar PSF taken with the star offset by 550~mas from 
the mask spot and using
the neutral density filter ND2  
to avoid heavy saturation. The non-coronagraphic
PSF was normalized to $10^6$~ct for an aperture with a diameter of
$3''$ and all the coronagraphic profiles were scaled to this 
ct$_{\rm n6}$-normalization considering the
exposure times and the attenuation of the neutral density filter 
ND2 (about a factor 130 for R\_PRIM and 105 for I\_PRIM) used for the 
non-coronagraphic observations. The two last columns in 
Table~\ref{Coropara} give the maximum counts of the
ct$_{\rm n6}$-normalized images.  

Depending on the spot size different features of the PSF are attenuated.
The smallest mask CLC-S-WF reduces the flux peak by about 
a factor $R_{\rm coro} \approx 110-150$, which corresponds to the 
ratio of the ${\rm max}({\rm ct}_{\rm 6n})$-counts between off-mask PSF 
and coronagraphic image given in Table~\ref{Coropara}.
This mask leaves a ring of strong residuals at a separation of 
about 60~mas just outside the coronagraphic spot.  
The next larger mask CLC-MT-WF ($\rho=81$~mas) attenuates the central peak 
completely $R_{\rm coro}\approx 300-600$, 
and the flux just outside the mask rim is comparable to
the strong features in the speckle ring at $\rho \approx 0.3''-0.4''$.    
For CLC-L-WF the speckles in the speckle ring are
the strongest emission features, while for CLC-XL-WF also all 
bright speckles in the ring are hidden and the
coronagraphic attenuation reaches $R_{\rm coro}\approx 1000-3000$.

Very useful for the determination of the stellar position in the coronagraphic 
images, in particular for the large
focal plane masks, are the interference features from the DM 
at $\rho \approx 0.6''-0.9''$ above,
below, left and right from the hidden star. They are essentially grating 
spectra of the central star created by the chessboard pattern from
the DM actuators which can be recognized in the pupil 
image (see Fig.~\ref{CoroPupilstop}).
 
Another focal plane mask in the visual coronagraph is the
astrometric mask CLC-MT-WF, which has a Lyot spot with a transmission
of about 0.1~\%, so that the central star can be seen in the 
science image as a faint emission peak inside the spot shadow. This 
is very useful for astrometric measurements of circumstellar
sources. This mask has no
suspension wires which is an advantage when compared to CLC-M-WF.
In addition, the mask has a Cartesian grid of smaller spots spaced by 
$1''$ for the astrometric
location of sources observed in off-axis settings 
where the central star is outside the detector field of view. 
A central spot is often also visible for the small opaque 
mask CLC\_S\_WF, but this spot is
produced by diffraction and should not be trusted for
astrometric purposes. 
\smallskip

The four quadrant phase masks (4QPM) have a geometry as shown in 
Fig.~\ref{CoroMaps} (top left). The plates produce a phase shift 
of $\pi$ through a small optical path difference in the left 
and right quadrants with respect to the upper and 
lower quadrants introducing a destructive interference at the 
interfaces and a particularly efficient nulling for the central 
crossing \citep{Rouan00}. The simple, monochromatic 4QPM 
used in ZIMPOL are designed for
one wavelength $\lambda_0$ and the attenuation is less 
efficient for wavelengths away from 
$\lambda_0$ \citep{Riaud03}. 4QPM1 with $\lambda_0=666$~nm 
is designed for the N\_R-band filter 
the H$\alpha$ line filters B\_Ha, N\_Ha and CntHa. 
The 4QPM2 with $\lambda_0=820$~nm is foreseen for the 
N\_I filter and the narrow band filter Cnt820.

The bar mask could be useful for the search of faint objects near
a small separation ($\rho<2.5''$) binary star. The orientation of
the binary can be aligned with the bar in field stabilized mode 
with an offset of the field position angle. We have not 
tested the performance of this mask.   

\begin{figure}
\includegraphics[width=8.8cm]{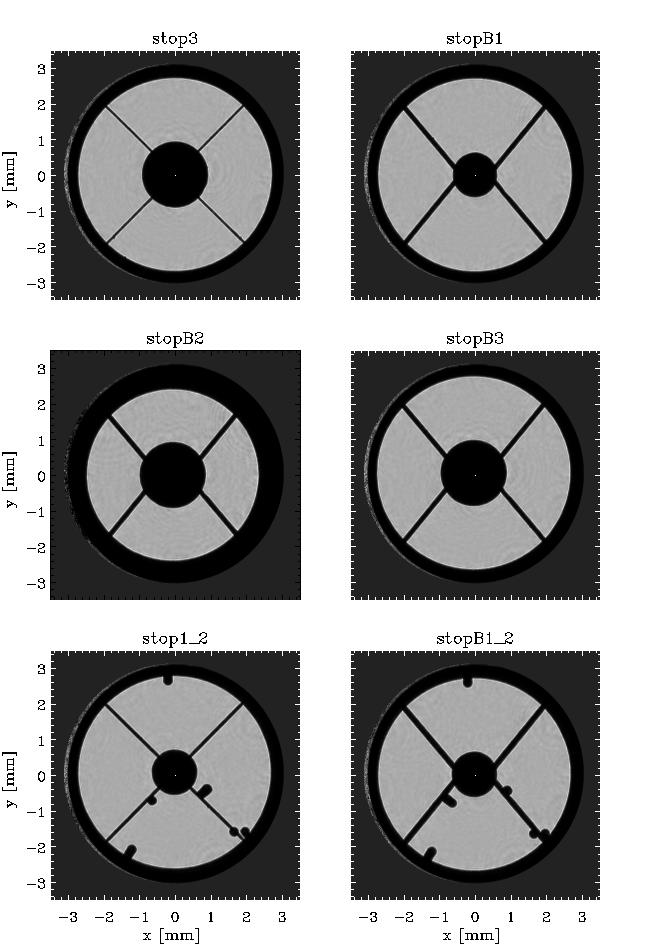}
\caption{Pupil stop images for the SPHERE visual coronagraph
taken with the internal point source and the pupil imaging lens.}
\label{pupilmaps}
\end{figure}

\subsection{Pupil plane stops}
An important part of classical Lyot
coronagraphs and 4QPMs are pupil plane stops which 
suppress the light diffracted by 
the sharp edges of the telescope pupil and the coronagraphic
mask in the focal plane \citep[e.g.,][]{Sivaramakrishnan01}. 
In the coronagraph pupil the diffracted light is located 
in two bright rings, one just outside the central M2-mirror obstruction 
and one along the outer pupil edge. In the coronagraphic image
the pupil masks reduce efficiently the strong inner diffraction rings 
and this effect is quantified in the following
for a test with the SPHERE visual coronagraph.
In addition, a pupil stop can also suppress the cross-shaped
pattern from the VLT M2-mirror spider and the ``unfocused''
light from bad actuators of the deformable mirror (DM).

\begin{table}
\caption{Geometric parameters for the telescope pupil, the
intermediate pupil, and the pupil stop masks in the SPHERE 
visual coronagraph.}
\label{Pupilpara}
\begin{tabular}{lccccc}
\noalign{\smallskip\hrule\smallskip}
            & $d_{\rm in}$ & $d_{\rm out}$ & $w_{\rm spid}$ & geom. & $T_{\rm geom}$  \\
            & [mm]       & [mm]        & [mm]     &           &           \\
\noalign{\smallskip\hrule\smallskip}
telescope     & 1148   & 8200    & 41  &  VLT    \\   
coro. pupil   & {\it 0.896}       
                       & {\it 5.97}     
                                 & {\it 0.037} 
                                          &  VLT  &      \\
\noalign{\smallskip pupil stops \smallskip}
clear        
             & --      & 6.60    & --     & --    & $100$~\% \\
\noalign{\smallskip}

stop3        & 1.80    & 5.40    & 0.036  & diag. & {\it 72.9~\%} \\
\noalign{\smallskip}
stopB1       & 1.20    & 5.41    & 0.110  & VLT   & {\it 75.5~\%}  \\
stopB2       & 1.81    & 4.79    & 0.107  & VLT   & {\it 53.5~\%}  \\
stopB3       & 1.81    & 5.40    & 0.106  & VLT   & {\it 70.9~\%}  \\
\noalign{\smallskip}
STOP1\_2     & 1.2$^{\rm a}$        
                      & 5.4$^{\rm a}$        
                                & {\it 0.084} & diag.+b   
                                                  & {\it 74.8~\%}  \\
{STOPB1\_2\hspace{-2mm}}    
             & 1.2$^{\rm a}$        
                      & 5.4$^{\rm a}$         
                                & {\it 0.18} & VLT+b & {\it 72.6~\%}  \\
\noalign{\smallskip}
SAM$^a$      &        &          &            & 7 holes   & \\
\noalign{\smallskip}
\noalign{\smallskip\hrule\smallskip}
\end{tabular}

\tablefoot{The columns indicate the pupil or stop name, the diameter 
$d_{\rm in}$ and $d_{\rm out}$ of the inner stop and the outer edge, 
the width $w_{\rm spid}$ and the geometry of the spider arms, and 
the geometric transmission $T_{\rm geom}$. Roman fonts 
give design values or values measured for the
mechanical components, while italic fonts are values measured from 
pupil images;
\tablefoottext{a}{The sparse aperture mask is described in 
\citet{Cheetham16} and was not tested by us.}}
\end{table}

With ZIMPOL, the pupil plane can be imaged with cam2 using a pupil 
lens located in FW2, while cam1 takes simultaneously a focal 
plane (PSF) image. Images of the stops of the SPHERE visual channel 
are shown in Fig.~\ref{pupilmaps}, while Table~\ref{Pupilpara} lists
geometric parameters of the components. The indicated geometric 
transmission $T_{\rm geom}$ is derived from the pupil images and 
corresponds to the open area of the pupil stop with 
respect to the geometry of the telescope aperture as imaged in
the coronagraphic pupil plane ($=100$~\%).  

There are two basic
types of pupil stops, simple masks which cover only the diffraction 
rings along the pupil rims for observing modes without pupil stabilization
(stop3 and stop1\_2) and pupil masks which hide in addition the 
telescope spiders for pupil stabilized observations (stopB1, stopB2, stopB3
and stopB1\_2). Because of the bad actuators 
of the deformable mirror, special stops with blockers for the scattered
light from these actuators located close to the pupil rim or spiders
were manufactured (stop1\_2 and stopB1\_2).
Different inner and outer stop diameters allow an optimization for 
high throughput (stopB1, stop1\_2, stopB1\_2), for a good rejection 
of the diffracted light (stopB2), or for an intermediate case (stop3, stopB3). 
The masks stop1\_2 and stopB1\_2 were used in 2015 as default masks.

\begin{figure}
\includegraphics[trim=0.1cm 0.1cm 1.5cm 0.4cm,clip, width=8.8cm]{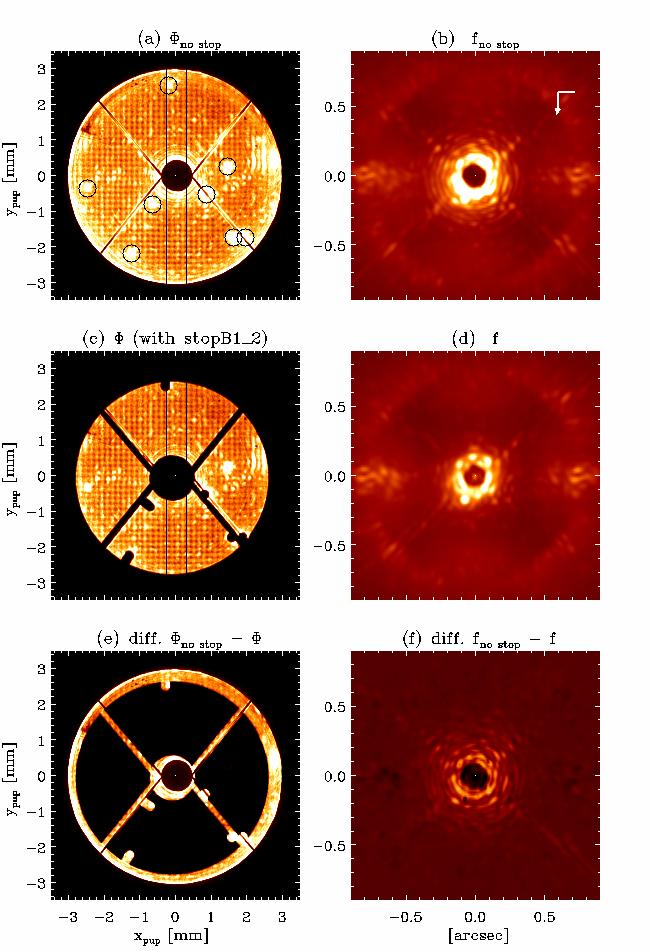}
\caption{Images of the pupil plane (left) and the center of the
focal plane (right) for a coronagraphic observation of $\alpha$ Hyi 
with the focal plane mask CLC-S-WF without pupil stop (top row)
and with pupil stop (middle row). The bottom panels are the
differences of the two upper panels.}
\label{CoroPupilstop}
\end{figure}

\begin{figure}
\includegraphics[trim=1.cm 13.cm 10.0cm 3.5cm,clip,width=8.8cm]{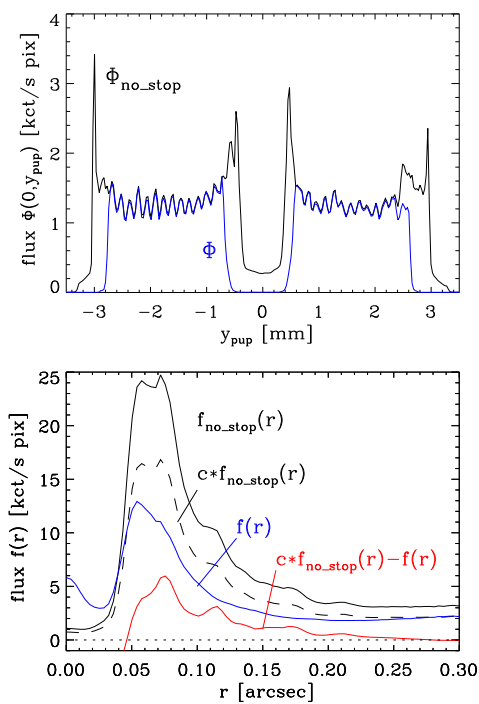}
\caption{Vertical pupil intensity profiles $\Phi$ (top) and
mean radial profiles $f(r)$ (bottom) for the coronagraphic 
observation (CLC\_S\_WF) of $\alpha$ Hyi without pupil stop (black)
and with stopB1\_2 (blue). The red curve in the bottom panel shows
the diffracted light in the focal plane which are suppressed 
by the pupil stop.} 
\label{CoroPupilProfiles}
\end{figure}

The effect of the stopB1\_2 for a coronagraphic
image taken with CLC\_S\_WF is demonstrated in Figs.~\ref{CoroPupilstop}
and \ref{CoroPupilProfiles} for the close binary $\alpha$ Hyi using
filter NB\_730 in FW0 and the pupil stabilized imaging mode. 
The panels in Fig.~\ref{CoroPupilstop} show in (a) the pupil 
image $\Phi_{\rm no\_stop}$ without stop and in (c) $\Phi$ with pupil 
stop, while panels (b) and (d) show the corresponding coronagraphic 
images $f_{\rm no\_stop}$ and $f$, respectively, where the faint companion B 
is located below and slightly left of the center
in these pupil-stabilized, non-derotated images. The bottom panels of 
Fig.~\ref{CoroPupilstop} show the differences of the two pupil
images $\Phi_{\rm no\_stop}-\Phi$ and the difference for 
the corresponding PSFs images $c\cdot f_{\rm no\_stop}- f$, where the
scaling factor $c=0.68$ accounts for 
the reduced off-axis flux transmission because of the geometric 
attenuation by the pupil stop.

The vertical cross section through the pupil and the radial
PSF profiles plotted in Fig.~\ref{CoroPupilProfiles} allow a 
more quantitative assessment of the different features in the pupil plane
and the focal plane. The profiles $\Phi(0,y)$ show clearly the
contribution of the extra light in the diffraction rings at 
$y\approx \pm 0.4$~mm and $\pm 3$~mm above the average level of about 
1300~ct/pix. The relative contribution of the
diffracted light is about $\Phi_{\rm diff}/\Phi_{\rm no\_stop}=5.0$~\%,
while the contribution of the 8 brightest maxima caused by bad actuators is 
$\Phi_{\rm dead}/\Phi_{\rm no\_stop}=1.5$~\%. In addition there
is also scattered light outside the pupil $r>3$~mm and
inside the central hole $r<0.4$~mm which
accounts for about $\Phi_{\rm scatt}/\Phi_{\rm no\_stop}=2.5$~\% of the
total light in the pupil.  

The pupil stop covers the diffracted light at the inner and outer
pupil edges, most of the dead actuator peaks, and the scattered 
light inside and outside the telescope aperture. Beside this,
the stop has a geometric transmission of $T_{\rm geom}=72.6$~\% of the 
telescope aperture. An approximate relation for the ratio 
of transmitted light with and without pupil stop is therefore
\begin{displaymath}
{\Phi\over \Phi_{\rm no\_stop}} \approx 
\Bigl(1 - {\Phi_{\rm diff}\over\Phi_{\rm no\_stop}} - 
{\Phi_{\rm dead}\over\Phi_{\rm no\_stop}} - 
{\Phi_{\rm scatt}\over\Phi_{\rm no\_stop}}\Bigr)\, \cdot \,
T_{\rm geom} \\
\approx 0.91\, T_{\rm geom}
\end{displaymath} 

The alignment of the pupil stop in Fig.~\ref{CoroPupilstop} is
offset by about 0.13~mm toward the 
left and slightly downwards. For this reason the telescope spiders 
on the upper left and lower right are not hidden by the pupil 
mask spiders. Fortunately, the misalignment is sufficiently small, 
so that the special light blockers still attenuate the 
unfocused light from the dead actuators. 
The noticed small misalignment is certainly not ideal, 
but also not devastating because the telescope spiders are not an 
important flux feature ($< 1$~\%) in the pupil. 

Important is the attenuation of the diffracted light in the pupil
for the reduction of the diffraction rings 
in the final coronagraphic science image. Because of 
the diffraction suppression the faint component B 
of $\alpha$ Hyi is much better visible in the coronagraphic image 
taken with pupil stop. Figure~\ref{CoroPupilProfiles} compares
the radial profiles for the coronagraphic images with $f(r)$
and without pupil stop $f_{\rm no\_stop}(r)$. This comparison
has to consider the reduced effective telescope aperture 
and therefore throughput $c=T_{\rm geom}=0.726$ because
of the geometric attenuation of the stopB1\_2.
The difference $c\cdot f_{\rm no\_stop}(r)-f(r)$ 
(red curve in Fig.~\ref{CoroPupilProfiles}) illustrates
the suppression of the diffracted light from the central star 
by the pupil stop. The effect is particularly large in the
radial range $\rho\approx 0.06''-0.20''$. The level of light 
in the coronagraphic image is reduced by factors $\approx 0.63$ and 
$0.55$ at the diffraction ring peak separation of 
76~mas and 115~mas, respectively, and $\approx 0.71$ for 
the minimum at 94~mas. Thus the effect is very significant
near the coronagraphic mask. 

The gain in the contrast performance is not only a function 
of the residual coronagraphic flux $f(r)$, but also on the 
speckle noise near the coronagraphic mask. It is difficult 
to give a general ``gain value'' for a coronagraphic system, 
because the speckles depend so much on the observing conditions. 
For the one case investigated here for $\alpha$ Hyi,
the gain in S/N for the measurement of the companion 
at $\rho=0.091''$ is thanks to the suppression of the diffracted 
light by stopB1\_2 a factor of $\approx 1.9$  higher
than without stop as listed in Table~\ref{CoroFlux} under
test A.

\subsection{Coronagraphic performance at small inner working angle.}
\label{CoroPerf}
\begin{figure}
\includegraphics[width=8.8cm]{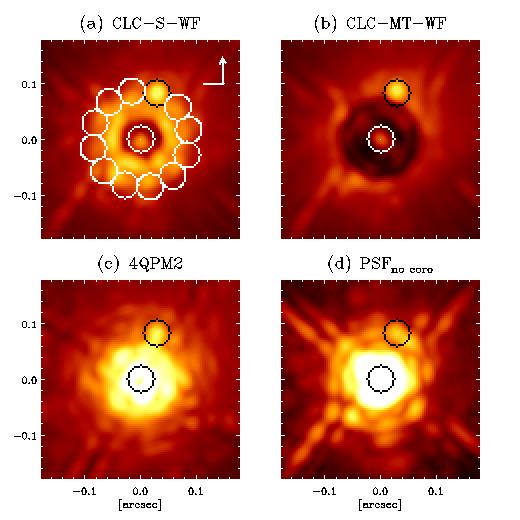}
\caption{\object{$\alpha$ Hyi} A and B observed in the I\_PRIM filter
with (a) the small and (b) the medium MT classical Lyot coronagraph, 
(d) without coronagraph , while (c) was taken in the N\_I band filter.
and the four quadrant phase mask 4QPM2. The circles are flux apertures 
for component A (center), component B at $\rho=91$~mas (black), while 
(a) shows also the concentric comparison apertures. North is up and
east to the left.}   
\label{CoroPSFcenter}
\end{figure}

Of great interest is a comparison of the different coronagraphs for the
detection and flux measurements of the faint companion to 
$\alpha$ Hyi separated by 91~mas.  
Figure~\ref{CoroPSFcenter} shows the innermost region
of the images obtained with (a) the small 
CLC-S-WF mask ($\rho=46.5$~mas), (b) the medium CLC-MT-WF
mask ($\rho=77.5$~mas), 
(c) the four quadrant phase mask 4QPM2, and (d) with 
non-coronagraphic observations. Apart from the mask, the 
instrument configuration for (a) and (b) is identical. For 
the non-coronagraphic image the primary star 
is offset from the focal plane mask (CLC-S-WF) by 550~mas and a ND2-filter 
is inserted in FW0 to avoid heavy saturation. These three data sets 
were taken with the I\_PRIM filter for cam1 
(shown in Fig.~\ref{CoroPSFcenter}) and the R\_PRIM filter for cam2. 
For the 4QPM2 images the N\_I filter for cam1 and the cnt820 filter
for cam2 are used which match the wavelength of the
phase mask.

\begin{table}
\caption{Measurements of the flux ratio $f_{\rm B}/f_{\rm A}$
and the $(S/N)_{\rm B}$ of \object{$\alpha$ Hyi} A and B, using different 
coronagraphic configurations of SPHERE/ZIMPOL.}
\label{CoroFlux}
\begin{tabular}{lllccl}
\noalign{\smallskip\hrule\smallskip}
\multispan{3}{\hfil Instrument configuration\hfil}       
                            & $f_B/f_A$   &  \hspace{-0.2cm}(S/N)$_{\rm B}$\hspace{-0.2cm} & Fig. \\
\multispan{3}{Coro / Angle, ~~~~~ Filter, $n_{\rm DIT}$$\times$$t_{\rm DIT}$[s]}          
                            & [$10^{-3}$] \\ 
\noalign{\smallskip\hrule\smallskip}
\noalign{\smallskip\noindent Test A: CLC-S-WF without / with pupil stop, 
                                         pupil stabilized \smallskip}
without stop &  NB730 & $10\times 8$  & $\phantom{i}$\tablefootmark{a} & 5.2    & \ref{CoroPupilstop}  \\   
with stop    &  NB730 & $10\times 8$  & $\phantom{i}$\tablefootmark{a} & 9.7    & \ref{CoroPupilstop}  \\   
\noalign{\smallskip\noindent 
Test B: different coronagraphic configurations \smallskip}
off coro,ND2\tablefootmark{b}  
                  &  I-PRIM  & 10$\times$1.1   & 5.6       & 4.8  &  \ref{CoroPSFcenter} \\   
CLC-S-WF         &  I-PRIM  & $10\times 3$     & 5.3       & 12.2  &  \ref{CoroPSFcenter} \\    
CLC-MT-WF        &  I-PRIM  & $10\times 3$     & 4.0       & 11.6  &  \ref{CoroPSFcenter} \\    
\noalign{\smallskip}
4QPM2            &   N\_I   & $20\times 3$     & $\phantom{i}$\tablefootmark{a}    
                                                       & 10.3   &  \ref{CoroPSFcenter} \\ 
4QPM2            &  Cnt820  & $20\times 3$     & $\phantom{i}$\tablefootmark{a}     
                                                      & 9.1     &     \\ 
\noalign{\smallskip}
off coro,ND2\tablefootmark{b}     
                 &  R-PRIM  & 10$\times$1.1 & 2.7       & 3.5    \\  
CLC-S-WF         &  R-PRIM  & $10\times 3$  & 3.2       & 8.2    \\ 
CLC-MT-WF        &  R-PRIM  & $10\times 3$  & 2.6       & 8.0    \\ 
\noalign{\smallskip\noindent 
Test C: different rotation angles with CLC-S-WF \smallskip}
0 degr     &  I-PRIM  & $20\times 3$   & 5.3           & 12.2   \\    
60 degr    &  I-PRIM  & $20\times 3$   & 3.7           & 5.2   \\     
120 degr   &  I-PRIM  & $20\times 3$   & 5.3           & 10.0  \\     
mean       &  I-PRIM  &                  & 
                \hspace{-0.1cm}4.7$\pm 0.8$\hspace{-0.1cm} 
                      & 10.0   \\   
median-subtr.  &  I-PRIM &               & 3.6\tablefootmark{c}         
                                                         & 15.9             \\ 
\noalign{\smallskip}
0 degr     &  R-PRIM  & $20\times 3$  & 2.7        & 5.8   & \ref{CoroRotation} \\     
60 degr    &  R-PRIM  & $20\times 3$  & 2.1        & 3.3   & \ref{CoroRotation} \\     
120 degr   &  R-PRIM  & $20\times 3$  & 3.1        & 4.6   & \ref{CoroRotation}\\     
mean       &  R-PRIM  &                 & 
              \hspace{-0.1cm}2.6$\pm 0.5$\hspace{-0.1cm} 
                            & 4.6      & \ref{CoroRotation} \\
median-subtr. & R-PRIM &                & 1.6\tablefootmark{c}       
                                                     & 10.5  & \ref{CoroRotation}  \\
\noalign{\smallskip\hrule\smallskip}
\end{tabular}
\tablefoot{
Commissioning data taken on 2014-10-10 between UT 7:40 and 8:50;
\tablefoottext{a}{no $f_A$ flux available; }
\tablefoottext{b}{PSF parameters for this test are 
given in Table~\ref{TabPSFAO} and they
are representative for all $\alpha$ Hyi data; }
\tablefoottext{c}{flux ratio affected by self-subtraction.}
}
\end{table}

In all these images the faint B component can be recognized, 
at least if its position is known. As illustrated in 
Fig.~\ref{CoroPSFcenter}(a), the flux $f_B$ is measured 
in the black aperture with a radius of $r=7$~pixels, subtracting 
the background level derived from the mean counts in the surrounding 
pixel ring with $7<r<8$~pix. This result is then corrected by the 
mean flux in the eleven white (``empty'') apertures 
at the same separation to account for a possible systematic effect 
introduced for example by diffraction rings. 
The standard deviation for the empty 
apertures is used as uncertainty $\sigma_{\rm B}$, and 
for the signal-to-noise ratio ${\rm S/N}=f_{\rm B}/\sigma_{\rm B}$ 
given in Table~\ref{CoroFlux}.
The flux of B is given as ratio $f_{\rm B}/f_{\rm A}$, where $f_A$ is
the flux of the primary measured in the non-coronagraphic PSF,
scaled with the transmission of the ND2-filter $T_{\rm ND2}=0.95~\%$, 
for I\_PRIM and $0.73~\%$ for R\_PRIM. 

The measured contrasts $f_{\rm B}/f_{\rm A}$ in Table~\ref{CoroFlux}
show a clear trend with wavelength, because the faint companion 
is redder than the primary star. For a given filter, similar
ratios $f_{\rm B}/f_{\rm A}$ are obtained for the different 
coronagraphs or instrument configurations and the 
$({\rm S/N})_{\rm B}$-values give a rough measure of the 
contrast performance. However, because
atmospheric seeing conditions and therefore AO performance are variable 
one should not give too much value on individual measurements. 

Table~\ref{CoroFlux} compares under test B the coronagraphic
detection performance of CLC-S-WF, CLC-MT-WF, and 4QPM2.
For all three coronagraphs the $({\rm S/N})_{\rm B}$ is about 
$2.0-2.5$ times higher than for non-coronagraphic images
taken at the same wavelength. This confirms the results of
the coronagraphic test A, which shows that a significant
fraction of the diffracted light is suppressed at small separations
by the coronagraph.  

Overall, the coronagraphic tests B show, 
that a faint companion at $0.091''$ separation with a contrast 
$f_{\rm B}/f_{\rm A}\approx 0.002-0.005$, which is
hardly or not visible in non-coronagraphic observations, is clearly
detected if a small or medium Lyot coronagraph, or a 4QPM is used. 

\subsection{High contrast observations and image rotation}

\begin{figure}
\includegraphics[trim=0.5cm 0.3cm 1.0cm 0.5cm,clip,width=8.8cm]{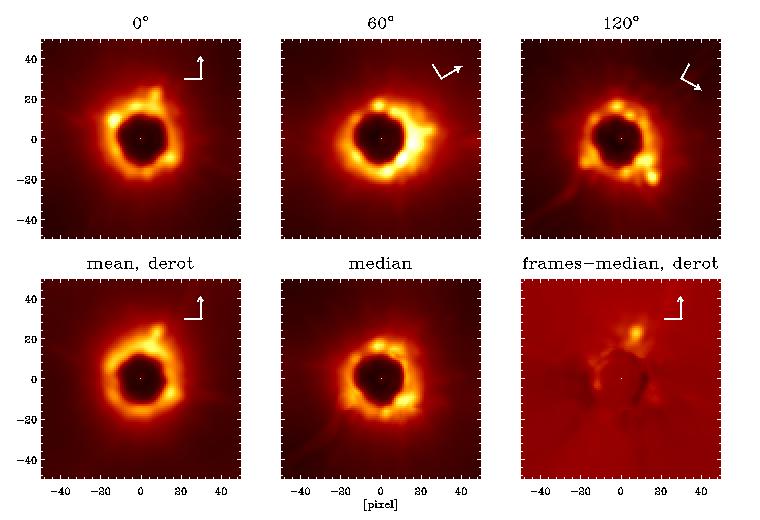}
\caption{Coronagraphic images of \object{$\alpha$ Hya}
taken in the R\_PRIM filter, the CLC-S-WF coronagraph and 
with image rotation 0$^\circ$, 60$^\circ$, and 120$^\circ$ 
(upper row). The lower row shows the mean of the derotated images (left),
the median of non-derotated image (middle), and the mean of the
rotated difference between non-derotated image and non-derotated median
(right) as result of a simple roll-angle angular differential imaging
procedure.}   
\label{CoroRotation}
\end{figure}

Field rotation is a very powerful strategy in high 
contrast imaging to correct for
quasi-static instrumental features in the PSF of the central star. With
so-called ``roll subtraction'' of two or several images 
taken with different sky orientations the stellar light can 
be significantly reduced and the companion becomes
more apparent \citep{Schneider03}. Even higher contrast levels
can be achieved with angular differential imaging which is 
based on a large number of frames taken
with continuous field rotation, where for each image or image section
a reference PSF is subtracted, which is derived from appropriately selected 
frames \citep{Marois06,Lafreniere07}, or a PSF fitting
procedure \citep{Amara12,Soummer12}.

Figure~\ref{CoroRotation} shows a very simple test for 
roll subtraction with SPHERE/ZIMPOL, based on three coronagraphic 
frames of $\alpha$ Hyi taken in field stabilized mode with
the different field orientation angles 
$\delta\theta=0^\circ$, $60^\circ$, and $120^\circ$ 
(see Fig.~\ref{CoroRotation}).
The individual frames show quite strong variations in the relative 
companion flux $f_B/f_A$ and signal-to-noise $({\rm S/N})_{\rm B}$
(Table~\ref{CoroFlux}, Test C), which are caused by atmospheric 
changes and the fact that $f_A$ was taken from a 
non-simultaneous measurement for the $f_B/f_A$ ratio.
Derotating and averaging the frames taken with  
$\delta\theta=0^\circ$, $60^\circ$, and $120^\circ$ does not enhance
much the $({\rm S/N})_{\rm B}$, most likely because the static 
noise features from the instrument are added to the mean
in three different orientations. Nonetheless, the mean flux ratio 
$f_B/f_A$ are good flux ratios for the $\alpha$ Hya system in the
I\_PRIM and R\_PRIM band, because these are means based on a total of
three minutes of observations.

Roll-subtraction is more efficient
in enhancing the contrast performance than simple averaging. 
For field-stabilized
observations, the telescope pupil rotates and therefore the
effects fixed to the telescope, like the spider pattern, are not
corrected. However, the orientation of the AO system and 
the coronagraph as seen by the detector remains unchanged 
in SPHERE/ZIMPOL because the derotator is
located further upstream. Figure~\ref{CoroRotation} shows
the median image of the three frames taken with 
$\delta\theta=0^\circ$, $60^\circ$, and $120^\circ$, which is then
subtracted from all three individual frames and the
resulting residual frames are derotated and averaged. 
Of course, the subtraction of a median image
from only 3 frames is only a very basic procedure and the
derived flux ratios suffer from self-subtraction 
effects (Table~\ref{CoroFlux}, Test C). However, this example 
shows that the final S/N-ratio is roughly doubled with
respect to an individual image and it
reveals in coronagraphic images of SPHERE/ZIMPOL quite a lot of fixed 
instrumental structures from the AO-system and the coronagraph, which
can be removed with image rotation. 

This kind of simple roll-subtraction can be achieved with only 
several, for example five exposures taken at different field 
position angles within a few minutes. Much more telescope time 
is required for higher contrast observations using the field rotation in
pupil stabilized mode or fixed derotator mode, 
because one needs to stay on target for a time of the order 
of an hour to achieve a sufficient sky rotation for 
an efficient ADI data reduction. 

\section{SPHERE/ZIMPOL imaging polarimetry}

ZIMPOL-polarimetry is a powerful differential imaging technique for high
contrast observations, because the opposite polarization 
directions are measured simultaneously and with the same detector 
pixel, that is through almost the same optical path yielding
images that are close to identical for an unpolarized source.
Therefore, the PSF of a bright, unpolarized central star,
including the variable residual AO speckles, is strongly suppressed 
in the differential signal, and a weak circumstellar polarization 
signal may become visible in the polarimetric images.

\subsection{Intrinsic polarimetric signal from the sky target}
\label{SectPolSky}
The SPHERE/ZIMPOL polarimetric imaging mode determines 
the linear polarization of a target
at each point in the field of view $(\alpha,\delta)$ described by the 
Stokes parameters, 
\begin{eqnarray}
Q(\alpha,\delta) &=& I_{0}(\alpha,\delta)- I_{90}(\alpha,\delta) 
                                  \quad {\rm and}  \\ 
U(\alpha,\delta) &=& I_{45}(\alpha,\delta)-I_{135}(\alpha,\delta)\,, 
\end{eqnarray}
where the indices of $I$ give the sky orientation for the 
electric field vector of the photons wave 
measured from North over East. The polarimetric images $Q$ or $U$ 
are vector components, unlike the intensity image $I$, and their 
signal can contain regions with 
positive and negative values. Therefore, the measured net signal
depends on the spatial resolution of the data, and 
strong cancelation can occur, if the intrinsic $+Q$ and $-Q$ 
quadrant pattern from circumstellar scattering is not well 
resolved \citep[see e.g.][]{Schmid06b}.
A quantitative polarimetric measurement should always consider 
these cancelation effects because of the
limited spatial resolution.  
\smallskip

Every polarimetric frame provides also a Stokes $I$-image or intensity 
frame which can be reduced and analyzed like ``normal'' imaging.
The intensity is simply the sum of the intensity components, either
\begin{eqnarray}
I_Q(\alpha,\delta) &=& I_{0}(\alpha,\delta) + I_{90}(\alpha,\delta) 
         \quad {\rm or} \\ 
I_U(\alpha,\delta) &=& I_{45}(\alpha,\delta) + I_{135}(\alpha,\delta)\,.
\end{eqnarray}
For a perfect instrument $I_Q=I_U=I$ and any deviations from this equality 
is a spurious measuring effect. 

The ZIMPOL $I_Q$- and $I_U$-frames obtained in polarimetric mode contain
in principle the same information like the frames taken in
imaging mode. The polarimetric $I_Q$ and $I_U$ have some properties which
can be beneficial for certain science applications, like:
\begin{itemize}
\item[--] a lower read-out noise for the 
slow polarimetry mode which is particularly suitable for faint sources,
\item[--] more accurate absolute and differential 
(2-channel) photometry, because the photometric throughput depends 
on instrument polarization effects which are controlled in 
polarimetric observations. This is an obvious problem for
an Nasmyth instrument, where the Al-coated VLT M3 mirror 
introduces in the ZIMPOL spectral range about $4~\%$ of polarization and
where channel splitting is done with a polarization beam splitter.  
\end{itemize}
Other features of the polarimetric mode can be disturbing 
for high contrast intensity imaging or
other high performance applications, in particular:
\begin{itemize}
\item[--] the presence of more detector pixel faults, caused by the 
charge shifting over pixels with non-optimal charge transfer efficiency,
requires a more sophisticated observing strategy and data 
reduction for bad pixel cleaning, 
\item[--] no pupil-stabilized polarimetry for angular differential 
imaging (ADI) -- instead there exists a static DROT-mode optimized 
for polarimetry. This provides a rotating field for ADI,
but also the pupil is rotating with a rotation rate different to
the field and therefore instrument features fixed to the pupil 
are less well suppressed,
\item[--] a reduced throughput ($\approx - 15~\%$) because of the additional
polarimetric components in the beam, 
\item[--] the presence of a few additional ghost features caused by
the polarimetric components.
\end{itemize}
These drawbacks might be minor when considering that
polarimetry provides intensity imaging with comparable quality
to ``standard'' imaging for many applications but 
delivers in addition a high performance polarization
measurement ``for free''. 
Therefore, it is always worthwhile to consider using the polarimetric
ZIMPOL mode for intensity imaging and profit from the additional
scientific information from the polarimetric signal.
\smallskip

The normalized Stokes parameters $Q/I$ and $U/I$ 
are important parameters for the characterization of astronomical sources,
but also for the determination of instrumental polarization effects from
the telescope and the instrument. In
SPHERE/ZIMPOL the instrumental polarization is to first order
field-independent and lower than 
$p_{\rm inst}=(Q_{\rm inst}^2+U_{\rm inst}^2)^{1/2}/I<1.0~\%$
as measured with zero polarization calibration stars (see next sections). 
In many science cases the central star can be 
a useful zero polarization calibration source for a correction
of the instrumental polarization assuming $Q_{\rm star}/I=U_{\rm star}/I=0$. 
This is achieved by applying a re-normalization, 
ensuring that $I_0=I_{90}$ and $I_{45}=I_{135}$ to the data.
If the polarization of the central star is indeed zero then
the residual polarization might be produced by a circumstellar 
polarization component $Q_{\rm cs}$ and $U_{\rm cs}$.
Such self-calibration procedures were often successfully applied 
to high contrast imaging polarimetry to disentangle 
the signal of a circumstellar polarization source from 
the instrumental polarization \citep[e.g.,][]{Quanz11,Avenhaus14}. 
This procedure corrects also 
for the interstellar polarization produced by the dust
along the line of sight. However, an accurate 
polarimetric measurement remains
an issue for self-calibrated data without a detailed assessment
of the instrumental polarization effects.

For a weak circumstellar emission of scattered light it is often 
much easier to measure the differential polarization signal
$Q_{\rm cs}$ and $U_{\rm cs}$ than the intrinsic intensity $I_{\rm cs}$. 
The measured intensity near a bright star is composed of an 
intrinsic circumstellar component $I_{\rm cs}(\alpha,\delta)$ and a 
dominating and strongly variable 
component $I_{\rm star}(\alpha,\delta,t)$ from the halo of the central star 
\begin{equation}
I(\alpha,\delta,t) \approx
I_{\rm star}(\alpha,\delta,t) + I_{\rm cs}(\alpha,\delta)\,.
\label{eqIvar}
\end{equation}
Often, there is $I_{\rm cs}(\alpha,\delta)\ll I_{\rm star}(\alpha,\delta)$ 
and therefore $I_{\rm cs}$ cannot be determined while the differential 
polarization $Q_{\rm cs}$ and $U_{\rm cs}$ can still be measured.
Without $I_{\rm cs}(\alpha,\delta)$, it is not possible 
to derive the fractional polarization 
$Q_{\rm cs}(\alpha,\delta)/I_{\rm cs}(\alpha,\delta)$ and  
$U_{\rm cs}(\alpha,\delta)/I_{\rm cs}(\alpha,\delta)$. 
For these cases, one can use the polarized surface brightness
contrasts $Q_{\rm cs}(\alpha,\delta)/I_{\rm star}$ and 
$U_{\rm cs}(\alpha,\delta)/I_{\rm star}$ which relate 
the measured circumstellar Stokes signals to the total flux
of the central star as measured in a large aperture, for example with
a diameter of $3''$. 
\smallskip

The Stokes fluxes $Q$ and $U$, and the corresponding
fractional Stokes parameters $Q/I$ and $U/I$ or contrast 
values $Q/I_{\rm star}$ and $U/I_{\rm star}$, are
the two components of a vector quantity, where the linearly polarized
flux $P(\alpha,\delta)$, or the corresponding fractional polarization
$p(\alpha,\delta)$, is the length of the vector and the
polarization position angle $\theta(\alpha,\delta)$ its
orientation (measured from N over E): 
\begin{eqnarray}
P(\alpha,\delta)      &=& \sqrt{Q^2(\alpha,\delta) + U^2(\alpha,\delta)}\,, 
\label{Poldef}\\
\theta(\alpha,\delta) &=& 0.5\cdot {\rm atan}\,\Biggl({U(\alpha,\delta)\over 
Q(\alpha,\delta)}\Biggr)\,.
\end{eqnarray}
The polarized flux $P$ and the 
polarization angle $\theta$ are well defined quantities, if 
the polarization signal $P$ is significantly larger than the noise.
For weak $Q$ or $U$ signals the individual pixel values will have
positive or negative signs because of the noise in the data. 
According to the definition of the polarized flux $P$ in Eq.~\ref{Poldef}
the $Q^2$ and $U^2$ terms count ``negative noise values'' like
positive values and a net polarization results even if the
mean $Q$- or $U$-signals are zero \citep[e.g.,][]{Clarke83}.
Therefore, one should use 
the individual Stokes $Q$- and $U$-values and images 
for the measurement of a weak polarization signal. 
The determination of the polarized flux $P$, the fractional 
polarization $p=P/I$, or the polarization contrast 
$C_p=P_{\rm cs}/I_{\rm star}$ needs to take the bias effect of noisy data
into account. For centro-symmetric linear polarization patterns, 
as expected for example for nearly centro-symmetric and 
optically thin circumstellar scattering, 
it can be very useful to define radial or azimuthal Stokes parameters 
with respect to the central object as proxy for the 
polarized flux, e.g. $Q_\phi\approx P$, as described in \citet{Schmid06b}. 

\subsection{Control of the polarimetric signal}
\label{Sectpolcontrol}
ZIMPOL measures according to Section~\ref{SectZIMprinciple} the differential 
polarimetric signal $P_Z(x,y) = I_\perp(x,y)- I_\parallel(x,y)$ (Eq.~\ref{EqPZ})  
at the position of the polarization modulator. This signal
includes the polarization from the sky target, but also the
polarization effects introduced by the telescope and the
SPHERE/ZIMPOL instrument. The system concept needs to control the 
polarization, so that the initial sky target signal can be reconstructed
from the measurements taken with different HWP2 orientations $i$ 
\begin{equation}
P_{i=1,2,..}^Z(x,y) \rightarrow  Q(\alpha,\delta)\,{\rm and} 
\, U(\alpha,\delta)
\end{equation}
using appropriate calibrations.

The polarimetric concept of SPHERE/ZIMPOL is complex, because
the instrument is fixed to the VLT Nasmyth platform. The strongly 
inclined M3 mirror of the telescope 
introduces already at the telescope focus a strong  
telescope polarization $I\rightarrow Q$ and 
polarization cross talks $U\leftrightarrow V$ 
\citep[see][]{Tinbergen07}. In addition there
is an image derotator in SPHERE, which introduces similarly 
strong polarization effects. 

For this reason, SPHERE/ZIMPOL uses an innovative concept 
which compensates and controls the instrumental polarization of 
the telescope and instrument with four steps. 
\smallskip

\begin{figure}
\includegraphics[trim=1.5cm 12.0cm 6.2cm 3.0cm,clip,width=8.8cm]{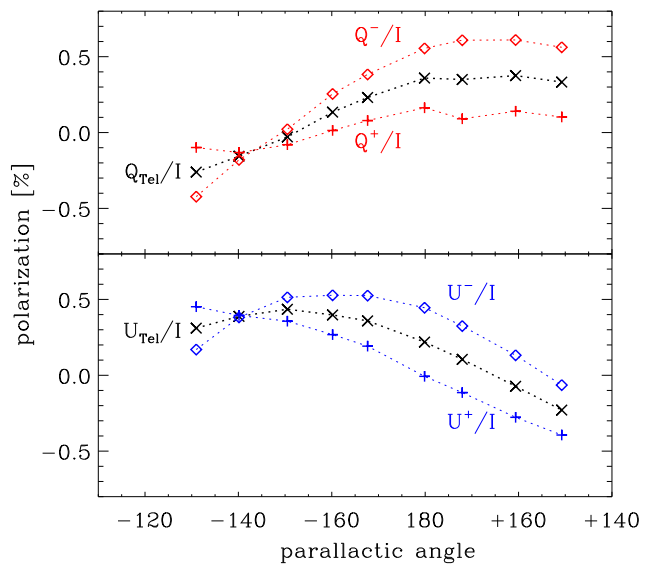}
\caption{Residual telescope polarization 
$q_{\rm tel}=Q_{\rm tel}/I$ and $u_{\rm tel}=U_{\rm tel}/I$ as function of
parallactic angle
for the unpolarized star $\epsilon$ Eri in the VBB filter.
Also shown are the measurements $q^+$, $q^-$, $u^+$, 
and $u^-$ of a polarimetric cycle, which include instrument 
polarization component $\pm p_{\rm SZ}$ for the P2-mode and a 
field position angle offset of $60^\circ$.}
\label{Figpoltel}
\end{figure}

\noindent
1. The M3 mirror of the VLT UT3 telescope
has an incidence angle of 45$^\circ$ and an Al-coating which produces 
in the $500-900$~nm range a polarization of about $3-5~\%$
and a strong $U\rightarrow V$ polarization cross talk. 
This polarization is compensated
with the rotatable, achromatic half-wave plate HWP1
and the following first mirror (PTTM) which has also   
an incidence angle of $45^\circ$ and a similar Al-coating.
HWP1 has an altitude $a$ dependent orientation 
$\theta_{\rm HWP1}=a/2+45^\circ$, which switches
the M3 polarization effects into a 
direction, like for a M3 mirror pointing in zenith direction. The following 
PTTM mirror acts then like a crossed mirror and compensates 
the M3 polarization effects to significantly smaller values, 
for the three components M3-HWP1-PTTM 
\citep{Joos07}. Figure~\ref{Figpoltel} shows 
for the unpolarized star $\epsilon$ Eri the
residual fractional telescope polarization
$q_{\rm tel}=Q_{\rm tel}/I$ and $u_{\rm tel}=U_{\rm tel}/I$
after M3-HWP1-PTTM. Hereafter, we call this also
 the telescope polarization because it is
fixed to the telescope and therefore changes with the parallactic
angle $\theta_{\rm para}$. The telescope polarization can be
corrected with calibrations of zero 
polarization standard stars (see Sect.~\ref{SectTelpol}).
A polarization compensation with a crossed-mirror
was already described by \citet{Cox76}
for an instrument fixed to the telescope tube, 
while \citet{MartinezPillet91} included also a rotating
half-wave plate for narrow-band applications with a Coud\'{e}-focus
instrument for solar physics. SPHERE/ZIMPOL introduces now
this approach for broad-band, high contrast imaging polarimetry
for an Nasmyth instrument at a stellar telescope 
\citep[see also][]{Tinbergen07}. 
\smallskip

\noindent
2. The fractional Stokes parameters for the 
instrument polarization of SPHERE/ZIMPOL $q_{\rm SZ}$ and $u_{\rm SZ}$
includes all components after the PTTM in CPI (Fig.~\ref{SPHEREblock}) 
up to the ZIMPOL polarization modulator. This polarization
is compensated with a measuring procedure, where
HWP2 is used as $Q^+$, $Q^-$, or $U^+$, $U^-$ polarization 
switch by applying to the HWP2 
position angle an offset cycle of $0^\circ$ and $45^\circ$, 
or $22.5^\circ$ and $67.5^\circ$, respectively. 
The $45^\circ$ offset, or $Q^-$ switch, reverses the
sign of the target and telescope polarization $Q + q_{\rm tel}I$, 
with respect to the $Q^+$ offset position, 
while the polarization effects introduced
by the following components $q_{\rm SZ}I$ and $u_{\rm SZ}I$ 
are unchanged. Subtracting the $Q^-$
from the $Q^+$ measurement cancels $q_{\rm SZ}$ 
after HWP2 and only $Q+q_{\rm tel}I$ remains 
\begin{eqnarray}
Q^+ - Q^- &=& [(Q+ q_{\rm tel} I) + q_{\rm SZ}I] 
                 - [-(Q + q_{\rm tel}I) + q_{\rm SZ}I] \\
         &=& 2\, (Q + q_{\rm tel}I)\,. \nonumber 
\end{eqnarray}
and similar for the $U$ polarization component. The polarization
switch is a well known technique to compensate instrumental
polarization to first order 
\citep[e.g.,][]{Appenzeller68,Kemp81,Scarrott83} and 
Fig.~\ref{Figpoltel} gives an example for SPHERE/ZIMPOL for
the unpolarized star $\epsilon$ Eri with an intrinsic polarization 
$Q/I$ and $U/I < 0.001$ \citep{Tinbergen79}. 
It is important to locate the HWP-switch as early as possible in 
the beam for keeping the non-compensated instrumental polarization
$q_{\rm tel}$ and $u_{\rm tel}$
introduced in front of the switch simple and easy to determine. 
\smallskip

\noindent
3. The derotator DROT in SPHERE, a three mirror system with
inclinations angles of $55^\circ$, $10^\circ$, and $55^\circ$, 
is a very critical component for polarimetric measurements
because it introduces an instrument polarization (diattenuation) 
$I\rightarrow Q$ of $m_{\rm 21}\approx 0.03$, 
and strongly wavelength dependent polarization
cross talks $U\leftrightarrow V$, which can be larger than $|m_{\rm 34}|>0.5$. 
To minimize the cross talk effects the linear polarization
direction to be measured by ZIMPOL is rotated by HWP2 into the 
$I_\perp$ and $I_\parallel$ orientation $\theta_{\rm DROT}$. 
This requires, that the HWP2 tracking law considers
the rotation of e.g. the $I_0$ sky orientation, 
compensates at the same time for the rotation effect of HWP1, 
and rotates the $I_0$ direction into the $I_\perp$ orientation 
of DROT depending on the derotator law. DROT
can either be in a fixed orientation for P1-mode, 
$\theta_{\rm DROT}=90^\circ$ 
or it moves like 
$\theta_{\rm DROT} = 0.5 ({\rm alt}-\theta_{\rm para})$ 
to stabilize the image on the detector in P2-mode. In these relation one
must consider the 180$^\circ$ angle periodicity of $\theta_{\rm DROT}$
for the image orientation.   

\smallskip

\noindent
4. The derotator polarization of about $p_{\rm DROT}\approx 3~\%$ 
is corrected with the polarization compensator PCOMP, 
an uncoated, co-rotating, inclined 
glass plate in ZIMPOL. This plate 
deflects more of the $I_\perp$-component than
the $I_\parallel$-component, so that $I_\perp-I_\parallel$ is reduced in the
transmitted beam (Sect.~\ref{Sectpolsetup}). 
For an inclination of 25$^\circ$, the
current value used for PCOMP, the derotator polarization and
other minor contributions from other components are
reduced to about 0.3~\%. 

In field-stabilized polarimetric imaging, or P2-mode, DROT
is not aligned with the orientation $\theta_{\rm Z}$ of 
the ZIMPOL polarimeter. Therefore, a rotatable half-wave plate 
within ZIMPOL (HWPZ), with an orientation 
$\theta_{\rm HWPZ}=\theta_{\rm DROT}/2$, is used to switch the
polarization to be measured into the $\theta_{\rm Z}$-orientation.
For P1-mode, the directions of $\theta_{\rm DROT}$ and $\theta_{\rm Z}$ are
identical and therefore HWPZ is not in the beam. 
The instrument polarization is essentially the same for 
the $Q$ and $U$ measurements, or $q_{\rm SZ} \approx u_{\rm SZ}$. 
(see Fig.~\ref{Figpoltel}), because the control  
of the polarization directions is the same after HWP2 . Typical values are 
$|q_{\rm SZ}|,\,|u_{\rm SZ}|\lapprox 0.3~\%$ 
for e.g. the V, N\_R, N\_I, or VBB-filters. 

\subsection{Calibrations for the polarimetric measurements}

The polarimetric measuring strategy includes   
four calibration steps:     
\begin{itemize}
\item{} c1 is  
the correction for the ZIMPOL 
modulation-demodulation efficiency $\epsilon_{\rm mod}$ 
as described in Sect.~\ref{Sectmodeff} (Table~\ref{Modeff}). For 
the determination of the fractional polarization in a large aperture 
also the frame transfer effect or $\epsilon_{\rm ft}$ (Eq.~\ref{Eqft})
needs to be considered. The calibration factors $\epsilon_{\rm mod}$ 
are obtained with a fully polarized illumination using the internal
flat field 
source and the polarizer in FW0. 
\item{} c2 is the subtraction of the telescope polarization  
which depends on pointing direction and filter. The telescope
polarization $p_{\rm tel}$ and $\theta_{\rm tel}$ 
is determined with zero-polarization standard star
calibrations. 
\item{} c3 is a small correction for the polarization efficiency loss
of the telescope mirrors M1 and M2 and that part (residual
polarization cross talk) of the M3-HWP1-PTTM  
configuration, which is not included in the modulation-demodulation 
efficiency calibration $\epsilon_{\rm mod}$. The effect is small 
with a correction factor close to one, which 
can be checked with high-polarization standard star calibrations.   
\item{} c4 is a position angle offset correction for the
effective HWP2 plate orientation to adjust the measured $Q,U$ parameters
to the North direction on sky. This polarization angle
offset $\delta_{\rm SZ}$ depends slightly on wavelengths
and is also calibrated with the high-polarization standard stars.  
\end{itemize}

The calibrations c1 and c2 are very important for all 
quantitative polarimetric measurements. The
corrections or adjustments c3 for $\epsilon_{\rm opt}$ and
c4 for $\delta_{\rm SZ}$ are only relevant for high signal-to-noise 
polarimetry ${\rm S/N}=p/\Delta p\gapprox 20$. 
However, all calibrations are essential for checking the proper working
of the instrument and one should be alarmed if unexpected 
calibration results are obtained. 

\begin{figure}
\includegraphics[trim=0.1cm 12.0cm 2.5cm 3.0cm,clip,width=8.8cm]{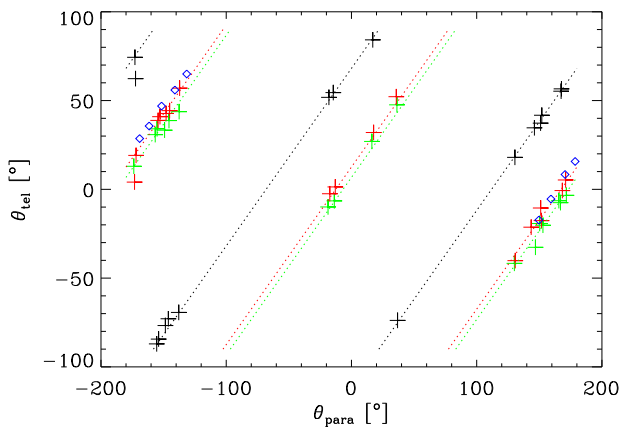}
\caption{Telescope polarization angle $\theta_{\rm tel}$ as 
function of the parallactic angle
for zero-polarization standard stars measured in the filters
V (green), N\_R (red), and N\_I (black) with dotted lines for the best
fits according to Eq.~\ref{Eqpoltel}. Blue diamonds are the VBB-filter
data from Fig.~\ref{Figpoltel}}.
\label{TelpolFig}
\end{figure}

\subsubsection{Calibration of the telescope polarization}
\label{SectTelpol}

The ``residual'' telescope polarization is regularly measured 
with zero-polarization
standard stars ($p_{\rm 0pol}\ll 0.1$~\%), as part of 
the ESO SPHERE instrument calibration plan. We use bright, nearby,
single solar type stars, preferentially from the
lists of \citet{Serkowski74} or \citet{Tinbergen79}.
These measurements consist 
typically of one full polarimetric cycle for each of the 
three filters V, N\_R, and N\_I in FW1 and FW2. Often ND-filters
are used in FW0 in order to avoid saturation. All these
calibration data are available in the ESO archive.  

For the analysis the measurements are bias subtracted
and corrected for the modulation-demodulation
efficiencies $\epsilon_{\rm mod}$ and the frame transfer effect
$\epsilon_{\rm ft}$. Integrated fluxes $Q,\,U$ and 
$I_Q,\, I_U$ are determined for synthetic 
apertures and fractional telescope polarization parameters
$q_{\rm tel} = Q/I_Q$, $u_{\rm tel} = U/I_U$, $p_{\rm tel}$ and the 
position angle $\theta_{\rm tel}$ is calculated. 

We have analyzed all useful zero-polarization standard 
star calibrations from 2015, which include 17 measurements for each
of the three filters V, N\_R, and N\_I. The 
derived telescope polarization values are in the range 
$p_{\rm tel} = 0.4-0.6$~\% with a scatter of only 
$\sigma_p\approx 0.04~\%$ for a given filter. As shown in Fig.~\ref{TelpolFig} 
the polarization position angle $\theta_{\rm tel}$
is proportional to the parallactic angle $\theta_{\rm para}$
with a wavelength dependent offset $\delta_{\rm tel}(\lambda_c)$
\begin{equation}
\theta_{\rm tel}(\theta_{\rm para},\lambda_c) = \theta_{\rm para} 
          + \delta_{\rm tel}(\lambda_c) \,,
\label{Eqpoltel}
\end{equation}    
where one must consider the 180$^\circ$ polarization angle periodicity. 
Table~\ref{TelpolTab}
lists for the zero-polarization standard stars the mean 
polarization $p_{\rm tel}$ and 
standard deviation $\sigma_{\rm p}$, and the mean position angle 
offset $\delta$ and $\sigma_{\rm \theta}$ from the best fit 
(Eq.~\ref{Eqpoltel}) also plotted in Fig.~\ref{TelpolFig}. 
The correlation between $\theta_{\rm tel}$ and $\theta_{\rm para}$
is expected because the compensation of the M3 mirror 
polarization with the HWP1 - PTTM combination yields
residuals which are to first order fixed to the telescope pupil. This
effect is also nicely seen in Fig.~\ref{Figpoltel} for the long 
data series of the unpolarized star $\epsilon$ Eri taken in Nov.~2016.

Table~\ref{TelpolTab} gives additional measurements for the 
wavelength dependence of
$p_{\rm tel}(\lambda_c),\,\delta_{\rm tel}(\lambda_c)$
based on multifilter data of \object{$\zeta$ Tuc} (HR 77)
taken during SPHERE commissioning in Oct.~2014.
The color dependence of $p_{\rm tel}$
and $\delta_{\rm tel}$ can be explained by  
differences in the M3 and PTTM mirror coatings
and some deviations of HWP1 from a perfect, broad-band half wave plate.  
The strong rotation of $\delta_{\rm tel}$ in the I-band is probably
related to the well known reflectivity minimum of Al-coatings in
this wavelength range. One might expect a slow temporal evolution 
of the telescope polarization 
because of the aging of the mirror coatings and perhaps a sudden change
associated with the M1 and M3 telescope mirror re-coating, which took
place in April 2017.     

Measurements of the fractional polarization $Q_{c1}/I$ and $U_{c1}/I$
of standard stars,
can now be corrected for the telescope polarization according to 
\begin{eqnarray}
Q_{c2}/I & = & Q_{c1}/I - p_{\rm tel}\, {\rm cos}\,
                    (2\,(\theta_{\rm para}+\delta_{\rm tel}))\, \label{Qtelcorr} \\
U_{c2}/I & = & U_{c1}/I - p_{\rm tel}\, {\rm sin}\,
                    (2\,(\theta_{\rm para}+\delta_{\rm tel})) \label{Utelcorr}\,.
\end{eqnarray}
The parameters $p_{\rm tel}$ and $\delta_{\rm tel}$ from Table~\ref{TelpolTab}
can be used for all observations from
2015 and possibly also for later observations. 
One should notice that
the corrected values $Q_{c1}/I$ and $U_{c1}/I$ for the standard star 
polarization parameters derived in large apertures must use
the factors $\epsilon_{\rm mod}\epsilon_{\rm ft}$ including the 
frame transfer smearing effect (Table~\ref{Modeff}).  

\begin{table}
\caption{Residual telescope polarization $p_{\rm tel}$ and
$\theta_{\rm tel}$ for the filters 
V, N\_R, and N\_I as measured from the standard star measurements 
taken in 2015 and the VBB filter measurements of $\epsilon$ Eri 
(Figs.~\ref{Figpoltel} and \ref{TelpolFig}.}
\label{TelpolTab}
\begin{tabular}{lcccccc}
\noalign{\smallskip\hrule\smallskip}
Filter   & $\lambda_c$ & $n_{\rm cal}$  
                 & $p_{\rm tel}$ 
                              & $\sigma_p$ 
                                         & $\delta_{\rm tel}$ 
                                                      & $\sigma_\delta$ \\
(F)      & [nm]  &       & [\%]       & [\%]     & [$^\circ$]  & [$^\circ$]    \\
\noalign{\smallskip\hrule\smallskip}
\noalign{\smallskip zero-polarization standard stars\smallskip}
V        & 554   & 17    & 0.57       & $\pm 0.03 $ & 6.7     & $\pm 3.2$    \\
N\_R     & 646   & 17    & 0.55       & $\pm 0.04 $ & 12.6    & $\pm 3.6$    \\
N\_I     & 817   & 17    & 0.42       & $\pm 0.04 $ & 68.1    & $\pm 3.8$    \\
\noalign{\smallskip $\zeta$ Tuc \smallskip}  
V\_S     & 532   & 1     & 0.53       &             & 5.8     \\
V        & 554   & 1     & 0.52       &             & 7.8     \\
V\_L     & 582   & 1     & 0.49       &             & 11.1     \\
N\_R     & 646   & 1     & 0.53       &             & 10.1     \\
NB730    & 733   & 1     & 0.46       &             & 16.9     \\
N\_I     & 817   & 1     & 0.44       &             & 71.5     \\
I\_L     & 871   & 1     & 0.52       &             & 72.1     \\
\noalign{\smallskip $\epsilon$ Eri \smallskip}  
VBB      & 735   & 9     & 0.41       & $\pm 0.02 $ & 16.7   & $\pm 1.6$  \\
\noalign{\smallskip\hrule\smallskip}
\end{tabular}
\end{table}

\subsubsection{Calibration of the polarization angle and efficiency}

Also high-polarization standard star measurements are taken
regularly in the V, N\_R, and N\_I filter as part of 
the ZIMPOL calibration plan. These data provide an
independent test for the corrections c1 and c2. 
For our analysis all well 
illuminated high polarization standard star calibrations from 
2015 are included. The measurements and data reduction
of high-polarization stars are identical to the zero-polarization
standards described above.

\begin{figure}
\includegraphics[trim=2.0cm 12.5cm 6.5cm 3.0cm,clip,width=8.8cm]{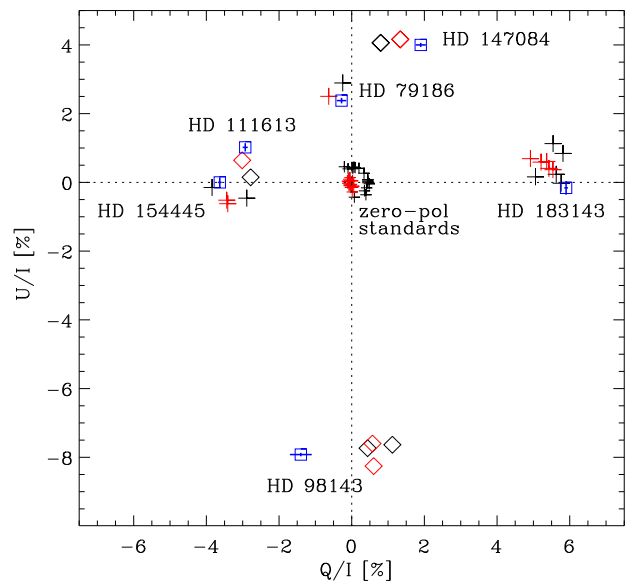}
\caption{Polarization of high polarization standard stars for the
   N\_R filter:
      black symbols give efficiency or c1 corrected values while
      red symbols
    are also corrected for the telescope polarization (c2-correction).
    For individual targets, either plus-signs or diamonds are used
    in order to avoid confusion. Corresponding literature values
    are plotted as blue squares with error bars. Also included
    are the zero-polarization standard stars from the previous subsection.}
\label{FigHighPol}
\end{figure}

Table~\ref{PolOffset} lists the obtained polarization 
values for the standard stars, after applying corrections
c1 and c2, and compares them to 
literature values. Figure~\ref{FigHighPol} shows the data for 
the N\_R-filter in the $Q/I-U/I$-plane
with black symbols for c1 corrected values,
and red symbols with the additional telescope correction c2 applied. 
The literature values are given in blue. 
Also shown are the 17 N\_R zero-polarization standard star 
data from the previous section located on a ring with
radius $r\approx 0.5$~\%, and the telescope corrected red points 
are clustered near the zero point with a small scatter
of $\sigma_p=0.12~\%$. A similar pattern of black 
points surrounding the telescope corrected red values is also present 
for the high-polarization standard HD 183143.

\begin{table}
\caption{Measured and corrected polarization $p_c$ [\%]
and $\theta_c$ [$^\circ$] of high-polarization standard stars and
comparisons with literature values $p_\ell,\theta_\ell$.}
\label{PolOffset}
\begin{tabular}{lcccccc}
\noalign{\smallskip\hrule\smallskip}
            & \multispan{2}{\hfil V \hfil} 
                       & \multispan{2}{\hfil N\_R \hfil} 
                                    & \multispan{2}{\hfil N\_I \hfil} \\
            & $p_{c,\ell}$  & $\theta_{c,\ell}$  
                       & $p_{c,\ell}$  & $\theta_{c,\ell}$   
                                    & $p_{c,\ell}$  & $\theta_{c,\ell}$  \\ 
\noalign{\smallskip\hrule\smallskip}
\noalign{\smallskip \object{HD 79186}, nobs = 1 \smallskip}
meas      & 2.75 & 51.1  & 2.58 & 52.1  & 2.18 & 52.5   \\
lit [1]   & 2.59 & 47.7  & 2.40 & 48.4  &      & (48.4) \\
$p_{\rm c}/p_{\rm \ell},\,\Delta\theta_{\rm c-\ell} $
          & 1.06 & +3.4  & 1.07 & +3.7  &      & (+4.1)   \\
\noalign{\medskip \object{HD 98143}, nobs = 2 \smallskip}
meas      & 7.90 & 135.6 & 7.95 & 137.1 & 7.29 & 137.1  \\
lit [2]
          & 8.03 &  133 & 8.04 & 130   & 7.10 & 130  \\
$p_{\rm c}/p_{\rm \ell},\,\Delta\theta_{\rm c-\ell} $
          & 0.98  & +2.6 & 0.99  & +7.1 & 1.03 & +7.1 \\
\noalign{\medskip \object{HD 111613}, nobs = 1 \smallskip}
meas      & 3.20  & 79.1 & 3.08 & 84.0 & 2.77 & 86.8  \\
lit [1]
          & 3.10  & 81.1 & 3.10 & 80.4 &      & (80.4)   \\
$p_{\rm c}/p_{\rm \ell},\,\Delta\theta_{\rm c-\ell} $
          & 1.03  & --2.0 & 0.99  & +3.6 &      & (+6.4) \\
%
\noalign{\medskip \object{HD 147084}, nobs = 1 \smallskip}
meas      &       &      & 4.37 & 36.1 &      &      \\
lit [1,3,4]
          & 4.15  & 31.7 & 4.44 & 32.2 & 4.40 & 31.5 \\
$p_{\rm c}/p_{\rm \ell},\,\Delta\theta_{\rm c-\ell} $
          &       &      & 0.98  & +3.9 \\
%
\noalign{\medskip \object{HD 154445}, nobs = 2 \smallskip}
meas      & 3.60 & 91.6 & 3.47 & 94.7 & 2.98 & 96.1 \\
lit [1,3,4] 
         & 3.72 & 89.9 & 3.63 & 89.5 & 3.29 & 90.5 \\
$p_{\rm c}/p_{\rm \ell},\,\Delta\theta_{\rm c-\ell} $
          & 0.97  & +1.7 & 0.96  & +5.2 & 0.91 & +5.6  \\
\noalign{\medskip \object{HD 183143}, nobs = 4 \smallskip}
meas     & 5.45 & 1.1   & 5.41 & 2.6  & 4.78   & 4.8    \\
lit [1,3,4] 
         & 6.15 & 179.2 & 5.81 & 178.9 & 5.36 & 179.0 \\
$p_{\rm c}/p_{\rm \ell},\Delta\theta_{\rm c-\ell} $
          & 0.89  & +1.9 & 0.93 & +3.7 & 0.89 & +5.8  \\
\noalign{\medskip {\bf mean values} \smallskip}
$\langle p_{\rm c}/p_{\rm \ell}\rangle$
              & {\bf 0.99}  & & {\bf 0.99} & & {\bf 0.94} & \\
$\langle\Delta\theta_{\rm c-\ell}\rangle$
              & & {\bf +1.5} & & {\bf +4.5}  & & {\bf +5.8} \\
$\sigma\,[\pm]$  & 0.07  & 2.1  & 0.05 & 1.4    & 0.08 & 1.1 \\
\noalign{\smallskip\hrule\smallskip}
\end{tabular}
\tablefoot{The data are corrected for the 
modulation efficiency $\epsilon_{\rm mod}$, the frame transfer smearing
$\epsilon_{\rm ft}$ and the telescope polarization. Literature values:
1: \citet{Serkowski75} for V- and R-band, $\theta({\rm R})$ adopted
also for $\theta({\rm I})$; 2: \citet{Whittet92}; 
3: \citet{Hsu82} with 0.75~$\mu$m band values extrapolated 
to N\_I-band (0.82~$\mu$m); 
4: \citet{Bailey82}.} 
\end{table}

The derived polarizations (red symbols in Fig.~\ref{FigHighPol}) for the 
high-polarization stars are slightly offset or
rotated in counter-clockwise direction 
when compared to the literature values of high polarization
stars. This is summarized in Table~\ref{PolOffset}, which compares 
for all three filter V, N\_R and N\_I the ratio $p_c/p_l$ between
the measured polarization $p_c$ and the literature value $p_\ell$, and 
the difference $\Delta\theta_{c-\ell}=\theta_c-\theta_\ell$ 
between the derived polarization angle $\theta_c$
and the corresponding literature value $\theta_\ell$.
The measurements were first corrected for the modulation efficiency
and frame transfer smearing $\epsilon_{\rm mod}\epsilon_{\rm ft}$ 
(Eq.~\ref{EqModeff}) and the telescope polarization (Eqs.~\ref{Qtelcorr}
and \ref{Utelcorr}). The bottom rows give as final result for 
each filter the average ratio 
$\langle p_c/p_\ell \rangle$ and the average angle difference 
$\langle \Delta\theta_{c-\ell} \rangle$ for the ZIMPOL 
V, N\_R, and N\_I filters. The average ratios 
$\langle p_c/p_\ell \rangle$ are compatible with 1, while 
$\langle \Delta\theta_{c-\ell} \rangle$ indicates at least for the
N\_R and N\_I filters significant polarization angle offsets 
$\delta_{\rm SZ}$ for SPHERE/ZIMPOL. 

The results of the high polarization standard star
  measurements should be considered as preliminary values.
  For HD 183143 it is clearly visible, that the scatter of $\sigma_p=0.52~\%$
  for the red corrected points in Fig.~\ref{FigHighPol}
  is significantly larger than
  the $\sigma_p=0.12~\%$ for the zero-polarization
  standard stars.

  Several effects contribute to this enhanced scatter
  in the high polarization standard
  star data. First, there could be instrumental cross talk effects
  $Q\rightarrow U,V$ or $U\rightarrow Q,V$, which may depend on
  instrument configuration, and become apparent 
  for objects like HD 183143 with a high linear polarization
  $p_{\rm star}\approx 5~\%$.
  Because these cross talks scale with $p$, they are much smaller
  for the zero-polarization standard stars. Second, the standard star
  measurements could be improved. They are not taken in a homogeneous
  way and would require a study on its own to minimize systematic measuring
  effects, e.g. differences between strongly and weakly illuminated data.
  Third, the
  polarization of the high polarization standard stars could be
  variable, and HD 183143 is apparently such a case \citep{Hsu82}.
  Achieving a smaller uncertainty $\Delta p < 0.5~\%$ in absolute
  polarimetric measurements of high polarization target with ZIMPOL
  is certainly possible but requires but requires an extensive
  analysis of data from well suited calibration targets which is beyond
  the scope of this paper. 

  We summarize here briefly the status of bright high polarization
  standard stars suitable for the calibration of ZIMPOL observations.
Only the more northern $\delta>-30^\circ$ 
objects \object{HD 147084} and
\object{HD 154445} were established as good calibration stars 
by \citet{Hsu82}, and their values are in good 
agreement $\Delta p\approx \pm 0.1~\%$, 
$\Delta\theta \approx \pm 1^\circ$ with \citet{Bailey82}, 
and \citet{Serkowski75}.  
As mentioned above \object{HD 183143} is known to show
some polarimetric variability 
\citep{Hsu82} and the same is true for several other high
polarization standard stars \citep{Bastien88,Bastien07}. 
The situation is even worse for the southern
Milky Way $\delta<-30^\circ$, because there exist essentially 
no well established bright high polarization standard stars 
suitable for SPHERE/ZIMPOL. The used objects 
\object{HD 79186}, \object{HD 98143}, and \object{HD 111613} have
been measured polarimetrically \citep{Serkowski75,Whittet92}, 
but the accuracy of the published values is unclear and the 
absence or the level of polarimetric 
variability has not been investigated.    

\subsubsection{Recommended polarimetric correction formulae}

The full correction of the measured $Q$ and $U$ signals of a science
target requires two steps: first the corrections c1 for 
the modulation efficiency $\epsilon_{\rm mod}$, c2 for the 
telescope polarization, and c3 for a small
efficiency loss because of the optical components:
\begin{eqnarray}
Q_{\rm c3} = \Bigl({1\over \epsilon_{\rm mod}}Q
       -I\, p_{\rm tel}(\lambda_c)\,{\rm cos}\,
       (2 (\theta_{\rm para}+\delta_{\rm tel}(\lambda_c)\Bigr)
          \,{1\over \epsilon_{\rm opt}}\,, \label{EqQc3} \\
U_{\rm c3} = \Bigl({1\over \epsilon_{\rm mod}}U
       -I\, p_{\rm tel}(\lambda_c)\,{\rm sin}\,(2 (\theta_{\rm para}+
          \delta_{\rm tel}(\lambda_c)\Bigr)
          \,{1\over \epsilon_{\rm opt}} \,. \label{EqUc3}
\end{eqnarray}
Note, that the polarization flux parameters $Q$ and $U$ are not affected
by the intensity dilution during the frame transfer, and therefore 
$\epsilon_{\rm ft}$ is not considered. 

In a second step the obtained 
values $Q_{\rm c3}$ and $U_{\rm c3}$ are then rotated ``polarimetrically''
to correct for the polarization angle offset of a few degrees,
\begin{eqnarray}
Q_{\rm c4} = Q_{\rm c3}\, {\rm cos}\,(2\delta_{\rm SZ})
                        +U_{\rm c3}\, {\rm sin}\,(2\delta_{\rm SZ})\,, \label{EqQc4}\\
U_{\rm c4} = U_{\rm c3}\, {\rm cos}\,(2\delta_{\rm SZ})
                        -Q_{\rm c3}\,{\rm sin}\,(2\delta_{\rm SZ})\,. \label{EqUc4}
\end{eqnarray}
Most important are the corrections c1 and c2 for the modulation
efficiency $\epsilon_{\rm mod}=0.75-0.91$, which 
depends on instrument parameters, and the additive
telescope polarization $\Delta(Q/I),\Delta(U/I)\approx -0.6$ to $+0.6$~\%. 
The corresponding parameters $\epsilon_{\rm mod}$ are given in
Table~\ref{Modeff} and $p_{\rm tel}(\lambda_c)$ and $\delta(\lambda_c)$ 
for the telescope polarization are listed in Table~\ref{TelpolTab}.
The corrections c3 and c4 for the optical efficiency 
$\epsilon_{\rm opt}$ and the position angle offset 
$\delta_{\rm SZ}$ are only important for strongly
polarized objects $Q/I,U/I\gapprox 2$~\% which can be measured
with a high polarimetric signal-to-noise  $S/N=\Delta p/p\gapprox 20$.
The value $\epsilon_{\rm opt}$ and $\delta_{\rm SZ}(\lambda)$
have not been determined yet with high precision and we 
recommend a value of 
$\epsilon_{\rm opt}=1/\langle p_c/p_\ell \rangle=0.98\,(^{+0.02}_{-0.06})$
and 
$\delta_{\rm SZ}(\lambda)=\langle\Delta\theta_{\rm c-\ell}\rangle$
according to Table~\ref{PolOffset}. 

We think that SPHERE/ZIMPOL provides a substantial progress
in quantitative polarimetry for AO assisted, high
resolution observations because the instrument polarization effects 
are well controlled and calibrated with accurate standard star measurements.
Therefore, quantitative measurements can also be obtained for
weak, non-axisymmetric and other demanding targets. 
Infrard instruments, like SPHERE/IRDIS
\citep{Langlois14,vanHolstein17} and
GPI \citep{Perrin15} provide also
calibrated polarization measurements so that the diagnostic
potential of multi-wavelength polarimetry can be exploited.

\subsection{Differential polarimetric beam shifts}
An important requirement for the performance of a 
high-contrast, differential polarimetric imager are small
aberrations between the measured $I_\perp$ and $I_\parallel$ images. 
In ZIMPOL, these two polarization modes are 
registered with the same detector pixels to minimize
optical aberrations. Small differential beam shifts between $I_\perp$ and
$I_\parallel$ can be produced by non-perfect half-wave plates and this
was considered in the ZIMPOL design. During instrument tests 
small beam shifts from the modulator assembly
were measured, but they were less than 0.1 pixels ($<0.4$~mas) 
and considered to be acceptable \citep[see][]{Roelfsema10}.

\begin{figure}
\includegraphics[width=8.8cm]{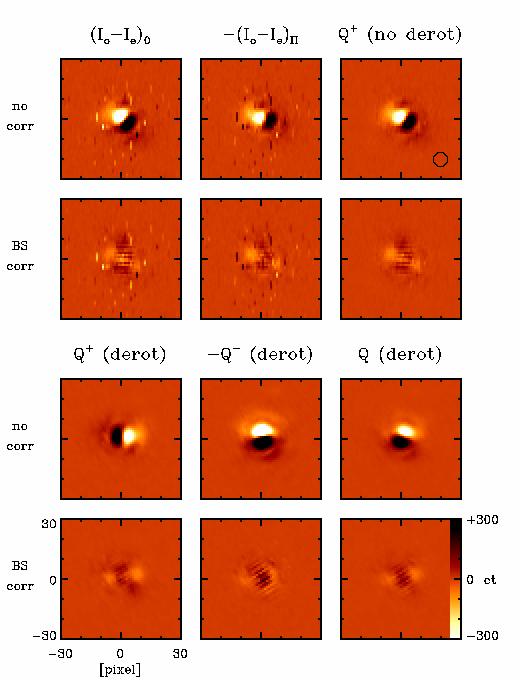}
\caption{Observed polarimetric beamshift for a stellar PSF
of $\alpha$ Cen B taken in June 2016 with the VBB and ND4 filters:  
The first row shows two consecutive polarization frames 
$(I_o-I_e)_0$ and $(I_o-I_e)_\Pi$ taken with opposite, 
zero and pi demodulation phase shifts and the resulting, 
charge-trap corrected, Q$^+$-frame. The third row illustrates 
the beamshift for the 
derotated Q$^+$-, Q$^-$-, and the final ``science'' Q-image. 
The second and bottom rows show the same but with
the beamshifts corrected between the odd-row $I_o$ and even-row 
$I_e$ sub-frames.}
\label{BSPSF}
\end{figure}

Unfortunately, we detected after the integration of ZIMPOL into 
SPHERE a substantially larger and unexpected beam shifts 
of up to 0.3 pixels (or $\approx 1$~mas) between $I_\perp$ 
and $I_\parallel$ caused by the inclined mirrors in CPI. 
These differential polarimetric beam shifts
are apparent in $Q$ or $U$ images as systematic positive and
negative features on opposite sides of a stellar PSF 
(see Fig.~\ref{BSPSF}) and this affects the speckle suppression
capabilities of the polarimetric mode. This subsection 
outlines the origin of this effect, illustrates the impact,
and gives a recipe for correcting the effect.

\subsubsection{Polarization aberrations from inclined mirrors}  
Extensive SPHERE/ZIMPOL testing showed that the main contributions 
to the differential polarimetric beam shifts
are caused by the strongly inclined mirrors M3 of the telescope, the
$45^\circ$ pupil tip-tilt mirror (PTTM), and the image derotator
mirrors, while the half wave plates are
at most minor contributers.  

PSF shifts between $I_\perp$ and $I_\parallel$ are introduced
by inclined metallic mirrors because 
the Fresnel coefficients for the phase shift are different 
for $I_\perp$ and $I_\parallel$ as described by \citet{Breckinridge15}
in a comprehensive analysis of polarimetric aberrations of telescopes. 
The same type of beam shifts causing deviations from the 
law of geometric optics on sub-wavelength scales are also known 
in optics as Goos-H\"ahnchen (GH) and Imbert-Federov shifts
\citep[e.g][]{Aiello09,Bliokh13}. 
These shifts are not well known for astronomical instruments, 
because the spatial resolution or/and polarimetric sensitivity 
of previous instruments did not achieve 
the SPHERE/ZIMPOL performance. Therefore, we describe here
the dominant effect and provide a simple calculation for the
expected PSF shift. 

\begin{figure}
\includegraphics[trim=3.3cm 13.5cm 3.8cm 3cm,clip,width=8.8cm]{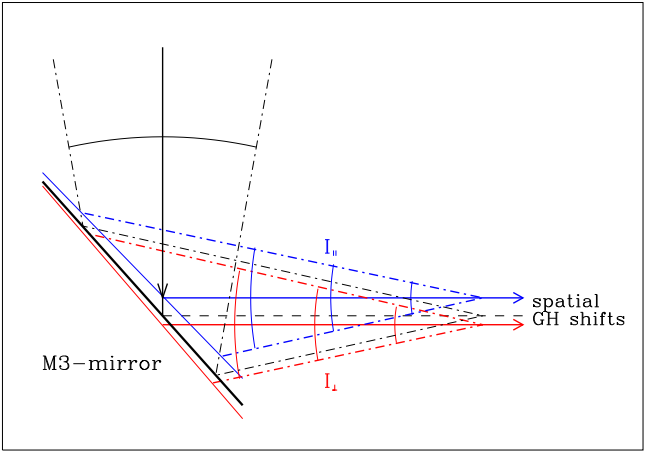}
\caption{Schematic and simplified illustration of the 
polarimetric beam shift effects for the M3 Nasmyth mirror. 
The incoming beam and the expected reflection according to 
geometric optics are plotted in black. The beam 
and wavefront displacements for $I_\perp$ and $I_\parallel$ caused by the
phase shifts, with the corresponding ''effective'' mirror surface 
location and tilts, are hugely exaggerated and drawn in 
red and blue respectively.}
\label{BSM3}
\end{figure}

A beamshift between the two polarization
components $I_\perp$ and $I_\parallel$ reflected from an inclined 
surface was first described by \citet{Goos47}. The
reflected $I_\perp$ and $I_\parallel$ components are 
shifted as expected for slightly offset ``effective'' mirror 
surfaces for the $I_\perp$ and $I_\parallel$ components. This 
is illustrated in Fig.~\ref{BSM3} with a simplified ray model
for the principal beam (solid black, blue and red line with
arrows) for 
two hugely exaggerated offsets for the ``effective'' mirror surfaces. 
The induced beam displacements are attributed to the phase shifts
$\phi$ introduced in the reflection, which depend on inclination
angle $\theta$. 
As shown by \citet{Artmann48} with an angular
spectrum decomposition and interference for a spatially limited beam,
the incidence angles of the incoming wave sections
vary around the incidence angle $\theta_0$ and therefore 
the waves receive an angle dependent phase shift in the reflection. This 
explains a spatial Goos-H\"ahnchen shift $\Delta$, which is proportional 
to the phase shift gradient $d\phi/d\theta$ 
\begin{equation}
\Delta_{\perp,\parallel} = 
-{\lambda\over 2\pi}\,{d\phi_{\perp,\parallel}\over d\theta}\Big|_{\theta_0}\,.
\label{EqGHshift}
\end{equation}
For metallic mirrors, positive and negative shifts are introduced 
for $I_\perp$ and $I_\parallel$, respectively, because 
$d\phi_{\perp}/d\theta$ and $d\phi_{\parallel}/d\theta$ have opposite 
signs. There exists also a transverse
beam-shift, called spatial Imbert-Fedorov shift \citep[e.g][]{Bliokh13},
which is also of some importance. The so-called angular shifts,
another type of the beam-shifts caused by the angle
dependent reflectivity of metallic mirrors, are not relevant 
for the image position in a focussed beam. 
It is beyond the scope of this paper to describe 
here in detail all these effects.

The beam shift can also be explained for a ``macroscopic'' view
of the converging beam of a Nasmyth telescope with an inclined mirror M3 
as sketched for the propagation of the wavefronts in Fig.~\ref{BSM3},  
and described in more detail in \citet{Breckinridge15}.
In this case, the angles of incidence $\theta$ on M3 are different, 
$45^\circ+\epsilon$ and $45^\circ-\epsilon$, for the ``left'' and ``right'' 
sections of the incoming spherical wave fronts in Fig.~\ref{BSM3}. 
Consequently, also the phase shift introduced 
by the reflection is different 
$\phi(45^\circ +\epsilon)$ and $\phi(45^\circ-\epsilon)$, so that the 
interfering
spherical wave sections converge along a slightly different angle, 
when compared to the angle predicted by geometric optics.  
When neglecting higher order terms, then the 
estimated focal point shift is about 
\begin{equation}
\Theta_{\perp,\parallel} \approx 
{\lambda\over 2\pi}\, {d\phi_{\perp,\parallel}\over d\theta}\Big|_{\theta_0} {1\over F\#}  {1\over D}
   = {\lambda\over 2\pi}\, {d\phi_{\perp,\parallel}\over d\theta}\Big|_{\theta_0} {1\over f} \,,
\end{equation}
where $F\#=f/D$ is the $F$-ratio for the converging or diverging beam, and 
${1/ F\#}$ the corresponding angular spread for the focused beam.
Because the phase shifts gradients $d\phi/d\theta$
are different and have even different signs for $I_\perp$ and $I_\parallel$
for metallic surfaces also positive and negative focal point
shifts $\Theta_\perp$ and $\Theta_\parallel$ result 
(Fig.~\ref{BSM3}).   
The focal point shifts are proportional to the inverse of
the focal length $f$ and therefore the focus displacement 
expressed in nm in the focal plane does not depend
on the telescope size and is just given (to first order) by
the phase shift gradient of the mirror coating for the incidence
angle $\theta_0$
\begin{displaymath}
\Delta_{\perp,\parallel} = f\cdot \Theta_{\perp,\parallel} = {\lambda\over 2\pi}\,
{d\phi_{\perp,\parallel}\over d\theta}\Big|_{\theta_0}  \,. 
\end{displaymath}
This is the same result as in Eq.~\ref{EqGHshift}, because both, 
a narrow beam and a macroscopic telescope beam are made of an 
angular spectrum of plane-parallel wavefronts impinging onto an
inclined mirror. 

The PSF-shifts $\Delta_{\perp,\parallel}$ do not depend on the $F$-ratio, 
the focal lengths, or the aperture diameter of the focussed beam,
but only on $\theta_0$ and the mirror surface. 
For an Al-mirror and $\lambda=800$~nm
the phase shift gradients is in units of wavelength 
per radian 
\begin{displaymath}
{d\phi_\perp\over d\theta}\Big|_{45^\circ} = 0.151 \quad {\rm and} \quad
{d\phi_\parallel\over d\theta}\Big|_{45^\circ} =-0.299 \,. 
\end{displaymath}
Multiplication with $\lambda/2\pi$ yields then the PSF-shifts
19.2~nm and $-38.1$~nm in the focal plane, respectively, and the 
differential PSF shift of 
\begin{displaymath}
\Delta_\perp - \Delta_\parallel = 57.3~{\rm nm} \,.
\end{displaymath}
For a telescope with the focal length of the VLT 
($f=120$~m) this differential PSF shift in the focal plane 
corresponding to an angle shift of 
$\Theta_\perp - \Theta_\parallel = 0.101$~mas on sky. 
The derived shift of 57.3~nm 
agrees with the result of \citet[][Table 2]{Breckinridge15}, who obtained 
for the same $\lambda$, $\theta_0$ and mirror surface a focal point
shift of $\Theta_\perp - \Theta_\parallel=0.625$~mas ($=3.15\cdot 10^{-9}$~rad)
for a telescope with $f=19.2$~m. This is essentially the 
same shift $\Delta_\perp - \Delta_\parallel = 60.5~{\rm nm}$ 
as calculated for the VLT, considering that the used
$d\phi_{\perp,\parallel}/d\theta$-values used by us could be slightly different.
 
The same principles for polarimetric beam shifts 
apply for all inclined mirror in a converging or diverging beam.
This are the Al-coated M3 mirror of 
the telescope, the Al-coated $45^\circ$ pupil tip-tilt mirror (PTTM),
and the image derotator mirrors. The beam-shift
introduced by the PTTM is essentially the same as for M3, except for the
different mirror orientation. Our measurements show
that the derotator is the dominant component for the beam shift in the
I- and R-band range, while the shift in the V-band is comparable 
to the M3 mirror. The strong effect from the derotator is
not surprising, because it consists of three mirrors, 
two 55$^\circ$-mirrors and one 20$^\circ$-mirror and the shifts 
of the two 55$^\circ$-mirrors add up. They have overcoated
silver surfaces and the strong wavelength dependence for the
beam shift is probably caused by this type of reflecting surface, for
which we do not know the $d\phi/d\theta$ phase shift properties. 

\subsubsection{Correction of the polarimetric beam shift}

The SPHERE/ZIMPOL beam shifts are the results of a 
complex combination of several contributing
components, mainly the inclined M3 mirror of the VLT,
the $45^\circ$ PTTM mirror, and the three mirrors in the 
image derotator. Moreover, the half wave plates HWP1, HWP2
and HWPZ rotate the polarization angle of the shifted 
components at different locations along the light paths. 
Several of these components rotate during
the observations and therefore each exposure requires
an ``individual'' correction which is only valid for a given
sky position and instrument configuration. 
In particular, the beam shifts for the four exposures of
a polarimetric cycle $Q^+$, $Q^-$, $U^+$ and $U^-$ are all different
(see $Q^+$ and $Q^-$ in Fig.~\ref{BSPSF}). Another significant
complication is that the beamshift from the derotator is
strongly wavelength dependent with a large, $\approx 1$~mas shift, in
the I-band, about 0.5~mas in the R-band and less in the V-band. 
Because of this complexity, there exists up to now no
instrument model for the correction of the differential beam shifts. 

However, a beamshift correction can be applied to the science data, 
if the offset between the science images in the odd-row and even-row 
subframes $I_o$ and $I_e$ can be determined, so that
the two frames can be accurately aligned
before the polarimetric combination of the frames is carried out 
(see Fig.~\ref{BSPSF}). 

\begin{figure}
\includegraphics[trim=0.1cm 0.1cm 1.5cm 0.5cm,clip,width=8.8cm]{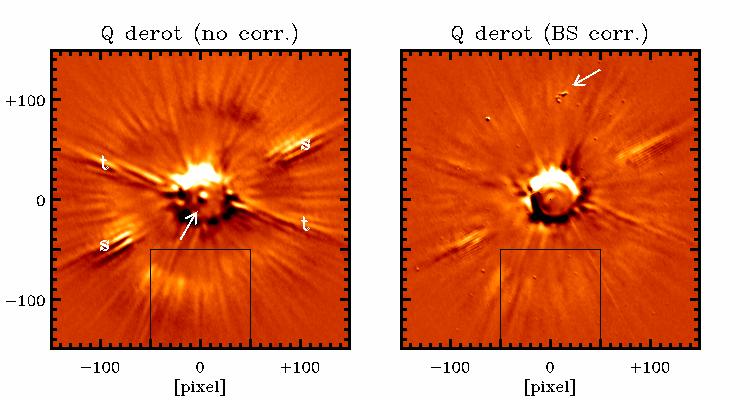}
\caption{Beam shift effect for $\alpha$ Cen B in the
VBB filter for a 
coronagraphic (CLC-MT-WF) Stokes $Q$ frame (left) and the beam shift 
corrected frame (right) for essentially the same instrument configuration
as the stellar PSF in Fig.~\ref{BSPSF}.
The arrow in the left panel points to the PSF peak 
transmitted through the mask. PSF residual features are marked
in the left panel with ``t'' and ``s'' 
according to Fig.~\ref{PSFVI}. In the right panel 
the arrow points to one of several spots from the camera, which 
show up because of the applied beam-shift correction. 
The squares at the bottom mark the area 
for the reported $\sigma(Q)$.}
\label{BScoro}
\end{figure}

It is more difficult to recenter frames
without well defined intensity peaks, e.g. frames with
low Strehl ratio, with saturated PSF, or coronagraphic 
images. For these cases, one can use beamshift corrections
derived from short polarimetric cycles of well defined stellar PSFs, 
taken with the same (or similar) filter, the same 
polarimetric configuration and telescope orientation. 
Because the beamshift changes only slowly with parallactic 
angle and altitude, one can take such beam shift calibrations
just before or/and after the science observation, using the 
PSFs of the same star, or another star with similar pointing 
directions (parallactic angle and altitude) 
$|\Delta p|$ and $|\Delta a| \lapprox 5^\circ$.

An example for a beamshift correction for a coronagraphic
image is shown in Fig.~\ref{BScoro}, where residuals in the 
uncorrected differential polarization image is at the level of up to 
$\approx 1~\%$ in $Q/I$ or $U/I$ near the coronagraphic mask. 
These data are corrected with the non-coronagraphic
PSF measurements of the same star (Fig.~\ref{BSPSF}) taken $10$~min later.
For this, the star was just offset from the coronagraphic mask
and observed with a short integration and a ND-filter. 
This calibration reduces the residuals in 
the polarimetric image significantly, for example the standard 
deviation in the box at the bottom of the frames in Fig.~\ref{BScoro} 
is 9.3~ct in the uncorrected frame and 4.4~ct in the corrected frame.
  
The residual noise is not completely removed by the beam shift
correction as can be seen for the strong ``s'' speckles. 
The positive/negative ring feature at the edge of the coronagraphic 
focal plane stop is even enhanced and new spurious positive/negative
spots are introduced, as indicated by
the arrow in the right panel of Fig.~\ref{BScoro}. 
These are intensity features originating from components 
located after the inclined mirrors. For example, the coronagraph, 
the deformable mirror, poorly corrected bad pixels, 
or dust on the micro-lens array of the ZIMPOL 
detector introduce spurious positive/negative polarimetric 
signals if a beam-shift correction is applied. Some of 
these new problems can be solved with detector dithering, 
but it is still unclear how some of these spurious features 
can be removed again.

The beamshift can also be measured in high quality coronagraphic
images, if the focal plane mask CLC-MT-WF is used. This
mask is by design slightly transparent and therefore 
the same polarization pattern is visible in Fig.~\ref{BScoro} (left)
for the attenuated star marked with an arrow, as for the
stellar PSF shown in Fig.~\ref{BSPSF} ($Q$, derot, no.corr).
Beamshift corrections based on the transmitted PSF using the CLC-MT-WF masks 
give good results, if the Strehl ratio for the observations is high and if the
star is well centered on the mask. Of course, frame selection is
important because a few good frames are sufficient to define the 
beamshift correction for many exposures of the same target
taken before and afterwards. 
 
\subsection{Polarized PSF}
\label{SectPSFpol}

\begin{figure}
\includegraphics[trim=0.1cm 0.1cm 1.5cm 0.5cm,clip,width=8.8cm]{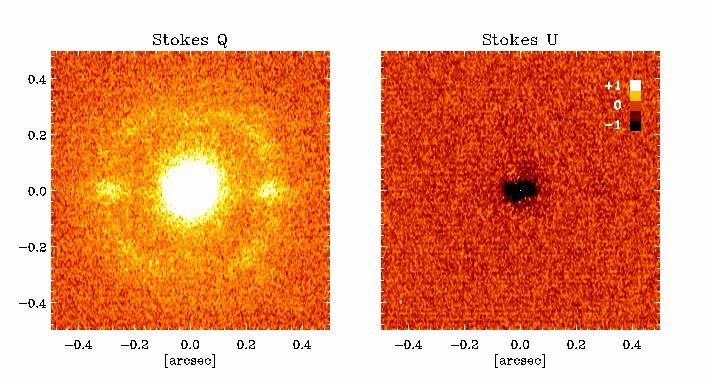}
\caption{Stokes Q and U images for the V-band filter of the
PSF standard HD 183143. These are the polarization images corresponding
to the intensity PSF shown in Fig.~\ref{PSFVI}.}
\label{PSFpolQU}
\end{figure}

The quality of the ZIMPOL polarimetric imaging can be assessed with
an unpolarized or polarized stellar PSF without circumstellar emission. 
Such an object should show over the whole image plane a constant 
fractional polarization. 
Fig.~\ref{PSFpolQU} shows the Stokes fluxes $Q$ and $U$ for the V-band
of HD 183143 from the same observations used for the characterization of
the intensity PSF in Fig.~\ref{PSFVI}. The PSF in
$Q$ is just a faint replica of the intensity PSF, with central
PSF peak, and speckle ring because the star has a 
high interstellar $Q/I$-polarization of
about 5~\%. In $U$, only a weak negative peak is visible because
of the small $U/I\approx -0.5$~\% telescope polarization.
These polarimetric PSFs are corrected for the beam-shift effect 
and therefore they show no positive/negative central features 
for $r<20$~pix. Outside $r>20$~pix, no difference is visible between 
beam-shift corrected and not corrected
profiles. The displayed PSFs were also not polarimetrically 
calibrated, because this has no impact on the polarization structure 
of the PSF.

\begin{figure}
\includegraphics[trim=2cm 12.5cm 2.5cm 3cm,clip,width=8.8cm]{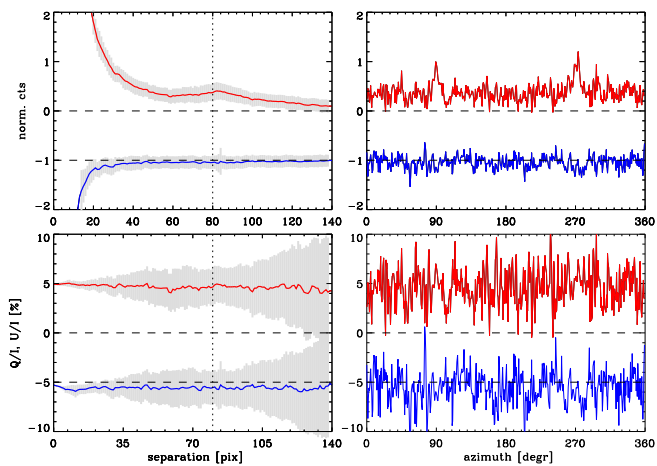}
\caption{HD 183143 mean radial V-band profiles (left) for the Stokes flux 
$Q(r)$ and $U(r)$ (top) and the fractional polarization $Q(r)/I(r)$  
and $U(r)/I(r)$ (bottom) and the corresponding azimuthal profiles
$Q(80,\phi)$, $U(80,\phi)$, and $Q(80,\phi)/I(80,\phi)$, 
$U(80,\phi)/I(80,\phi)$ (left). The $U$-data are shifted
downwards in for better visibility and the dashed lines indicate the offset. } 
\label{PSFpolVprof}
\end{figure}

Figure~\ref{PSFpolVprof} shows the corresponding mean 
radial $Q(r)$ and the azimuthal profile $Q(\phi,80)$ 
for $r=80$~pix through the speckle ring and the same for Stokes $U$. 
The lower panels show the corresponding fractional
polarization signal, which are essentially constant
in the radial and azimuthal profiles without
instrumental structures in $Q(x,y)/I(x,y)$ and 
$Q(x,y)/I(x,y)$ despite the large dynamic range with
$> 3000$~ct/pix for the PSF peak and
only a few counts per pix and frame further out. 

This polarimetric fidelity is a particular advantage of the
ZIMPOL technique which registers the opposite polarization modes
$I_\perp$ and $I_\parallel$ with the same pixels. The differential
polarimetric signal is therefore essentially independent of
the exact flatfielding factors $a^{\rm ff}$ 
\begin{displaymath}
a_\perp^{\rm ff}I_\perp - a_\parallel^{\rm ff}I_\parallel 
= a^{\rm ff}(I_\perp - I_\parallel)
\end{displaymath}
or bias level subtraction values $c^{\rm b}$
\begin{displaymath}
(I_\perp-c_\perp^{\rm bias}) - (I_\parallel-c_\parallel^{\rm bias}) = I_\perp - I_\parallel\,
\end{displaymath}
because the flat-fielding factors 
$a_\perp^{\rm ff}=a_\parallel^{\rm ff}=a^{\rm ff}$ or the bias levels 
are $c_\perp^{\rm bias}=c_\parallel^{\rm bias}=c^{\rm bias}$ are 
very close to identical for the 
two modes. This beneficial property of ZIMPOL is very useful for
the analysis of polarization signals from circumstellar sources.  

\subsection{\object{R Aqr}: an example for circumstellar 
polarization measurements}
\label{SectRAqr}
We used R Aqr as test source for polarimetric imaging during the 
SPHERE/ZIMPOL commissioning. This object is bright, $m_I=4.4^m$ 
at the time of our observations,
and it showed in previous studies with aperture polarimetry  
strong, and highly variable, linear polarization 
\citep[e.g][]{Serkowski01,Joshi12,Aspin85}. 
R Aqr is a nearby $d=220$~pc symbiotic binary with a 
Mira variable undergoing heavy mass loss and binary interactions 
\citep[][and references therein]{Schmid17}. 

\begin{figure*}
\includegraphics[width=18cm]{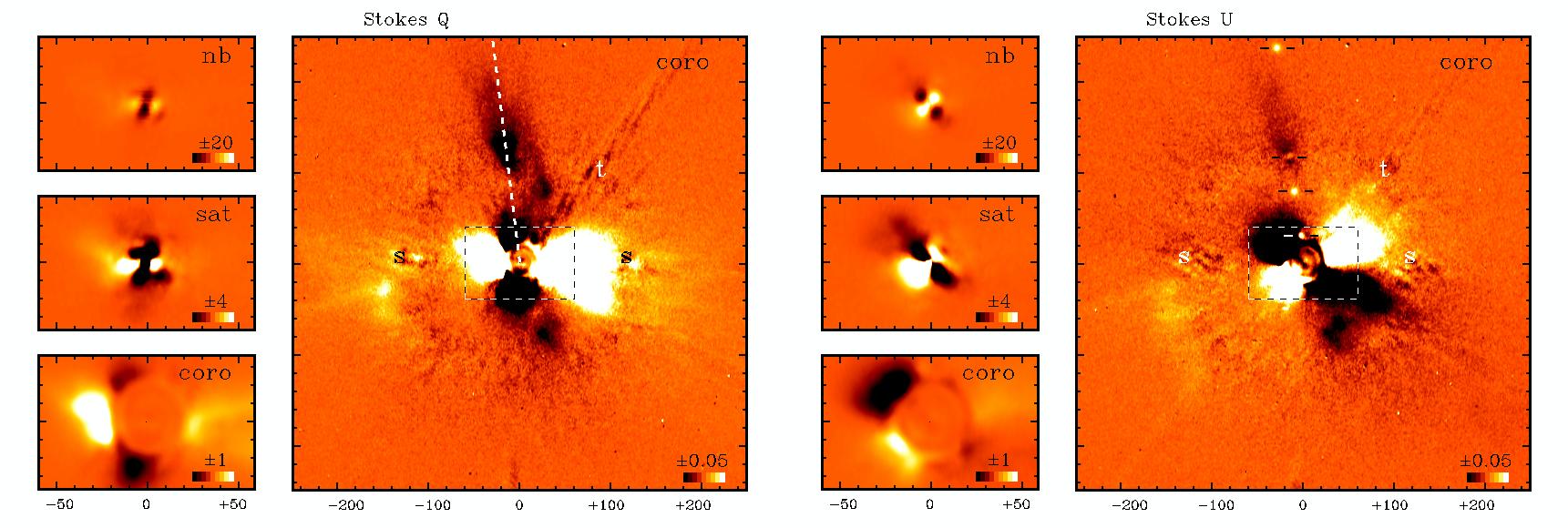}
\caption{\object{R Aqr} I-band Stokes Q (left) and U (right) signal measured
for the unsaturated PSF in the narrow-band (nb) Cnt820 filter 
(upper left), for the saturated broad band I\_PRIM filter 
(middle left), and deep coronagraphic observations for the
I\_PRIM filter and slow polarimetry mode (lower left and large frame). 
The cut for the count profiles of Fig.~\ref{RAqrcut} is
shown in the large Q-frame. Units are in pixels of 3.6~mas.}
\label{RAqrQU}
\end{figure*}

Polarimetric data for R Aqr were taken in many different 
filters \citep{Schmid17}, but we focus here on the 
I-band data taken obtained three different instrument modes: \\
(i) in fast modulation using the 20~nm narrow-band Cnt820 filter 
for the PSF peak (hereafter called ``nb''-image), \\
(ii) fast polarimetry in the 150~nm wide broad-band I\_PRIM filter where
the PSF peak saturated (``sat''-image), and \\
(iii) deep I\_PRIM, slow-modulation polarimetry
in coronagraphic mode (``coro''-image) for the ``outer'' regions. \\
The parameters of the different observations are summarized in 
Table~\ref{TabRAqrobs}, and the Stokes $Q$ and $U$
images are displayed in Fig.~\ref{RAqrQU}, and Table~\ref{TabPSFAO} 
lists atmospheric and PSF parameters for the $I_{\rm nb}$
image, which are representative for all R Aqr data.  

\subsubsection{Data reduction}
The data were first reduced in a standard way, with 
an overscan bias subtraction, flatfielding, polarimetric signal 
extraction and combination. Already these basic steps allow a
qualitative analysis of all extended polarization
features described below. 

In a second step, the reduction was rerun, but with applying
a polarimetric beam shift correction before the combination 
of the polarization frames. The beam shift correction was derived from the
unsaturated ``nb''-data and also 
applied to the saturated and coronagraphic 
I\_PRIM filter data. Applying the beam shift correction
reveals also a $Q$- and $U$-quadrant pattern for the PSF-peak described below,
and it reduces spurious polarization residuals at the position of
strong speckles. 

For an accurate quantitative analysis we correct our data with
an image rotation of $\theta_0=2^\circ$ in clockwise direction to account for 
the astrometric offset from the North direction according 
to Eq.~\ref{Eqthoffset1}. Then we apply the polarimetric calibrations,
deriving first $Q_{\rm c1},U_{\rm c1}$ (Eq.~\ref{EqModeff}) using
the modulation-demodulation efficiencies $\epsilon_{\rm mod}$
(Table~\ref{Modeff}), the corrections for the
telescope polarization $p_{\rm tel}=0.4$~\%, $\delta_{\rm tel}=68^\circ$ from
Table~\ref{PolOffset}, which yield for the parallactic angle of 
the observations ($\theta_{\rm para}\approx -114^\circ$ to $-122^\circ$) roughly 
$Q_{\rm c3}/I\approx Q_{\rm c1}/I$ and $U_{\rm c3}/I\approx U_{\rm c1}/I+0.004$
(Eqs.~\ref{EqQc3},\ref{EqUc3}  with $\epsilon_{\rm opt}=1$). As last step
we apply a polarization angle offset correction of $6^\circ$ according
to Eqs.~\ref{EqQc4},\ref{EqUc4}. Both, the astrometric correction 
of $2^\circ$ and the polarimetric correction of $6^\circ$ rotate
the polarimetric $Q$- and $U$-quadrant pattern in clockwise directions
so that the slightly tilted patterns from the basic 
reduction get an orientation which is
essentially undistinguishible from the expected up-down / left-right
$Q$-pattern and the corresponding diagonal $\pm 45^\circ$ $U$-pattern
shown in Fig.~\ref{RAqrQU}.    

Despite all these corrections, there remain ambiguities in the
measured circumstellar polarization of R Aqr because of two effects.
The limited resolution produces polarimetric cancellation 
between the $+Q$ and $-Q$ or the $+U$ and $-U$ regions 
\citep[see e.g.][]{Schmid06b}, and the unresolved emission 
of the star is also intrinsically 
polarized and contributes to the measured polarization signal as 
can be inferred from the available multi-wavelength polarimetry
taken during the same night. A more detailed analysis would 
be required to disentangle the intrinsic polarization
of the central star from the circumstellar 
polarization signal. This is of much scientific interest, 
but must be deferred to a future paper focussing
on the mass loss of R Aqr. The scope of this section 
are the measuring capabilities 
of SPHERE/ZIMPOL for high performance polarimetry and therefore
we just analyze the calibrated measurements. 

All data
are count normalized ${\rm ct}_{\rm n6}$ as in previous sections 
by setting the total counts of the star within an aperture 
diameter of $3''$ to $10^6$ counts. For the saturated and coronagraphic 
images, where the PSF peak counts are not available, the halo-flux 
was adjusted to the normalized, unsaturated ''nb''-image for $r$ in the range 
$100~{\rm pix}<r<250~{\rm pix}$. This procedure assumes, 
that the PSF halo flux remained unchanged 
and provides a good relative calibration for the three images. 

\begin{table}
\caption{Polarimetric test data for R Aqr from 2014-10-11 used in
this work}
\label{TabRAqrobs}
\begin{tabular}{lccc}
\noalign{\smallskip\hrule\smallskip}
Parameter       & \multispan{3}{\hfil Image \hfil} \\
                & ''nb''  & ''sat''  & ''coro'' \\
\noalign{\smallskip\hrule\smallskip}
data sets\tablefootmark{a}       
                & 0047-50 & 0059-62 & 0071-74 \\
filter          & Cnt820  &  I\_PRIM & I\_PRIM  \\
coronagraph     & --      &  --      & CLC-MT-WF \\
exp. $n_{\rm DIT}\times t_{\rm DIT}$
                & $10\times 1.2$~s & $10\times 1.2$~s  & $10\times 20$~s \\
QU-cycles       & 1                & 1                 & 1               \\
$t_{\rm total}$   & 48~s             & 48~s              & 800~s           \\ 
modulation      & fast             & fast              & slow            \\
ct frame$^{-1}$
                & $\approx 4.2\cdot 10^6$        
                                  & $\approx 1.5\cdot 10^7$                  
                                          &  $\approx 2.1\cdot 10^8$               \\  
\noalign{\smallskip\hrule\smallskip}
\end{tabular}
\tablefoot{
\tablefoottext{a}{Identifications corresponds to the fits-file 
header keyword ``origfile'' without prefix ``SPHERE\_ZIMPOL\_OBS284''.}
}
\end{table}

\subsubsection{Description of the Stokes $Q$ and $U$ images}

The polarimetric data show the typical Stokes $Q$ and Stokes $U$ 
quadrant pattern of a circumstellar 
scattering region. The distribution of the scattering dust is a spherical
and clumpy causing geometric features in $Q$ and $U$ which deviate clearly from 
a symmetric, smooth quadrant structure. 

\paragraph{Polarization of the PSF peak.}
The ''nb''-images for $Q$ and $U$ 
in Fig.~\ref{RAqrQU} show the polarization in the PSF 
peak of the mira variable. Clear $Q$ and $U$ 
quadrant patterns are visible within $r\lapprox 15$~pix with maxima 
and minima located at a separation of about 
$r\approx 5$ pixels ($18$~mas or $\approx 3.6$~AU)
from the intensity peak center.

This polarization originates from scatterings in 
immediate surroundings of the central star. For an unresolved, 
strictly centro-symmetric 
scattering case the $+Q,\,-Q$ or $+U,-U$ components
would just cancel, and such a zero is indeed seen in 
the center. The radius of R Aqr measured
by interferometry is about 7 -- 8~mas \citep[e.g.,][]{Ragland08} 
or about 2~pixels, thus the peak intensities in the
$Q$ and $U$ parameters
originates from within $1-3$ stellar radii. 
The measurements at small separations depend strongly on 
the self-cancellation because the intrinsic positive and 
negative regions of $+Q,\,-Q$ or $+U,-U$ overlap 
substantially when convolved with the instrument PSF. 
Therefore, the fractional polarization inside 5~pix reaches
extreme values of only $Q/I,U/I\approx \pm 0.25$~\%, lower than
$\approx \pm 1.0~\%$ in the surrounding anulii $r=5-10$~pix, or
$\approx \pm 1.5~\%$ for $r=10-30$~pix. The intrinsic value for 
this circumstellar polarization is certainly
much larger, but hard to quantify without a detailed
simulations of the signal convolution with the instrument
PSF. 

\paragraph{Polarimetry for $0.03''< \rho\lapprox 0.20''$.} 
The ''sat''-images taken with the $150$~nm broad-band filter 
I\_PRIM yielded about four times higher photon counts (for the same
exposure time) than the ``nb''-image, roughly as expected considering
the larger filter width but also the much lower stellar flux in
the $750 - 800$~nm region. Several pixels in the PSF peak are saturated 
and it is difficult to measure the beamshift. Therefore
we applied the same beamshift correction as for the unsaturated
''nb''-image. 

The ``sat''-images are ideal for 
polarimetric measurements of the separation range 
30 mas $<\rho <$ 200~mas filling the gap between the
saturated peak and the inner working angle 
of the coronagraphic data. 
In the ''sat''-data of R Aqr (Fig.~\ref{RAqrQU}),
the polarized intensity $Q$ from the circumstellar dust scattering is
stronger on the East-side of the Mira variable than on the West side,
most likely because of the hot companion located 45~mas to the
West of the mira \citep{Schmid17}.    

Sensitive polarimetric observations in the radial range 
30 mas $<\rho <$ 200~mas are demanding because of 
the huge intensity gradient near the PSF peak with 
a flux ratio $f(0)/f(\rho)\approx 1000$ for $\rho=0.20''$ 
(55~pix). Photon noise limited observations are
achievable for the separation range $\rho\approx 0.07''-0.20''$
with exposures where the PSF-peak is just
at the saturation limit or in saturation by a factor
of a few. For lower illumination the read-out noise can dominate
already around $0.1''$. For strong saturation of more than a factor of $>5$ 
the charges start to ``bleed'' into neighboring pixels mainly along 
the detector column direction and this affects the inner working 
angle of the observation.

\paragraph{Coronagraphic polarimetry for $r\gapprox 0.1''$.} 
The small bottom panels give the center of the 
I\_PRIM-band ''coro''-image. Clearly visible is 
the attenuation of the mask CLC-MT-WF, 
which has a spot radius of $\rho=77$~mas. The immediate surroundings 
of the coronagraph show a continuation of the structure seen in
the non-coronagraphic images. 
The two large panels in Fig.~\ref{RAqrQU} are based on the same 
''coro''-images as the small bottom panels, but the images are
displayed with plotting and color scales emphasizing the weak 
polarization intensities at larger separation. 

The integration time for the ``coro''-images 
is 16.7 times longer than for the ``sat''-images. 
In addition, the ``coro''-images are
much more sensitive because they were taken with  
the low gain 1.5~e$^-$/ct slow modulation mode with a low read-out noise.
In slow modulation the spurious features from pixels with
reduced charge transfer efficiency are weaker. Therefore, 
this mode is well suited for high precision polarimetry with long 
integrations outside the speckle ring $\rho>0.6''$ where the 
temporal variations in the PSF structure are small. 
Some quasi-static speckle features 
are visible in Fig.~\ref{RAqrQU} and they are marked 
according to Sect.~\ref{SectPSFAO} (Fig.~\ref{PSFVI}). 

For R Aqr, the high sensitivity of the 
$Q_{\rm coro}$-, $U_{\rm coro}$-frames reveal
well defined polarization features from
individual dust clouds at separations of about 
120~pix ($\approx$ 430 mas) in the North
and at about 80~pix (290~mas) in the southwest, as well as other fainter
clouds. In addition, there are many extended structures and clear
asymmetries in the distribution of the $Q$ and $U$ signal apparent out
to separations of about 150~pix (600~mas).  

\paragraph{Polarization signal at separations $\gapprox$ 1 arcsec.}

\begin{figure}
\includegraphics[trim=0.5cm 0.5cm 1cm 0.5cm,clip,width=8.8cm]{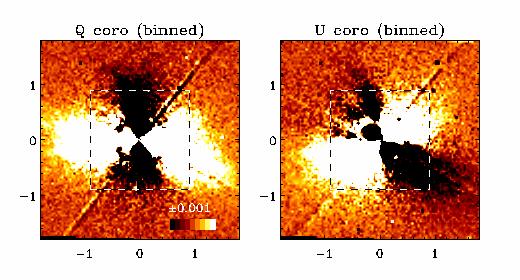}
\caption{Binned \object{R Aqr} 
I-band Stokes $Q$ (left) and $U$ (right) map extending
over the whole detector of $3.6''\times 3.6''$ detector area. The dashed
boxes are the $1.8''\times 1.8''$ areas shown in the large panels
of Fig.~\ref{RAqrQU}.}
\label{RAqrQUbinned}
\end{figure}

The detection limits for extended polarimetric emission can be pushed
further by averaging the signal in larger measuring areas. 
Figure~\ref{RAqrQUbinned} shows the coronagraphic R Aqr data, but 
binned by $10 \times 10$~pixels and the color scale sharpened to 
$ct_{\rm n6}\pm 0.001$, by a factor of 50 when compared to the large 
frames in Fig.~\ref{RAqrQU}. Both, the $Q$- and $U$-quadrant
patterns extend now to the edge of the detector at a separation of 
$\rho=1.8''$. At this level of sensitivity the correction for
field dependent instrumental polarization effects and  
the intrinsic polarization of the central star and its light halo
become important issues for the accurate measurement of the ``large'' scale 
polarization signal.    

SPHERE/ZIMPOL allows for polarimetric observations
with field angle offsets or polarization angle offset, which
are both powerful methods to disentangle instrumental
effects from weak but real sky signals. 

\begin{figure}
\includegraphics[trim=2cm 12.5cm 6.5cm 3cm,clip,width=8.8cm]{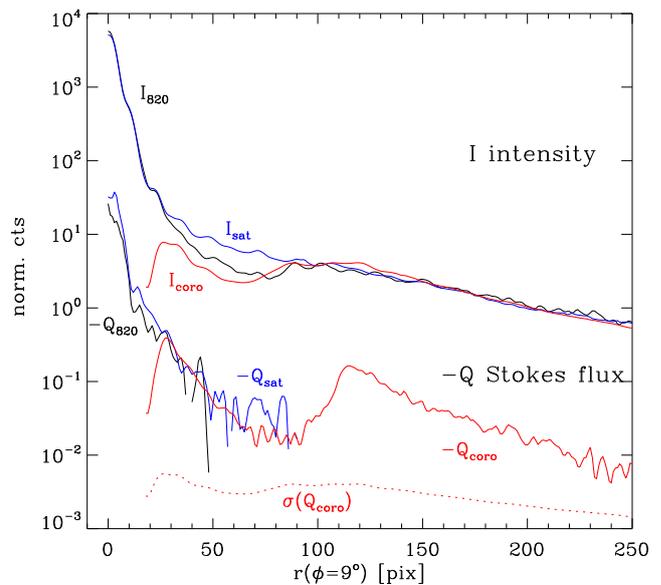}
\caption{Radial profiles for R Aqr for the $\phi=7^\circ$-direction
through the bright scattering clouds in the north as indicated
in Fig.~\ref{RAqrQU}.} 
\label{RAqrcut}
\end{figure} 

\subsubsection{Quantitative analysis for a radial cut}
 
This section gives a quantitative analysis for the
R Aqr polarization signal along the dashed line plotted in the 
large $Q_{\rm coro}$-image (Fig.~\ref{RAqrQU}) through the strong 
dust feature in the North. In the corresponding $U_{\rm coro}$-frame, 
we inserted along the same direction 
artificial point signals, marked with horizontal dashes, 
for the evaluation of the detection
limits of faint, polarized point sources. 

\paragraph{Measured surface brightness and sensitivity limit.}
Figure~\ref{RAqrcut} shows 
the radial cuts $I(r,7^\circ)$ and $Q(r,7^\circ)$
for the ``nb''-, ``sat''- and ``coro''-images. For this 
1-dim. profile extraction the reduced images were rotated by $\-7^\circ$, 
so that the cuts are simple mean profiles of 
5 vertical columns from the central source 
at $r=0$ to the upper edge at $r=500$~pix. 
This cut through the northern dust cloud at $r\approx 120$~pix 
avoids the strong vertical frame transfer trail in 
the non-coronagraphic images. The cloud, is not visible in 
the intensity profile, but appears as very strong,
high signal-to-noise polarization source in polarimetric 
differential imaging. 
The profiles in Fig.~\ref{RAqrcut} are given in 
flux normalized counts ${\rm ct}_{\rm n6}$ like in 
Figs.~\ref{PSFVI} and \ref{PSFprofgoodbad} and 
the dotted curve is the 1-$\sigma$ 
photon noise limit $\sigma(Q_{\rm coro})$ for the ``coro''-profile. 
\begin{table*}
\caption{Normalized mean counts ct$_{\rm n6}$ for the PSF 
surface brightness of R Aqr and the polarized surface brightness of 
the circumstellar region along the position angle $\phi=9^\circ$ for
the frames for the intensity $I_{\rm nb}$, $I_{\rm sat}$, and $I_{\rm coro}$ 
and the polarized flux $Q_{\rm nb}$, $Q_{\rm sat}$, and $Q_{\rm coro}$.}
\label{TabRAqrpol}
\begin{tabular}{lcccccccc}
\noalign{\smallskip\hrule\smallskip}
Parameter  & {\hfil $\rho [''] =$} & 0 & 0.018 & 0.108 & 0.288 & 0.432 & 0.864 & 1.728 \\
          & {\hfil ${\rm r\,[pix]\,=}$}  
                    & 0   & 5  & 30 & 80  & 120 & 240 & 480  \\
        & area [pix]& $1\times 1$ & $1\times 5$ & $10\times 10$ & $10\times 10$ 
                             &  $10\times 10$ & $10\times 10$ & $20\times 20$ \\
\noalign{\smallskip\hrule\smallskip}
\noalign{\smallskip intensity \smallskip}
$I_{\rm nb}(r)$   & ct$_{\rm n6}$ 
                  & {\bf 5775}                  & 1846  & 14.9 & $2.9::$  \\ 
$I_{\rm sat}(r)$  & ct$_{\rm n6}$ 
                  & 5116\tablefootmark{a} & {\bf 1823}  & {\bf 18.7} & 4.7
                              & 3.2 & $0.67::$ &                 \\
$I_{\rm coro}(r)$ & ct$_{\rm n6}$ 
                  & 15.4\tablefootmark{b} & 1.60$^b$  & 7.13 & {\bf 3.39} 
                              & {\bf 3.82}  & {\bf 0.61}  & {\bf 0.073}             \\
\noalign{\smallskip polarized flux \smallskip}
$-Q_{\rm nb}(r)$  & ct$_{\rm n6}$ 
                  & {\bf 26.2}       & 12.1  & 0.24  & 0.013::    \\ 
$-Q_{\rm sat}(r)$ & ct$_{\rm n6}$ 
                  & 32.3\tablefootmark{a} & {\bf 21.1}  & {\bf 0.368}  & 0.028 
                              & 0.086 & $5.4\cdot 10^{-3}$ \\
$-Q_{\rm coro}(r)$& ct$_{\rm n6}$ 
                  & 0.025\tablefootmark{b} 
                     & 0.0292\tablefootmark{b} & {\bf $0.319$} & {\bf 0.0216}
                               & {\bf 0.126}  
                                    &  $\mathbf{7.4\cdot 10^{-3}}$ 
                                             & $\mathbf{0.14\cdot 10^{-3}}$ \\  
\noalign{\smallskip fractional polarization  \smallskip}
$-Q(r)/I(r)$       & & 0.45~\%        & 1.16~\%  & 1.98~\% &  0.64~\%   
                               & 3.3~\% & 1.2~\% & 0.19~\% \\
\noalign{\smallskip surface brightness contrast [$\Delta$mag/arcsec$^2$] \smallskip}
$C_{\rm SB}(r)$\tablefootmark{c} & mag  & $-6.6$  &  $-5.4$ &  0.6  &    1.5 &    1.3 &    3.3 &   5.6 \\ 
$C_{\rm SBpol}(r)$ (R Aqr)  
                   & mag & $-0.8$  & $-0.5$  & 4.0  & 6.9     & 5.0    & 8.1    & 12.4 \\
\noalign{\smallskip artifical point source contrast [$\Delta$mag] \smallskip}
$C_{\rm pol}$\tablefootmark{d} (b,c,d,e)      & mag  &       &          & 10.8  &  12.5    & 12.5   & 12.5     \\
$C_{\rm pol}(r)$ $5\sigma$-limit 
        & mag  &       &     & 8.8   &  12.6    & 13.0    & 14.9   & 15.6  \\
\noalign{\smallskip\hrule\smallskip}
\end{tabular}
\tablefoot{
\tablefoottext{a}{saturated PSF peak;}
\tablefoottext{b}{coronagraphically attenuated PSF peak;}
\tablefoottext{c}{surface brightness contrast $C_{\rm SB}=SB-m_{\rm star}$;}
\tablefoottext{d}{polarized flux point source contrast 
$C_{\rm pol}=m_{\rm pol}-m_{\rm star}$.}}
\end{table*}                                

Table~\ref{TabRAqrpol} lists for selected areas along the
profile the measured ${\rm ct}_{\rm n6}$-values for $I(r)$ and
$Q(r)$, the fractional polarization $Q(r)/I(r)$,
and the corresponding surface brightness contrast $C_{\rm PSF}(r)$
for the intensity and $C_{\rm SB}$ for the
polarization in magnitudes. 
Note, that the sign of the $Q$-Stokes flux 
signal is negative in the North of the central source.

The three intensity profiles $I_{\rm nb}$, $I_{\rm sat}$, and $I_{\rm coro}$
(Fig.~\ref{RAqrcut} and Table~\ref{TabRAqrpol})
show quite some differences at small
separations which reflect most likely the atmospheric and AO variations
between the $I_{\rm nb}$ and $I_{\rm sat}$ frames, and the suppression 
of the PSF diffraction rings for $I_{\rm coro}$. 
Despite this,
the three $Q$-profiles agree very well at $r=30$~pix and 
for $Q_{\rm sat}$ and $Q_{\rm coro}$ 
also at 80~pix ($Q_{\rm nb}(80~{\rm pix})$ is dominated by read-out noise). 
This confirms the finding 
from Fig.~\ref{RAqrQU}, that the ``nb''-, ``sat''-, 
and ``coro''-frames are matching very well and provide a 
continuous mapping of the 
differential polarization signal from the center 
$r=0$ to large separations $r > 250$ pix ($\rho>0.9''$).

Table~\ref{TabRAqrpol} lists in bold letters the $I(r)$- 
and $Q(r)$-values with the highest signal-to-noise, which 
are not affected by saturation or attenuation by 
the coronagraph. The polarized surface brightness 
of the circumstellar dust is about 50 to 500 times lower than 
the intensity of the stellar PSF as follows from the fractional
polarization $Q(r)/I(r)$. This ratio depends
on the PSF and therefore the variable AO performance 
and is not very useful for the characterization of the circumstellar
scattering. 

The polarized surface brightness contrast $C_{\rm SBpol}$ is a better 
quantity to describe the polarized radiation from the dust
(Sect.~\ref{SectPolSky}). For a region north of the central 
light source we can approximate 
$C_{\rm SBpol}(r)\approx -Q(r)/I_{\rm star}$ where $I_{\rm star}$ is the
total stellar intensity as measured in an aperture of $3''$-diameter.
The obtained values range from
${\rm SB}_{\rm pol}-m_{\rm star}=4.0$ to $12.4$~mag~arcsec$^{-2}$ 
for the separation range from 
$\rho=0.11''$ to 1.73$''$ (Table~\ref{TabRAqrpol}). This surface
brightness in polarized flux is up to $19^m$ fainter than the
surface brightness of the PSF 
intensity peak $C_{\rm SB}(0)=-6.6^m$.  

The R Aqr ``coro''-images presented here are not really deep data, 
because they consist only of $n_{\rm DIT}=40$~frames or 
$t_{\rm total}=400$~s of integration for $Q$ and the same for $U$. 
A comparison of R Aqr with other SPHERE/ZIMPOL polarimetric
data in the literature shows that $C_{\rm SBpol}$ is about $1-2$ 
mag higher (relative emission $Q_{\rm cs}/I_{\rm star},U_{\rm cs}/I_{\rm star}$ 
fainter) than the bright rims of bright proto-planetary
disks, like HD 135344B with $C_{\rm SBpol}\approx 5$~mag/arcsec$^2$ at
$\rho = 0.2''$ \citep{Stolker16}, or HD 142527 with 
$\approx 7$~mag/arcsec$^2$ at $\rho = 0.9''$ \citep{Avenhaus17}. 
Compared to these disk systems, the R Aqr central star is a 
bright source and therefore a comparable or better contrast 
is achieved with a short integration time.
An about 2~mag higher contrast (fainter signals) of 
$C_{\rm SBpol}\approx 7.6$~mag/arcsec$^2$ at $\rho = 0.3''$ or
$9.5$~mag/arcsec$^2$ at $\rho = 1.2''$ is measured for the debris 
disk HIP 79977 \citep{Engler17}.

\paragraph{Detection limits for point sources}

For the determination of the
point source contrast limit, we use the $U$-image, 
which serves as low background observation 
for the $\phi=7^\circ$ profile direction. 
We insert in the $U$-frames artifical point sources at $r=30$ 80, 120, 
240, and 480 pixels and the $U$-polarization signals shown in
Fig.~\ref{RAqrQU} are ${\rm ct}_{n6}=0.058$~ct/pix 
corresponding to a polarization contrast of 
$C_{\rm pol} = f_{\rm pol}/f_{\rm star}=10^{-5}$ ($\Delta m=12.5^m$)
for $r=80$, 120 and 240~pix and ${\rm ct}_{n6}=0.29$~ct/pix
at $r=30$ (see Table~\ref{TabRAqrpol}).

The 5~$\sigma$ contrast limits $C_{\rm pol}(r) = U_{5\sigma}(r)/I_{\rm star}$
for polarized point sources are then derived by measuring 
the extracted signal similar to the measurements of 
$\alpha$ Hya B in Sect.~\ref{CoroPerf} 
for artifical point sources with a range of $U$-fluxes. 
This yields the $5\sigma$ limits of 
$\Delta m\approx 12.6^m, 13.0^m$ or 
$14.9^m$ at separations of $0.29''$, $0.43''$, or $0.86''$, respectively
(Table~\ref{TabRAqrpol}). 
 
These are contrast limits comparable to the limits reported
by \citet{vanHolstein17} for polarimetric observations of the 
HR 8799 planets in the near-IR with SPHERE/IRDIS. For 
separations $\rho\gapprox 0.6''$
the ZIMPOL and IRDIS polarimetric modes are for bright
targets essentially photon noise limited. Thus, deeper polarimetric
contrast limits can be achieved, if more photons are collected with
longer integrations or with observations of brighter targets. At small 
separations $\rho\lapprox 0.6''$, there is in the R Aqr 
polarimetry data presented here, some residual speckle noise besides 
the photon noise, because our simple test data are not optimized for
deep contrast limits. ZIMPOL was designed for 
the photon noise limited search of polarized light from 
extra-solar planets around very bright stars $\lapprox 3^m$
based on a large number of well illuminated frames
$n_{\rm DIT} > 5000$ and long total integration times
$t_{\rm total} > 3$~h \citep{Schmid06a,Thalmann08}. 
Thus, the real polarized point source contrast limit of SPHERE/ZIMPOL 
can only be demonstrated with such long integration 
of very bright stars using coronagraphy and fast
modulation polarimetry and more
elaborate observing strategies than the simple 
``one HWP2-cycle'' test used for our R Aqr. 

A non-polarimetric ZIMPOL point source observation using angular differential
imaging (ADI) of PZ Tel A and B is described in \citet{Maire16a}. This system
with a separation of $\rho=0.48''$ has in the R\_PRIM band a contrast 
of $\Delta {\rm mag}=9.7^m$. The companion is detected with
30~min of integration in one ZIMPOL arm, with only 14$^\circ$ of field
rotation and under mediocre $>1.1''$ seeing conditions. Companions
with much higher contrast of at least $\Delta {\rm mag}\approx 12.5^m$ 
should be observable with ZIMPOL ADI if data are taken under better
conditions and a better observing strategy.

\section{Discussion}

This paper presents a detailed description of the
SPHERE/ZIMPOL instrument to support and promote scientific 
investigations based on new or archival observations.
The provided information should also be helpful to optimize
observing and data reduction strategies, for
an evaluation of the performance of this instrument
for possible upgrades \citep[e.g.,][]{Lovis17}, or the 
design of more advanced systems for future
telescopes like the ELT \citep{Kasper13,Keller10}.

High contrast and high 
resolution observations from the ground require the 
combination of an adaptive optics
system with coronagraphy, and a differential imager 
for speckle noise suppression. All these methods are tricky, 
because the resulting performance depends on the target brightness, 
the atmosphere, the selection of the coronagraph, the
mode of the differential imager, and on the appropriate 
observing strategy. It is not possible to describe such a 
complex system and its performance in a single paper, or a 
compact user manual. This paper provides a comprehensive 
hardware description, and highlight three important aspects, which are
particularly special for the SPHERE/ZIMPOL instrument when compared
to other AO-imaging systems: (i) the characterization of the 
PSF-properties in the visual provided by the SPHERE AO system 
and the VLT telescope, (ii) the performance of the visual
coronagraph, and (iii) the polarimetric measuring strategy
and data characteristics of the ZIMPOL system. 

More technical information about ZIMPOL will become available
in the future: a paper on the astrometric 
calibration is in preparation, and a description of photometric parameters 
is planned. \citet{Schmid17} describe
many technical aspects on the H$\alpha$ imaging and absolute 
photometric measurements in several filters.
Further, we expect that useful performance
characteristics of SPHERE/ZIMPOL can be extracted from 
science papers, which aim for accurate measurements or push 
the performance of this instrument to the limits.

\subsection{Key performance properties of ZIMPOL}

SPHERE/VLT is one of the new generation extreme AO-systems 
available at large telescopes. 
The quality of the obtained PSFs depend strongly on the AO-system, 
the atmospheric conditions and the guide star magnitude $m_R$
and Table~\ref{TabPSFAO} characterizes the visible performance of the AO 
system for several typical cases using simple measuring parameters, 
e.g. the normalized PSF peak flux 
ct$_{\rm n6}(0)$. Good corrections
with ct$_{\rm n6}(0)/10^6>0.3~\%$ (Strehl ratio $\gapprox 15$~\% in 
the R-band) are achieved for stars down to $m_{\rm R}\approx 8^m$ with
an atmospheric seeing of $\approx 1''$. 
For bright stars $m_{\rm R}<7^m$ and good seeing conditions 
$<0.8''$ the relative peak flux is twice as good 
ct$_{\rm n6}(0)/10^6>0.6~\%$ and this corresponds to Strehl ratios
of $\approx 30-50~\%$ in the R-band \citep[see also][]{Fusco16}. This
provides for SPHERE/ZIMPOL a point source contrast performance of
$\Delta m \approx 12.5^m$ outside $0.15''$  
for the polarized flux (Table~\ref{TabRAqrpol}). 
The existing ADI-data of the PZ Tel binary
\citet{Maire16a} indicate, that similar contrasts performance are
also possible for the intensity signal of a point source.
For extended polarized emission, the system reaches outside of 
$0.15''$ surface brightness contrast limits better than 
$C_{\rm SBpol} > 7.5^m{\rm arcsec}^{-2}$. 

The most outstanding property of SPHERE/ZIMPOL is the spatial resolution
of up to 20~mas surpassing all other imaging instruments available at the VLT
\citep{deZeeuw16}. This limit can even be improved with sparse aperture
masking, which is one of the most recent upgrades of the SPHERE/ZIMPOL system 
\citep{Cheetham16}. Observations with even higher resolution requires currently
interferometric observations. 

Visual extreme AO-systems 
are also available at other observatories, e.g. 
the pioneering MagAO-system at the Magellan telescope \citep{Close13},  
or SCExAO at the Subaru telescope \citep{Jovanovic15}. 
Scientifically important features 
shared by all these visible AO systems is the access to the 
strong H$\alpha$ emission line, and the extension of high 
resolution observations from the traditional near/mid-IR to 
the visible wavelength range 
\citep[see][]{Close16}. There will be some healthy competition between 
these systems, but much more important is the mutual benefit in establishing
common calibration targets, improving and checking measuring
strategies and data reduction, and enhancing the science return 
thanks to complementary performance characteristics. 

Many properties of the ZIMPOL system are quite typical for high resolution
imagers using AO and they are not repeated here. We list here the
special features of ZIMPOL:
\begin{itemize}
\item{} a small detector pixel scale of $3.6~{\rm mas}\times 3.6~{\rm mas}$, 
giving 77160~pix arcsec$^{-2}$ for each detector and a high full well 
photo-electron capacity of $7\cdot 10^5$ e$^-$/pix, allowing for 
efficient high contrast observations of the circumstellar regions of 
bright stars $m_R<8^m$ in broad band filters, 
\item{} dual beam imaging and polarimetry for simultaneous observation
with two cameras with many different options for the combination of
filter bands, for example the combination of 1~nm or 5~nm H$\alpha$ 
filters with a H$\alpha$ continuum filter for spectral differential imaging.
Alternatively, also narrow band filters can be
combined with broad band filters 
for simultaneous, and therefore accurate flux measurements of the
bright component in a high contrast system. 
For example, a bright star can be observed with the narrow Cnt820 
or Cnt\_Ha filters in one arm, and the faint companion or 
circumstellar features in the second arm with the corresponding broad band 
filter I\_PRIM or R\_PRIM, respectively,    
\item{} a high resolution imaging polarimeter based on a fast 
modulation-demodulation technique for speckle noise suppression allowing
high precision, $\Delta p \lapprox 10^{-4}$, broad-band polarimetry,  
\item{} an innovative concept for the compensation and control of
the instrument polarization effects providing absolutely
calibrated linear polarization measurements with an accuracy
of about $\pm 0.1~\%$ for the relative Stokes parameters $Q/I$
and $U/I$, despite the fact that SPHERE/ZIMPOL is a polarimetrically
complex instrument at the Nasmyth focus. 
\item{} the status of SPHERE/ZIMPOL as an ESO/VLT facility instrument, 
which ensures a well monitored and characterized system, steady 
improvements of the instrument operation, high standards for
the execution of the observations and data calibration, 
a well established and user-friendly data archiving, and 
hopefully a long life-time. 
Another huge benefit is the large ESO user community using 
this system for a wide range of science targets with innovative
observing strategies. 
\end{itemize}

\subsection{Possible instrument upgrades}
Some evolution of the ZIMPOL system capabilities can
be expected in the near future, because they can be realized 
without much effort. Very desirable would be the availability
of the ZIMPOL off-axis fields shown in Fig.~\ref{Figfields}, 
which are currently not offered as user mode. This would provide
the possibility to extend the field of view for more extended 
objects to a diameter of $8''$, matching better the 
$11''\times 12.5''$ field offered by the infrared channel SPHERE/IRDIS. 

A very useful  
upgrade would be the availability of a low read-out noise mode
for imaging, equivalent to the slow modulation polarimetric mode
with low detector gain. Especially, low flux measurements taken
in narrow band and line filters, 
as well as observations in the
off-axis fields outside the light halo of the bright central star would
profit because they are often read-out noise limited. Just 
taking long integrations $> 5$~min with a running AO system is 
critical because a few seconds of strongly reduced AO-performance
or even an AO open loop, caused for example by a particularly 
bad atmospheric turbulence event, can degrade an 
entire long integration with an enhanced background of light 
from the central star. 
Another problem of long integrations are small pointing drifts 
of $\gapprox 10$~mas which cause image smearing and this 
would be avoidable with shorter integrations and a realignment
of individual frames in the data reduction. 

Another type of quite easy instrument upgrades are 
new filters or coronagraphic masks in the exchange wheels
of ZIMPOL or the visual coronagraph, respectively. New science cases 
may emerge, which call for filter changes, or better coronagraphic 
concepts could be implemented \citep[e.g.,][]{Patapis18}.

Of course, most relevant would be any upgrades to the AO system or
any other effort to improve the AO performance, like the suppression of 
the low wind effect with changes to the telescope \citep{Sauvage16b}. 
Better wavefront corrections for SPHERE based on new software or hardware
would be particularly beneficial for the short-wavelength ZIMPOL 
subsystem where typical Strehl ratios are $\lapprox 50$~\% with quite 
some room for improvements. 

\subsection{New research opportunities offered by SPHERE/ZIMPOL}

The special technical properties of SPHERE/ZIMPOL offer many new
research opportunities, some are unique to this instrument
and some are shared with the other visible light AO systems 
mentioned above. At least, the visual ZIMPOL observations  
are complementary to the data from the SPHERE near-IR focal plane 
instruments IRDIS and IFS, or other near-IR AO systems like 
for example Gemini/GPI \citep{Macintosh14} or Subaru 
SCExAO \citep{Jovanovic15}, by extending the wavelength domain of
high resolution and high contrast imaging towards shorter wavelengths.
In the following, we provide an incomplete list 
of science topics where SPHERE/ZIMPOL
is already providing or will provide interesting or even 
very important contributions in high contrast imaging. In this 
discussion on science opportunities one should not forget the 
technical requirement that SPHERE/ZIMPOL observations need 
a bright central star $m_R\lapprox 9^m$ for the AO wavefront 
sensing and that the system provides only a quite limited field 
of view $\rho<4''$.

\subsubsection{Search for extra-solar planets in reflected light.}
ZIMPOL was selected by the European Southern Observatory 
for the SPHERE VLT ``planet finder'' instrument with the mandate
to explore the detection limits of high contrast
polarimetric imaging for the search of reflected light from
extrasolar planets \citep{Schmid06a}. The aim of this
unique instrument is to reach a contrast limit between 
the polarized flux of the planet  $p_{\rm pl}\times I_{\rm pl}$ and 
the total flux of a star $I_{\rm star}$ of 
\begin{displaymath}
C_{\rm pol}={{p_{\rm pl}\times I_{\rm pl}}\over I_{\rm star}}=10^{-8} 
\end{displaymath}
within an angular separation smaller than $\rho<1''$. 
This limit would allow a detection of a Jupiter-sized giant
planet or even terrestrial planets 
with a physical separation of $0.5-1$~AU around one of the nearest
bright stars \citep{Thalmann08,Milli13}. This aim defined many
of the ZIMPOL design decisions and therefore 
this imager is tuned for the detection of 
very faint polarization sources near very bright stars. 

The SPHERE-team carries out as part of their guaranteed
time observations obtained for building the instrument, 
an investigation of the achievable detection limits of
for the search of extra-solar planets. 
Very high contrast observations of a small number of targets
are currently taken. The data confirm that ZIMPOL can reach 
at least for separations $\rho>0.5''$ the above mentioned 
detection limit. The short test observations of R Aqr 
described in the previous section give already an impression
of the high contrast performance in polarimetric imaging. 
SPHERE/ZIMPOL is pioneering this technique and the achievable 
contrast limits are certainly of interest 
for other high contrast search programs targeting the 
reflected light from extra-solar planets and for the 
planning of future instruments.  
Whether a successful detection will be possible with
SPHERE/ZIMPOL depends on the presence of favorable
planets within about 5~pc and further progress in the
observations and data analysis. 

\subsubsection{Differential polarimetric imaging of 
circumstellar disk.}
Disks around young stars are a primary science case for the
SPHERE instrument and differential polarimetric imaging
is a powerful technique for high contrast disk observations
\citep[e.g.,][]{Kuhn01,Perrin09,Quanz11,Muto12}.
The high spatial resolution and the high polarimetric sensitivity 
of ZIMPOL are ideal for the mapping of faint circumstellar disks in 
polarized light. The visual ZIMPOL data can be combined with 
near-IR observations, for example
from SPHERE/IRDIS, to study the color dependence of the reflected 
light from protoplanetary disks around young stars like for example
HD 135344B or TW Hya \citep{Stolker16,vanBoekel17}.  

Let us compare for this important science case the pro and 
cons for the ZIMPOL and IRDIS polarimetry 
modes for the differential polarimetric imaging of 
circumstellar disk, focussing on detecting a disk and mapping
structural features. For fainter central stars $m_R>8^m$
IRDIS polarimetry provides as important advantage a significantly
better AO performance, because all light
in the 500-900~nm range can be used for the wave front sensing.
The ZIMPOL science channel shares this light with the WFS and therefore
only 21~\% of the flux is available for the WFS, if the gray beam splitter
is used, or about 80~\% for the dichroic beam splitter. Another advantage
of IRDIS is the larger field of view of $11''\times 11''$, about 
ten times larger than the $3.6''\times 3.6''$ detector field 
of view of ZIMPOL, or much more efficient than using multiple 
field observations inside
the $8''$ diameter instrument field of view. On the other side, 
the main advantages of ZIMPOL polarimetry when compared to
IRDIS or other IR-polarimeters are the higher
spatial resolution and the very good speckle noise suppression by the 
fast-modulation polarimetry techniques as described in this paper
for R Aqr. Therefore, a higher contrast can be obtained 
for circumstellar disk around bright stars in the separation range
$\rho\approx 0.02''-0.2''$ and a good example is the detection
of the inner disk of HD 142527 by \citet{Avenhaus17}. 
For disks around bright stars, ZIMPOL is at least competitive for 
the separation range $\rho\approx 0.2-2.0$ as demonstrated 
for the faint debris disks HIP 79977 \citep{Engler17}, 
but also for the innermost regions 
of bright protoplanetary disks, like HD 100457 \citep{Benisty17}.
In addition, ZIMPOL provides a more advanced polarimetric concept 
than IRDIS or other AO-assisted polarimetric imagers, 
allowing an easier calibration of polarimetric data and
a quantitative analysis of the polarized reflectivity of the 
scattering dust 

\subsubsection{Mass loss from red giants.} The combination of 
high spatial resolution and sensitive polarimetry
is ideal for the mapping of the light scattering 
from the circumstellar dust with SPHERE/ZIMPOL
\citep[e.g.,][]{Kervella15,Khouri16,Ohnaka17a}. As shown
in Sect.~\ref{SectRAqr} for R Aqr, the polarized light from
dust scattering can be measured over a very wide separation
range, and asymmetries, clumps and their evolution 
can be investigated in much detail. This important information
from high resolution polarimetry, which was pioneered
with interferometric observations by \citet{Ireland05},
or sparse aperture masking by \citep{Norris12}, is 
now also available with ``simple'' imaging. 
Such observations could be particularly useful for 
investigations and the modeling 
of the complex dust formation process in pulsating AGB stars 
\citep[e.g.,][]{Aronson17,Hoefner16,Hoefner08}. 

Light scattering observations are highly complementary to 
observations of the thermal emission of the circumstellar dust
in the mid-IR with e.g. the VLTI/MIDI interferometer \citep{Paladini17}. 
AO observations in the visual achieves a comparable or even
better resolution than mid-IR interferometry and is therefore
well suited for the mapping of the complex distribution of 
circumstellar dust near the mass losing star. The visible
range is ideal for the small dust particles formed around
mass-losing red giants because they scatter much
more efficiently short wavelength light.
A technical challenge for very bright red giants are detector
saturation issues. Red giants are much fainter in the visual and
the ZIMPOL system is designed especially for high contrast observations 
of very bright targets and therefore also for the investigation
of the brightest, most extended, nearby red giants.
 
\subsubsection{Emission lines in stellar jets and outflows.} 
Stellar jets and outflows produce often H$\alpha$ and other emission 
lines from shocks or photoionized regions. ZIMPOL line filters are
available for the H$\alpha$ 656~nm, [O\,I] 630~nm, 
and HeI/NaI 588/589~nm lines which may serve as tracer of different
types of ionized or partially ionized gas. 

Achieving high contrast and high resolution observations in
line filters is important for stellar jets from young 
stars \citep{Frank14} to access the innermost 10~AU where
the outflow is not yet significantly perturbed 
by the interaction with the ambient medium. This requires line
observations at separations below 70~mas for sources
in nearby star-forming regions. Observing the innermost
flow morphology is important to pinpoint the initial ejection
site of the matter, e.g. a stellar wind, an ``X-wind''  from the inner 
edge of the disk, or a disk wind, which is then further accelerated
and collimated by the combined action of magnetic fields and rotation
\citep{Ferreira06}. A first demonstration of the ZIMPOL potential on
this topic is provided in \citet{Antoniucci16} with H$\alpha$ and 
[O\,I] observations of the young binary Z CMa. The authors
could trace the collimated jet from one of the components 
down to $\sim 70$~mas from the driving source, revealing a
jet wiggling on time-scales of a few years, which may be
induced by a non-detected close-in companion. 

For many astronomical objects countless 
imaging data allready exists of the circumstellar line emission
in the visual wavelength 
region taken with ground-based or space telescopes. 
The line observations of R Aqr provide a good example for 
the complementarity of emission line imaging with the 
SPHERE/ZIMPOL AO system \citep{Schmid17}, with HST 
imaging \citep{Melnikov18}, and with seeing limited imaging
\citep{Liimets18}.

\subsubsection{Resolving the atmosphere of red giants.}
The most extended red giant stars can be resolved with
``simple ZIMPOL imaging'' with a resolution of up to 20~mas
as demonstrated for example 
for R Dor \citep{Khouri16} or $\alpha$ Ori \citep{Kervella16}. 
``Simple imaging'' because one can take many images in
several filters within a few minutes,
select the best data and measure wavelength dependencies.
Geometric features, such as large spots or polar and equatorial
zones, can be investigated and flux ratio maps can be obtained
with simultaneous differential imaging, in e.g. the 
TiO\_717 and Cnt748 band filters which sample
cold and hot surfaces regions, or the N\_Ha and CntHa  
filter pair for possible signs of shock heating. 
The ``simple imaging'' is also ideal for a monitoring
program of the temporal evolution of surface features in these
stars. Even better spatial resolution $\rho< 20$~mas 
should be achievable with sparse aperture masking 
or with advanced data analysis techniques. 
For these reasons the ``simple'', $\approx 20$~mas resolution, 
SPHERE/ZIMPOL imaging of extended red giant atmospheres provides 
useful complementary information with respect to the higher 
resolution, but much harder to obtain interferometric data 
\citep[e.g.,][]{Haniff95,vanBelle96,Ohnaka17b}. 

\subsubsection{Close binary stars.}
For very close binary stars, the $\approx 20$~mas spatial resolution 
of SPHERE/ZIMPOL is of course very useful for orbit determinations and
the photometry of the individual components 
\citep[e.g.,][]{Janson18}. 
A particular niche for the visual ZIMPOL instrument, when 
compared to near-IR AO instruments,  
are faint, hot companions to red stars, like
white dwarfs companions to Ba-star, or H$\alpha$ emitting active
components to M-giants like the symbiotic system R Aqr shown
in Fig.~\ref{rawcounts} \citep[see also][]{Schmid17}. 
The relative position between roughly equal
flux ($|\log (f_1/f_2)|\lapprox 1$) hot and cold binary components can
certainly be determined for separations of $\approx 10$~mas or 
even smaller with simultaneous measurements
of the combined binary PSF in a visual and a red filter. Photocenter 
differences between two bands might even be measurable at the 
milli-arcsec level, if other stars in the field or the features 
of a coronagraphic mask can be used as relative 
astrometric reference. 

\subsubsection{Solar system objects.} 
The SPHERE AO system is capable to lock on moving solar
system objects if they are bright enough $m_{\rm R}\lapprox 10^m$
and not too extended $\lapprox 2''$.
This was demonstrated during the
SPHERE commissioning for 
Titan\footnote{ESO press release www.eso.org/public/news/eso1417}
which has a diameter of $\rho=0.8''$  
and even Neptune with $\rho=2.4''$ \citep{Fusco16}. 
Thus, many bright asteroids, the Galilean moons, and Saturn's moon 
Titan can be imaged in the visual. Sizes, shapes and 
surface structures can be investigated in much detail 
\citep{Vernazza17}, and with enhanced resolution when
compared to the near-IR range \citep[e.g.,][]{Marchis06},
including the polarimetric properties of the 
reflecting terrains.

\subsection{Conclusions} 
SPHERE/ZIMPOL is a very versatile adaptive optics
instrument and we therefore
expect many exciting new scientific results from 
this instrument. The above listed technical performances, 
upgrade options, and science topics give only a few examples 
of possible observational projects with this instrument. 

Observational results based on adaptive optics profit
a lot from the much enhanced spatial resolution and this provides 
since many years a continuous string of new detections
\citep[see][]{Davies12}. The description of SPHERE/ZIMPOL 
given in this paper should help to define the best observing 
strategy for reaching deeper detection limits for new 
discoveries with AO observations at very high spatial resolution
in the visual, using polarimetric imaging, 
angular or differential imaging with broad band, narrow band, or 
line filters.  

On the other side, accurate quantitative measurements
with AO systems are often difficult, because of the strongly variable 
atmospheric conditions and the resulting system performance. 
Particularly problematic is the photometry for very faint 
companions or extended circumstellar features for which
simultaneous or quasi-simultaneous differential measurements
are impossible. This makes the accurate characterization
of high contrast objects in different wavelength bands, taken
often with different instruments and usually under different
atmospheric conditions very difficult and often uncertain. 
A lot of effort is required to describe 
AO observations accuratly and in a reproducible way but this is
required for a detailed characterization of high contrast
objects. This paper provides therefore a lot of technical information
for the accurate characterization and calibration of 
SPHERE/ZIMPOL measurements.

\begin{acknowledgements}
HMS thanks Wolfgang L\"{o}ffler, Universiteit Leiden, for very
helpful discussions and comments about the Goos-H\"ahnchen shift and 
related effects. 
SPHERE is an instrument designed and built by a consortium consisting
of IPAG (Grenoble, France), MPIA (Heidelberg, Germany), LAM (Marseille,
France), LESIA (Paris, France), Laboratoire Lagrange (Nice, France), 
INAF – Osservatorio di Padova (Italy), Observatoire de Gen`eve 
(Switzerland), ETH Zurich (Switzerland), NOVA (Netherlands), ONERA (France) 
and ASTRON (Netherlands), in collaboration with ESO. 
SPHERE was funded by ESO, with additional contributions from CNRS (France), 
MPIA (Germany), INAF (Italy), FINES (Switzerland) and NOVA (Netherlands). 
SPHERE also received funding from the European Commission Sixth and 
Seventh Framework Programmes as part of the Optical Infrared 
Coordination Network for Astronomy (OPTICON)
under grant number RII3-Ct-2004-001566 for FP6 (2004–2008), grant number
226604 for FP7 (2009–2012) and grant number 312430 for FP7 (2013–2016).

HMS, SH and NE acknoledge financial support from SNSF 
through grant 200020\_162630. Part of this work has been carried out 
within the framework of the National Centre for Competence in 
Research PlanetS supported by the Swiss National Science Foundation. H.A.
and S.P.Q. acknowledge the financial support of the SNSF.  

\end{acknowledgements}

\bibliographystyle{aa} 
\bibliography{ZIMPOL_ArXiv.bib} 

\begin{thebibliography}{134}
\expandafter\ifx\csname natexlab\endcsname\relax\def\natexlab#1{#1}\fi

\bibitem[{{Aiello} {et~al.}(2009){Aiello}, {Merano}, \& {Woerdman}}]{Aiello09}
{Aiello}, A., {Merano}, M., \& {Woerdman}, J.~P. 2009, \pra, 80, 061801

\bibitem[{{Akiyama} {et~al.}(2015){Akiyama}, {Muto}, {Kusakabe}, {Kataoka},
  {Hashimoto}, {Tsukagoshi}, {Kwon}, {Kudo}, {Kandori}, {Grady}, {Takami},
  {Janson}, {Kuzuhara}, {Henning}, {Sitko}, {Carson}, {Mayama}, {Currie},
  {Thalmann}, {Wisniewski}, {Momose}, {Ohashi}, {Abe}, {Brandner}, {Brandt},
  {Egner}, {Feldt}, {Goto}, {Guyon}, {Hayano}, {Hayashi}, {Hayashi}, {Hodapp},
  {Ishi}, {Iye}, {Knapp}, {Matsuo}, {Mcelwain}, {Miyama}, {Morino},
  {Moro-Martin}, {Nishimura}, {Pyo}, {Serabyn}, {Suenaga}, {Suto}, {Suzuki},
  {Takahashi}, {Takato}, {Terada}, {Tomono}, {Turner}, {Watanabe}, {Yamada},
  {Takami}, {Usuda}, \& {Tamura}}]{Akiyama15}
{Akiyama}, E., {Muto}, T., {Kusakabe}, N., {et~al.} 2015, \apjl, 802, L17

\bibitem[{{Amara} \& {Quanz}(2012)}]{Amara12}
{Amara}, A. \& {Quanz}, S.~P. 2012, \mnras, 427, 948

\bibitem[{{Antoniucci} {et~al.}(2016){Antoniucci}, {Podio}, {Nisini},
  {Bacciotti}, {Lagadec}, {Sissa}, {La Camera}, {Giannini}, {Schmid},
  {Gratton}, {Turatto}, {Desidera}, {Bonnefoy}, {Chauvin}, {Dougados},
  {Bazzon}, {Thalmann}, \& {Langlois}}]{Antoniucci16}
{Antoniucci}, S., {Podio}, L., {Nisini}, B., {et~al.} 2016, \aap, 593, L13

\bibitem[{{Appenzeller}(1968)}]{Appenzeller68}
{Appenzeller}, I. 1968, \apj, 151, 907

\bibitem[{{Aronson} {et~al.}(2017){Aronson}, {Bladh}, \&
  {H{\"o}fner}}]{Aronson17}
{Aronson}, E., {Bladh}, S., \& {H{\"o}fner}, S. 2017, \aap, 603, A116

\bibitem[{{Artmann}(1948)}]{Artmann48}
{Artmann}, K. 1948, Annalen der Physik, 437, 87

\bibitem[{{Aspin} {et~al.}(1985){Aspin}, {Schwarz}, {McLean}, \&
  {Boyle}}]{Aspin85}
{Aspin}, C., {Schwarz}, H.~E., {McLean}, I.~S., \& {Boyle}, R.~P. 1985, \aap,
  149, L21

\bibitem[{{Avenhaus} {et~al.}(2017){Avenhaus}, {Quanz}, {Schmid}, {Dominik},
  {Stolker}, {Ginski}, {de Boer}, {Szul{\'a}gyi}, {Garufi}, {Zurlo},
  {Hagelberg}, {Benisty}, {Henning}, {M{\'e}nard}, {Meyer}, {Baruffolo},
  {Bazzon}, {Beuzit}, {Costille}, {Dohlen}, {Girard}, {Gisler}, {Kasper},
  {Mouillet}, {Pragt}, {Roelfsema}, {Salasnich}, \& {Sauvage}}]{Avenhaus17}
{Avenhaus}, H., {Quanz}, S.~P., {Schmid}, H.~M., {et~al.} 2017, \aj, 154, 33

\bibitem[{{Avenhaus} {et~al.}(2014){Avenhaus}, {Quanz}, {Schmid}, {Meyer},
  {Garufi}, {Wolf}, \& {Dominik}}]{Avenhaus14}
{Avenhaus}, H., {Quanz}, S.~P., {Schmid}, H.~M., {et~al.} 2014, \apj, 781, 87

\bibitem[{{Bailey} \& {Hough}(1982)}]{Bailey82}
{Bailey}, J. \& {Hough}, J.~H. 1982, \pasp, 94, 618

\bibitem[{{Bastien} {et~al.}(1988){Bastien}, {Drissen}, {Menard}, {Moffat},
  {Robert}, \& {St-Louis}}]{Bastien88}
{Bastien}, P., {Drissen}, L., {Menard}, F., {et~al.} 1988, \aj, 95, 900

\bibitem[{{Bastien} {et~al.}(2007){Bastien}, {Vernet}, {Drissen}, {M{\'e}nard},
  {Moffat}, {Robert}, \& {St-Louis}}]{Bastien07}
{Bastien}, P., {Vernet}, E., {Drissen}, L., {et~al.} 2007, in Astronomical
  Society of the Pacific Conference Series, Vol. 364, The Future of
  Photometric, Spectrophotometric and Polarimetric Standardization, ed.
  C.~{Sterken}, 529

\bibitem[{{Bazzon} {et~al.}(2012){Bazzon}, {Gisler}, {Roelfsema}, {Schmid},
  {Pragt}, {Elswijk}, {de Haan}, {Downing}, {Salasnich}, {Pavlov}, {Beuzit},
  {Dohlen}, {Mouillet}, \& {Wildi}}]{Bazzon12}
{Bazzon}, A., {Gisler}, D., {Roelfsema}, R., {et~al.} 2012, in \procspie, Vol.
  8446, Ground-based and Airborne Instrumentation for Astronomy IV., 93

\bibitem[{{Bellini} {et~al.}(2014){Bellini}, {Anderson}, {van der Marel},
  {Watkins}, {King}, {Bianchini}, {Chanam{\'e}}, {Chandar}, {Cool}, {Ferraro},
  {Ford}, \& {Massari}}]{Bellini14}
{Bellini}, A., {Anderson}, J., {van der Marel}, R.~P., {et~al.} 2014, \apj,
  797, 115

\bibitem[{{Benisty} {et~al.}(2017){Benisty}, {Stolker}, {Pohl}, {de Boer},
  {Lesur}, {Dominik}, {Dullemond}, {Langlois}, {Min}, {Wagner}, {Henning},
  {Juhasz}, {Pinilla}, {Facchini}, {Apai}, {van Boekel}, {Garufi}, {Ginski},
  {M{\'e}nard}, {Pinte}, {Quanz}, {Zurlo}, {Boccaletti}, {Bonnefoy}, {Beuzit},
  {Chauvin}, {Cudel}, {Desidera}, {Feldt}, {Fontanive}, {Gratton}, {Kasper},
  {Lagrange}, {LeCoroller}, {Mouillet}, {Mesa}, {Sissa}, {Vigan}, {Antichi},
  {Buey}, {Fusco}, {Gisler}, {Llored}, {Magnard}, {Moeller-Nilsson}, {Pragt},
  {Roelfsema}, {Sauvage}, \& {Wildi}}]{Benisty17}
{Benisty}, M., {Stolker}, T., {Pohl}, A., {et~al.} 2017, \aap, 597, A42

\bibitem[{{Beuzit} {et~al.}(2008){Beuzit}, {Feldt}, {Dohlen}, {Mouillet},
  {Puget}, {Wildi}, {Abe}, {Antichi}, {Baruffolo}, {Baudoz}, {Boccaletti},
  {Carbillet}, {Charton}, {Claudi}, {Downing}, {Fabron}, {Feautrier},
  {Fedrigo}, {Fusco}, {Gach}, {Gratton}, {Henning}, {Hubin}, {Joos}, {Kasper},
  {Langlois}, {Lenzen}, {Moutou}, {Pavlov}, {Petit}, {Pragt}, {Rabou}, {Rigal},
  {Roelfsema}, {Rousset}, {Saisse}, {Schmid}, {Stadler}, {Thalmann}, {Turatto},
  {Udry}, {Vakili}, \& {Waters}}]{Beuzit08}
{Beuzit}, J.-L., {Feldt}, M., {Dohlen}, K., {et~al.} 2008, in \procspie, Vol.
  7014, Ground-based and Airborne Instrumentation for Astronomy II., 18

\bibitem[{{Bliokh} \& {Aiello}(2013)}]{Bliokh13}
{Bliokh}, K.~Y. \& {Aiello}, A. 2013, Journal of Optics, 15, 014001

\bibitem[{{Boccaletti} {et~al.}(2008){Boccaletti}, {Abe}, {Baudrand}, {Daban},
  {Douet}, {Guerri}, {Robbe-Dubois}, {Bendjoya}, {Dohlen}, \&
  {Mawet}}]{Boccaletti08}
{Boccaletti}, A., {Abe}, L., {Baudrand}, J., {et~al.} 2008, in \procspie, Vol.
  7015, Adaptive Optics Systems, 70151B

\bibitem[{{Bonnefoy} {et~al.}(2016){Bonnefoy}, {Zurlo}, {Baudino}, {Lucas},
  {Mesa}, {Maire}, {Vigan}, {Galicher}, {Homeier}, {Marocco}, {Gratton},
  {Chauvin}, {Allard}, {Desidera}, {Kasper}, {Moutou}, {Lagrange}, {Antichi},
  {Baruffolo}, {Baudrand}, {Beuzit}, {Boccaletti}, {Cantalloube}, {Carbillet},
  {Charton}, {Claudi}, {Costille}, {Dohlen}, {Dominik}, {Fantinel},
  {Feautrier}, {Feldt}, {Fusco}, {Gigan}, {Girard}, {Gluck}, {Gry}, {Henning},
  {Janson}, {Langlois}, {Madec}, {Magnard}, {Maurel}, {Mawet}, {Meyer},
  {Milli}, {Moeller-Nilsson}, {Mouillet}, {Pavlov}, {Perret}, {Pujet}, {Quanz},
  {Rochat}, {Rousset}, {Roux}, {Salasnich}, {Salter}, {Sauvage}, {Schmid},
  {Sevin}, {Soenke}, {Stadler}, {Turatto}, {Udry}, {Vakili}, {Wahhaj}, \&
  {Wildi}}]{Bonnefoy16}
{Bonnefoy}, M., {Zurlo}, A., {Baudino}, J.~L., {et~al.} 2016, \aap, 587, A58

\bibitem[{{Born} \& {Wolf}(1999)}]{Born99}
{Born}, M. \& {Wolf}, E. 1999, {Principles of Optics, Cambridge University
  Press}

\bibitem[{{Brandner} \& {Hormuth}(2016)}]{Brandner16}
{Brandner}, W. \& {Hormuth}, F. 2016, in Astrophysics and Space Science
  Library, Vol. 439, Astronomy at High Angular Resolution, ed. H.~M.~J.
  {Boffin}, G.~{Hussain}, J.-P. {Berger}, \& L.~{Schmidtobreick}, 1

\bibitem[{{Breckinridge} {et~al.}(2015){Breckinridge}, {Lam}, \&
  {Chipman}}]{Breckinridge15}
{Breckinridge}, J.~B., {Lam}, W.~S.~T., \& {Chipman}, R.~A. 2015, \pasp, 127,
  445

\bibitem[{{Cheetham} {et~al.}(2016){Cheetham}, {Girard}, {Lacour}, {Schworer},
  {Haubois}, \& {Beuzit}}]{Cheetham16}
{Cheetham}, A.~C., {Girard}, J., {Lacour}, S., {et~al.} 2016, in \procspie,
  Vol. 9907, Optical and Infrared Interferometry and Imaging V, 99072T

\bibitem[{{Clarke} {et~al.}(1983){Clarke}, {Stewart}, {Schwarz}, \&
  {Brooks}}]{Clarke83}
{Clarke}, D., {Stewart}, B.~G., {Schwarz}, H.~E., \& {Brooks}, A. 1983, \aap,
  126, 260

\bibitem[{{Claudi} {et~al.}(2008){Claudi}, {Turatto}, {Gratton}, {Antichi},
  {Bonavita}, {Bruno}, {Cascone}, {De Caprio}, {Desidera}, {Giro}, {Mesa},
  {Scuderi}, {Dohlen}, {Beuzit}, \& {Puget}}]{Claudi08}
{Claudi}, R.~U., {Turatto}, M., {Gratton}, R.~G., {et~al.} 2008, in \procspie,
  Vol. 7014, Ground-based and Airborne Instrumentation for Astronomy II, 70143E

\bibitem[{{Close}(2016)}]{Close16}
{Close}, L.~M. 2016, in \procspie, Vol. 9909, Adaptive Optics Systems V, 99091E

\bibitem[{{Close} {et~al.}(2013){Close}, {Males}, {Morzinski}, {Kopon},
  {Follette}, {Rodigas}, {Hinz}, {Wu}, {Puglisi}, {Esposito}, {Riccardi},
  {Pinna}, {Xompero}, {Briguglio}, {Uomoto}, \& {Hare}}]{Close13}
{Close}, L.~M., {Males}, J.~R., {Morzinski}, K., {et~al.} 2013, \apj, 774, 94

\bibitem[{{Collett}(1992)}]{Collett92}
{Collett}, E. 1992, {Polarized light. Fundamentals and applications, New York:
  Dekker}

\bibitem[{{Cox}(1976)}]{Cox76}
{Cox}, L.~J. 1976, \mnras, 176, 525

\bibitem[{{Davies} \& {Kasper}(2012)}]{Davies12}
{Davies}, R. \& {Kasper}, M. 2012, \araa, 50, 305

\bibitem[{{de Juan Ovelar} {et~al.}(2012){de Juan Ovelar}, {Diamantopoulou},
  {Roelfsema}, {van Werkhoven}, {Snik}, {Pragt}, \& {Keller}}]{deJuanOvelar12}
{de Juan Ovelar}, M., {Diamantopoulou}, S., {Roelfsema}, R., {et~al.} 2012, in
  \procspie, Vol. 8449, Modeling, Systems Engineering, and Project Management
  for Astronomy V, 844912

\bibitem[{{de Zeeuw}(2016)}]{deZeeuw16}
{de Zeeuw}, T. 2016, The Messenger, 166, 2

\bibitem[{{Dixon} {et~al.}(1995){Dixon}, {Davidsen}, \& {Ferguson}}]{Dixon95}
{Dixon}, W.~V.~D., {Davidsen}, A.~F., \& {Ferguson}, H.~C. 1995, \apjl, 454,
  L47

\bibitem[{{Dohlen} {et~al.}(2008){Dohlen}, {Langlois}, {Saisse}, {Hill},
  {Origne}, {Jacquet}, {Fabron}, {Blanc}, {Llored}, {Carle}, {Moutou}, {Vigan},
  {Boccaletti}, {Carbillet}, {Mouillet}, \& {Beuzit}}]{Dohlen08}
{Dohlen}, K., {Langlois}, M., {Saisse}, M., {et~al.} 2008, in \procspie, Vol.
  7014, Ground-based and Airborne Instrumentation for Astronomy II, 70143L

\bibitem[{{Dohlen} {et~al.}(2016){Dohlen}, {Vigan}, {Mouillet}, {Wildi},
  {Sauvage}, {Fusco}, {Beuzit}, {Puget}, {Le Mignant}, {Roelfsema}, {Pragt},
  {Schmid}, {Gratton}, {Mesa}, {Claudi}, {Langlois}, {Costille}, {Hugot},
  {O'Neil}, {Guerra}, {N'Diaye}, {Girard}, {Mawet}, \& {Zins}}]{Dohlen16}
{Dohlen}, K., {Vigan}, A., {Mouillet}, D., {et~al.} 2016, in \procspie, Vol.
  9908, Ground-based and Airborne Instrumentation for Astronomy VI, 99083D

\bibitem[{{Engler} {et~al.}(2017){Engler}, {Schmid}, {Thalmann}, {Boccaletti},
  {Bazzon}, {Baruffolo}, {Beuzit}, {Claudi}, {Costille}, {Desidera}, {Dohlen},
  {Dominik}, {Feldt}, {Fusco}, {Ginski}, {Gisler}, {Girard}, {Gratton},
  {Henning}, {Hubin}, {Janson}, {Kasper}, {Kral}, {Langlois}, {Lagadec},
  {M{\'e}nard}, {Meyer}, {Milli}, {Mouillet}, {Olofsson}, {Pavlov}, {Pragt},
  {Puget}, {Quanz}, {Roelfsema}, {Salasnich}, {Siebenmorgen}, {Sissa},
  {Suarez}, {Szulagyi}, {Turatto}, {Udry}, \& {Wildi}}]{Engler17}
{Engler}, N., {Schmid}, H.~M., {Thalmann}, C., {et~al.} 2017, \aap, 607, A90

\bibitem[{{Ferreira} {et~al.}(2006){Ferreira}, {Dougados}, \&
  {Cabrit}}]{Ferreira06}
{Ferreira}, J., {Dougados}, C., \& {Cabrit}, S. 2006, \aap, 453, 785

\bibitem[{{Frank} {et~al.}(2014){Frank}, {Ray}, {Cabrit}, {Hartigan}, {Arce},
  {Bacciotti}, {Bally}, {Benisty}, {Eisl{\"o}ffel}, {G{\"u}del}, {Lebedev},
  {Nisini}, \& {Raga}}]{Frank14}
{Frank}, A., {Ray}, T.~P., {Cabrit}, S., {et~al.} 2014, Protostars and Planets
  VI, 451

\bibitem[{{Fusco} {et~al.}(2016){Fusco}, {Sauvage}, {Mouillet}, {Costille},
  {Petit}, {Beuzit}, {Dohlen}, {Milli}, {Girard}, {Kasper}, {Vigan}, {Suarez},
  {Soenke}, {Downing}, {N'Diaye}, {Baudoz}, {Sevin}, {Baruffolo}, {Schmid},
  {Salasnich}, {Hugot}, \& {Hubin}}]{Fusco16}
{Fusco}, T., {Sauvage}, J.-F., {Mouillet}, D., {et~al.} 2016, in \procspie,
  Vol. 9909, Adaptive Optics Systems V, 99090U

\bibitem[{{Fusco} {et~al.}(2014){Fusco}, {Sauvage}, {Petit}, {Costille},
  {Dohlen}, {Mouillet}, {Beuzit}, {Kasper}, {Suarez}, {Soenke}, {Fedrigo},
  {Downing}, {Baudoz}, {Sevin}, {Perret}, {Barrufolo}, {Salasnich}, {Puget},
  {Feautrier}, {Rochat}, {Moulin}, {Deboulb{\'e}}, {Hugot}, {Vigan}, {Mawet},
  {Girard}, \& {Hubin}}]{Fusco14}
{Fusco}, T., {Sauvage}, J.-F., {Petit}, C., {et~al.} 2014, in \procspie, Vol.
  9148, Adaptive Optics Systems IV, 1

\bibitem[{{Gandorfer}(1999)}]{Gandorfer99}
{Gandorfer}, A.~M. 1999, Optical Engineering, 38

\bibitem[{{Gandorfer} \& {Povel}(1997)}]{Gandorfer97}
{Gandorfer}, A.~M. \& {Povel}, H.~P. 1997, \aap, 328, 381

\bibitem[{{Garufi} {et~al.}(2016){Garufi}, {Quanz}, {Schmid}, {Mulders},
  {Avenhaus}, {Boccaletti}, {Ginski}, {Langlois}, {Stolker}, {Augereau},
  {Benisty}, {Lopez}, {Dominik}, {Gratton}, {Henning}, {Janson}, {M{\'e}nard},
  {Meyer}, {Pinte}, {Sissa}, {Vigan}, {Zurlo}, {Bazzon}, {Buenzli}, {Bonnefoy},
  {Brandner}, {Chauvin}, {Cheetham}, {Cudel}, {Desidera}, {Feldt}, {Galicher},
  {Kasper}, {Lagrange}, {Lannier}, {Maire}, {Mesa}, {Mouillet}, {Peretti},
  {Perrot}, {Salter}, \& {Wildi}}]{Garufi16}
{Garufi}, A., {Quanz}, S.~P., {Schmid}, H.~M., {et~al.} 2016, \aap, 588, A8

\bibitem[{{Gisler} {et~al.}(2003){Gisler}, {Feller}, \& {Gandorfer}}]{Gisler03}
{Gisler}, D., {Feller}, A., \& {Gandorfer}, A.~M. 2003, in \procspie, Vol.
  4843, Polarimetry in Astronomy, 45--54

\bibitem[{{Gisler} {et~al.}(2004){Gisler}, {Schmid}, {Thalmann}, {Povel},
  {Stenflo}, {Joos}, {Feldt}, {Lenzen}, {Tinbergen}, {Gratton}, {Stuik},
  {Stam}, {Brandner}, {Hippler}, {Turatto}, {Neuhauser}, {Dominik}, {Hatzes},
  {Henning}, {Lima}, {Quirrenbach}, {Waters}, {Wuchterl}, \&
  {Zinnecker}}]{Gisler04}
{Gisler}, D., {Schmid}, H.~M., {Thalmann}, C., {et~al.} 2004, in \procspie,
  Vol. 5492, Ground-based Instrumentation for Astronomy, 463--474

\bibitem[{{Goos} \& {H{\"a}nchen}(1947)}]{Goos47}
{Goos}, F. \& {H{\"a}nchen}, H. 1947, Annalen der Physik, 436, 333

\bibitem[{{Haniff} {et~al.}(1995){Haniff}, {Scholz}, \& {Tuthill}}]{Haniff95}
{Haniff}, C.~A., {Scholz}, M., \& {Tuthill}, P.~G. 1995, \mnras, 276, 640

\bibitem[{{H{\"o}fner}(2008)}]{Hoefner08}
{H{\"o}fner}, S. 2008, \aap, 491, L1

\bibitem[{{H{\"o}fner} {et~al.}(2016){H{\"o}fner}, {Bladh}, {Aringer}, \&
  {Ahuja}}]{Hoefner16}
{H{\"o}fner}, S., {Bladh}, S., {Aringer}, B., \& {Ahuja}, R. 2016, \aap, 594,
  A108

\bibitem[{{Hsu} \& {Breger}(1982)}]{Hsu82}
{Hsu}, J.-C. \& {Breger}, M. 1982, \apj, 262, 732

\bibitem[{{Ireland} {et~al.}(2005){Ireland}, {Tuthill}, {Davis}, \&
  {Tango}}]{Ireland05}
{Ireland}, M.~J., {Tuthill}, P.~G., {Davis}, J., \& {Tango}, W. 2005, \mnras,
  361, 337

\bibitem[{{Janson} {et~al.}(2018){Janson}, {Durkan}, {Bonnefoy}, {Rodet},
  {K\"{o}hler}, {Lacour}, {Brandner}, {Henning}, \& {Girard}}]{Janson18}
{Janson}, M., {Durkan}, S., {Bonnefoy}, M., {et~al.} 2018, \aap, submitted

\bibitem[{{Joos}(2007)}]{Joos07}
{Joos}, F. 2007, {Polarimetry of Gas Planets, PhD Thesis, ETH Zurich, No.
  17051,}

\bibitem[{{Joshi} {et~al.}(2012){Joshi}, {Ganesh}, \& {Baliyan}}]{Joshi12}
{Joshi}, U.~C., {Ganesh}, S., \& {Baliyan}, K.~S. 2012, in AIP Conference
  Proceedings, Vol. 1429, Stellar polarimetry: from birth to death, 222--225

\bibitem[{{Jovanovic} {et~al.}(2015){Jovanovic}, {Martinache}, {Guyon},
  {Clergeon}, {Singh}, {Kudo}, {Garrel}, {Newman}, {Doughty}, {Lozi}, {Males},
  {Minowa}, {Hayano}, {Takato}, {Morino}, {Kuhn}, {Serabyn}, {Norris},
  {Tuthill}, {Schworer}, {Stewart}, {Close}, {Huby}, {Perrin}, {Lacour},
  {Gauchet}, {Vievard}, {Murakami}, {Oshiyama}, {Baba}, {Matsuo}, {Nishikawa},
  {Tamura}, {Lai}, {Marchis}, {Duchene}, {Kotani}, \& {Woillez}}]{Jovanovic15}
{Jovanovic}, N., {Martinache}, F., {Guyon}, O., {et~al.} 2015, \pasp, 127, 890

\bibitem[{{Kasper} {et~al.}(2012){Kasper}, {Beuzit}, {Feldt}, {Dohlen},
  {Mouillet}, {Puget}, {Wildi}, {Abe}, {Baruffolo}, {Baudoz}, {Bazzon},
  {Boccaletti}, {Brast}, {Buey}, {Chesneau}, {Claudi}, {Costille},
  {Delboulb{\'e}}, {Desidera}, {Dominik}, {Dorn}, {Downing}, {Feautrier},
  {Fedrigo}, {Fusco}, {Girard}, {Giro}, {Gluck}, {Gonte}, {Gojak}, {Gratton},
  {Henning}, {Hubin}, {Lagrange}, {Langlois}, {Mignant}, {Lizon}, {Lilley},
  {Madec}, {Magnard}, {Martinez}, {Mawet}, {Mesa}, {M{\"u}ller-Nilsson},
  {Moulin}, {Moutou}, {O'Neal}, {Pavlov}, {Perret}, {Petit}, {Popovic},
  {Pragt}, {Rabou}, {Rochat}, {Roelfsema}, {Salasnich}, {Sauvage}, {Schmid},
  {Schuhler}, {Sevin}, {Siebenmorgen}, {Soenke}, {Stadler}, {Suarez},
  {Turatto}, {Udry}, {Vigan}, \& {Zins}}]{Kasper12}
{Kasper}, M., {Beuzit}, J.-L., {Feldt}, M., {et~al.} 2012, The Messenger, 149,
  17

\bibitem[{{Kasper} {et~al.}(2013){Kasper}, {Verinaud}, \& {Mawet}}]{Kasper13}
{Kasper}, M., {Verinaud}, C., \& {Mawet}, D. 2013, in Proceedings of the Third
  AO4ELT Conference, ed. S.~{Esposito} \& L.~{Fini}, 8

\bibitem[{{Keller} {et~al.}(2010){Keller}, {Schmid}, {Venema}, {Hanenburg},
  {Jager}, {Kasper}, {Martinez}, {Rigal}, {Rodenhuis}, {Roelfsema}, {Snik},
  {Verinaud}, \& {Yaitskova}}]{Keller10}
{Keller}, C.~U., {Schmid}, H.~M., {Venema}, L.~B., {et~al.} 2010, in \procspie,
  Vol. 7735, Ground-based and Airborne Instrumentation for Astronomy III,
  77356G

\bibitem[{{Kemp} \& {Barbour}(1981)}]{Kemp81}
{Kemp}, J.~C. \& {Barbour}, M.~S. 1981, \pasp, 93, 521

\bibitem[{{Kervella} {et~al.}(2008){Kervella}, {Domiciano de Souza}, \&
  {Bendjoya}}]{Kervella08}
{Kervella}, P., {Domiciano de Souza}, A., \& {Bendjoya}, P. 2008, \aap, 484,
  L13

\bibitem[{{Kervella} {et~al.}(2016){Kervella}, {Lagadec}, {Montarg{\`e}s},
  {Ridgway}, {Chiavassa}, {Haubois}, {Schmid}, {Langlois}, {Gallenne}, \&
  {Perrin}}]{Kervella16}
{Kervella}, P., {Lagadec}, E., {Montarg{\`e}s}, M., {et~al.} 2016, \aap, 585,
  A28

\bibitem[{{Kervella} {et~al.}(2015){Kervella}, {Montarg{\`e}s}, {Lagadec},
  {Ridgway}, {Haubois}, {Girard}, {Ohnaka}, {Perrin}, \&
  {Gallenne}}]{Kervella15}
{Kervella}, P., {Montarg{\`e}s}, M., {Lagadec}, E., {et~al.} 2015, \aap, 578,
  A77

\bibitem[{{Khouri} {et~al.}(2016){Khouri}, {Maercker}, {Waters}, {Vlemmings},
  {Kervella}, {de Koter}, {Ginski}, {De Beck}, {Decin}, {Min}, {Dominik},
  {O'Gorman}, {Schmid}, {Lombaert}, \& {Lagadec}}]{Khouri16}
{Khouri}, T., {Maercker}, M., {Waters}, L.~B.~F.~M., {et~al.} 2016, \aap, 591,
  A70

\bibitem[{{Kuhn} {et~al.}(2001){Kuhn}, {Potter}, \& {Parise}}]{Kuhn01}
{Kuhn}, J.~R., {Potter}, D., \& {Parise}, B. 2001, \apjl, 553, L189

\bibitem[{{Lafreni{\`e}re} {et~al.}(2007){Lafreni{\`e}re}, {Marois}, {Doyon},
  {Nadeau}, \& {Artigau}}]{Lafreniere07}
{Lafreni{\`e}re}, D., {Marois}, C., {Doyon}, R., {Nadeau}, D., \& {Artigau},
  {\'E}. 2007, \apj, 660, 770

\bibitem[{{Langlois} {et~al.}(2014){Langlois}, {Dohlen}, {Vigan}, {Zurlo},
  {Moutou}, {Schmid}, {Mili}, {Beuzit}, {Boccaletti}, {Carle}, {Costille},
  {Dorn}, {Gluck}, {Hubin}, {Feldt}, {Kasper}, {Lizon}, {Madec}, {Le Mignant},
  {Mouillet}, {Puget}, {Sauvage}, \& {Wildi}}]{Langlois14}
{Langlois}, M., {Dohlen}, K., {Vigan}, A., {et~al.} 2014, in \procspie, Vol.
  9147, Ground-based and Airborne Instrumentation for Astronomy V, 91471R

\bibitem[{{Law} {et~al.}(2009){Law}, {Mackay}, {Dekany}, {Ireland}, {Lloyd},
  {Moore}, {Robertson}, {Tuthill}, \& {Woodruff}}]{Law09}
{Law}, N.~M., {Mackay}, C.~D., {Dekany}, R.~G., {et~al.} 2009, \apj, 692, 924

\bibitem[{{Liimets} {et~al.}(2018){Liimets}, {Corradi}, {Jones}, {Verro},
  {Santander-Garc{\'{\i}}a}, {Kolka}, {Sidonio}, {Kankare}, {Kankare},
  {Pursimo}, \& {Wilson}}]{Liimets18}
{Liimets}, T., {Corradi}, R.~L.~M., {Jones}, D., {et~al.} 2018, \aap, 612, A118

\bibitem[{{Lovis} {et~al.}(2017){Lovis}, {Snellen}, {Mouillet}, {Pepe},
  {Wildi}, {Astudillo-Defru}, {Beuzit}, {Bonfils}, {Cheetham}, {Conod},
  {Delfosse}, {Ehrenreich}, {Figueira}, {Forveille}, {Martins}, {Quanz},
  {Santos}, {Schmid}, {S{\'e}gransan}, \& {Udry}}]{Lovis17}
{Lovis}, C., {Snellen}, I., {Mouillet}, D., {et~al.} 2017, \aap, 599, A16

\bibitem[{{Macintosh} {et~al.}(2014){Macintosh}, {Graham}, {Ingraham},
  {Konopacky}, {Marois}, {Perrin}, {Poyneer}, {Bauman}, {Barman}, {Burrows},
  {Cardwell}, {Chilcote}, {De Rosa}, {Dillon}, {Doyon}, {Dunn}, {Erikson},
  {Fitzgerald}, {Gavel}, {Goodsell}, {Hartung}, {Hibon}, {Kalas}, {Larkin},
  {Maire}, {Marchis}, {Marley}, {McBride}, {Millar-Blanchaer}, {Morzinski},
  {Norton}, {Oppenheimer}, {Palmer}, {Patience}, {Pueyo}, {Rantakyro},
  {Sadakuni}, {Saddlemyer}, {Savransky}, {Serio}, {Soummer},
  {Sivaramakrishnan}, {Song}, {Thomas}, {Wallace}, {Wiktorowicz}, \&
  {Wolff}}]{Macintosh14}
{Macintosh}, B., {Graham}, J.~R., {Ingraham}, P., {et~al.} 2014, Proceedings of
  the National Academy of Science, 111, 12661

\bibitem[{{Maire} {et~al.}(2016{\natexlab{a}}){Maire}, {Bonnefoy}, {Ginski},
  {Vigan}, {Messina}, {Mesa}, {Galicher}, {Gratton}, {Desidera}, {Kopytova},
  {Millward}, {Thalmann}, {Claudi}, {Ehrenreich}, {Zurlo}, {Chauvin},
  {Antichi}, {Baruffolo}, {Bazzon}, {Beuzit}, {Blanchard}, {Boccaletti}, {de
  Boer}, {Carle}, {Cascone}, {Costille}, {De Caprio}, {Delboulb{\'e}},
  {Dohlen}, {Dominik}, {Feldt}, {Fusco}, {Girard}, {Giro}, {Gisler}, {Gluck},
  {Gry}, {Henning}, {Hubin}, {Hugot}, {Jaquet}, {Kasper}, {Lagrange},
  {Langlois}, {Le Mignant}, {Llored}, {Madec}, {Martinez}, {Mawet}, {Milli},
  {M{\"o}ller-Nilsson}, {Mouillet}, {Moulin}, {Moutou}, {Orign{\'e}}, {Pavlov},
  {Petit}, {Pragt}, {Puget}, {Ramos}, {Rochat}, {Roelfsema}, {Salasnich},
  {Sauvage}, {Schmid}, {Turatto}, {Udry}, {Vakili}, {Wahhaj}, {Weber}, \&
  {Wildi}}]{Maire16a}
{Maire}, A.-L., {Bonnefoy}, M., {Ginski}, C., {et~al.} 2016{\natexlab{a}},
  \aap, 587, A56

\bibitem[{{Maire} {et~al.}(2016{\natexlab{b}}){Maire}, {Langlois}, {Dohlen},
  {Lagrange}, {Gratton}, {Chauvin}, {Desidera}, {Girard}, {Milli}, {Vigan},
  {Zins}, {Delorme}, {Beuzit}, {Claudi}, {Feldt}, {Mouillet}, {Puget},
  {Turatto}, \& {Wildi}}]{Maire16b}
{Maire}, A.-L., {Langlois}, M., {Dohlen}, K., {et~al.} 2016{\natexlab{b}}, in
  \procspie, Vol. 9908, Ground-based and Airborne Instrumentation for Astronomy
  VI, 990834

\bibitem[{{Malbet}(1996)}]{Malbet96}
{Malbet}, F. 1996, \aaps, 115, 161

\bibitem[{{Marchis} {et~al.}(2006){Marchis}, {Kaasalainen}, {Hom}, {Berthier},
  {Enriquez}, {Hestroffer}, {Le Mignant}, \& {de Pater}}]{Marchis06}
{Marchis}, F., {Kaasalainen}, M., {Hom}, E.~F.~Y., {et~al.} 2006, \icarus, 185,
  39

\bibitem[{{Marois} {et~al.}(2006){Marois}, {Lafreni{\`e}re}, {Doyon},
  {Macintosh}, \& {Nadeau}}]{Marois06}
{Marois}, C., {Lafreni{\`e}re}, D., {Doyon}, R., {Macintosh}, B., \& {Nadeau},
  D. 2006, \apj, 641, 556

\bibitem[{{Martinez Pillet} \& {Sanchez Almeida}(1991)}]{MartinezPillet91}
{Martinez Pillet}, V. \& {Sanchez Almeida}, J. 1991, \aap, 252, 861

\bibitem[{{McLaughlin} {et~al.}(2006){McLaughlin}, {Anderson}, {Meylan},
  {Gebhardt}, {Pryor}, {Minniti}, \& {Phinney}}]{McLaughlin06}
{McLaughlin}, D.~E., {Anderson}, J., {Meylan}, G., {et~al.} 2006, \apjs, 166,
  249

\bibitem[{{Melnikov} {et~al.}(2018){Melnikov}, {Stute}, \&
  {Eisl{\"o}ffel}}]{Melnikov18}
{Melnikov}, S., {Stute}, M., \& {Eisl{\"o}ffel}, J. 2018, \aap, 612, A77

\bibitem[{{Milli} {et~al.}(2017){Milli}, {Mouillet}, {Fusco}, {Girard},
  {Masciadri}, {Pena}, {Sauvage}, {Reyes}, {Dohlen}, {Beuzit}, {Kasper},
  {Sarazin}, \& {Cantalloube}}]{Milli17}
{Milli}, J., {Mouillet}, D., {Fusco}, T., {et~al.} 2017, ArXiv e-prints

\bibitem[{{Milli} {et~al.}(2013){Milli}, {Mouillet}, {Mawet}, {Schmid},
  {Bazzon}, {Girard}, {Dohlen}, \& {Roelfsema}}]{Milli13}
{Milli}, J., {Mouillet}, D., {Mawet}, D., {et~al.} 2013, \aap, 556, A64

\bibitem[{{Momany} {et~al.}(2012){Momany}, {Saviane}, {Smette}, {Bayo},
  {Girardi}, {Marconi}, {Milone}, \& {Bressan}}]{Momany12}
{Momany}, Y., {Saviane}, I., {Smette}, A., {et~al.} 2012, \aap, 537, A2

\bibitem[{{Muto} {et~al.}(2012){Muto}, {Grady}, {Hashimoto}, {Fukagawa},
  {Hornbeck}, {Sitko}, {Russell}, {Werren}, {Cur{\'e}}, {Currie}, {Ohashi},
  {Okamoto}, {Momose}, {Honda}, {Inutsuka}, {Takeuchi}, {Dong}, {Abe},
  {Brandner}, {Brandt}, {Carson}, {Egner}, {Feldt}, {Fukue}, {Goto}, {Guyon},
  {Hayano}, {Hayashi}, {Hayashi}, {Henning}, {Hodapp}, {Ishii}, {Iye},
  {Janson}, {Kandori}, {Knapp}, {Kudo}, {Kusakabe}, {Kuzuhara}, {Matsuo},
  {Mayama}, {McElwain}, {Miyama}, {Morino}, {Moro-Martin}, {Nishimura}, {Pyo},
  {Serabyn}, {Suto}, {Suzuki}, {Takami}, {Takato}, {Terada}, {Thalmann},
  {Tomono}, {Turner}, {Watanabe}, {Wisniewski}, {Yamada}, {Takami}, {Usuda}, \&
  {Tamura}}]{Muto12}
{Muto}, T., {Grady}, C.~A., {Hashimoto}, J., {et~al.} 2012, \apjl, 748, L22

\bibitem[{{Norris} {et~al.}(2012){Norris}, {Tuthill}, {Ireland}, {Lacour},
  {Zijlstra}, {Lykou}, {Evans}, {Stewart}, \& {Bedding}}]{Norris12}
{Norris}, B.~R.~M., {Tuthill}, P.~G., {Ireland}, M.~J., {et~al.} 2012, \nat,
  484, 220

\bibitem[{{Ohnaka} {et~al.}(2017{\natexlab{a}}){Ohnaka}, {Weigelt}, \&
  {Hofmann}}]{Ohnaka17a}
{Ohnaka}, K., {Weigelt}, G., \& {Hofmann}, K.-H. 2017{\natexlab{a}}, \aap, 597,
  A20

\bibitem[{{Ohnaka} {et~al.}(2017{\natexlab{b}}){Ohnaka}, {Weigelt}, \&
  {Hofmann}}]{Ohnaka17b}
{Ohnaka}, K., {Weigelt}, G., \& {Hofmann}, K.-H. 2017{\natexlab{b}}, \nat, 548,
  310

\bibitem[{{Paladini} {et~al.}(2017){Paladini}, {Klotz}, {Sacuto}, {Lagadec},
  {Wittkowski}, {Richichi}, {Hron}, {Jorissen}, {Groenewegen}, {Kerschbaum},
  {Verhoelst}, {Rau}, {Olofsson}, {Zhao-Geisler}, \& {Matter}}]{Paladini17}
{Paladini}, C., {Klotz}, D., {Sacuto}, S., {et~al.} 2017, \aap, 600, A136

\bibitem[{{Patapis} {et~al.}(2018){Patapis}, {K\"{u}hn}, \&
  {Schmid}}]{Patapis18}
{Patapis}, P., {K\"{u}hn}, J., \& {Schmid}, H.~M. 2018, in \procspie, Vol.
  10706, Advances in Optical and Mechanical Technologies for Telescopes and
  Instrumentation III, 107065J

\bibitem[{{Perrin} {et~al.}(2015){Perrin}, {Duchene}, {Millar-Blanchaer},
  {Fitzgerald}, {Graham}, {Wiktorowicz}, {Kalas}, {Macintosh}, {Bauman},
  {Cardwell}, {Chilcote}, {De Rosa}, {Dillon}, {Doyon}, {Dunn}, {Erikson},
  {Gavel}, {Goodsell}, {Hartung}, {Hibon}, {Ingraham}, {Kerley}, {Konapacky},
  {Larkin}, {Maire}, {Marchis}, {Marois}, {Mittal}, {Morzinski}, {Oppenheimer},
  {Palmer}, {Patience}, {Poyneer}, {Pueyo}, {Rantakyr{\"o}}, {Sadakuni},
  {Saddlemyer}, {Savransky}, {Soummer}, {Sivaramakrishnan}, {Song}, {Thomas},
  {Wallace}, {Wang}, \& {Wolff}}]{Perrin15}
{Perrin}, M.~D., {Duchene}, G., {Millar-Blanchaer}, M., {et~al.} 2015, \apj,
  799, 182

\bibitem[{{Perrin} {et~al.}(2009){Perrin}, {Schneider}, {Duchene}, {Pinte},
  {Grady}, {Wisniewski}, \& {Hines}}]{Perrin09}
{Perrin}, M.~D., {Schneider}, G., {Duchene}, G., {et~al.} 2009, \apjl, 707,
  L132

\bibitem[{{Povel} {et~al.}(1990){Povel}, {Aebersold}, \& {Stenflo}}]{Povel90}
{Povel}, H., {Aebersold}, H., \& {Stenflo}, J.~O. 1990, \ao, 29, 1186

\bibitem[{{Povel}(1995)}]{Povel95}
{Povel}, H.-P. 1995, Optical Engineering, 34

\bibitem[{{Pragt} {et~al.}(2012){Pragt}, {Roelfsema}, {Gisler}, {Wildi},
  {Schmid}, {Rigal}, {Elswijk}, {de Haan}, {Bazzon}, {Dohlen}, {Costille}, \&
  {Dominik}}]{Pragt12}
{Pragt}, J., {Roelfsema}, R., {Gisler}, D., {et~al.} 2012, in \procspie, Vol.
  8446, Ground-based and Airborne Instrumentation for Astronomy IV, 844697

\bibitem[{{Quanz} {et~al.}(2011){Quanz}, {Schmid}, {Geissler}, {Meyer},
  {Henning}, {Brandner}, \& {Wolf}}]{Quanz11}
{Quanz}, S.~P., {Schmid}, H.~M., {Geissler}, K., {et~al.} 2011, \apj, 738, 23

\bibitem[{{Ragland} {et~al.}(2008){Ragland}, {Le Coroller}, {Pluzhnik},
  {Cotton}, {Danchi}, {Monnier}, {Traub}, {Willson}, {Berger}, \&
  {Lacasse}}]{Ragland08}
{Ragland}, S., {Le Coroller}, H., {Pluzhnik}, E., {et~al.} 2008, \apj, 679, 746

\bibitem[{{Rapson} {et~al.}(2015){Rapson}, {Kastner}, {Millar-Blanchaer}, \&
  {Dong}}]{Rapson15}
{Rapson}, V.~A., {Kastner}, J.~H., {Millar-Blanchaer}, M.~A., \& {Dong}, R.
  2015, \apjl, 815, L26

\bibitem[{{Riaud} {et~al.}(2003){Riaud}, {Boccaletti}, {Baudrand}, \&
  {Rouan}}]{Riaud03}
{Riaud}, P., {Boccaletti}, A., {Baudrand}, J., \& {Rouan}, D. 2003, \pasp, 115,
  712

\bibitem[{{Roelfsema} {et~al.}(2014){Roelfsema}, {Bazzon}, {Schmid}, {Pragt},
  {Gisler}, {Dominik}, {Baruffolo}, {Beuzit}, {Costille}, {Dohlen}, {Downing},
  {Elswijk}, {de Haan}, {Hubin}, {Kasper}, {Keller}, {Lizon}, {Mouillet},
  {Pavlov}, {Puget}, {Salasnich}, {Sauvage}, \& {Wildi}}]{Roelfsema14}
{Roelfsema}, R., {Bazzon}, A., {Schmid}, H.~M., {et~al.} 2014, in \procspie,
  Vol. 9147, Ground-based and Airborne Instrumentation for Astronomy V, 91473W

\bibitem[{{Roelfsema} {et~al.}(2016){Roelfsema}, {Bazzon}, {Schmid}, {Pragt},
  {Govaert}, {Gisler}, {Dominik}, {Baruffolo}, {Beuzit}, {Costille}, {Dohlen},
  {Downing}, {Elswijk}, {de Haan}, {Hubin}, {Kasper}, {Keller}, {Lizon},
  {Mouillet}, {Pavlov}, {Puget}, {Salasnich}, {Sauvage}, \&
  {Wildi}}]{Roelfsema16}
{Roelfsema}, R., {Bazzon}, A., {Schmid}, H.~M., {et~al.} 2016, in \procspie,
  Vol. 9909, Adaptive Optics Systems V, 990927

\bibitem[{{Roelfsema} {et~al.}(2011){Roelfsema}, {Gisler}, {Pragt}, {Schmid},
  {Bazzon}, {Dominik}, {Baruffolo}, {Beuzit}, {Charton}, {Dohlen}, {Downing},
  {Elswijk}, {Feldt}, {de Haan}, {Hubin}, {Kasper}, {Keller}, {Lizon},
  {Mouillet}, {Pavlov}, {Puget}, {Rochat}, {Salasnich}, {Steiner}, {Thalmann},
  {Waters}, \& {Wildi}}]{Roelfsema11}
{Roelfsema}, R., {Gisler}, D., {Pragt}, J., {et~al.} 2011, in \procspie, Vol.
  8151, Techniques and Instrumentation for Detection of Exoplanets V, 81510N

\bibitem[{{Roelfsema} {et~al.}(2010){Roelfsema}, {Schmid}, {Pragt}, {Gisler},
  {Waters}, {Bazzon}, {Baruffolo}, {Beuzit}, {Boccaletti}, {Charton}, {Cumani},
  {Dohlen}, {Downing}, {Elswijk}, {Feldt}, {Groothuis}, {de Haan}, {Hanenburg},
  {Hubin}, {Joos}, {Kasper}, {Keller}, {Kragt}, {Lizon}, {Mouillet}, {Pavlov},
  {Rigal}, {Rochat}, {Salasnich}, {Steiner}, {Thalmann}, {Venema}, \&
  {Wildi}}]{Roelfsema10}
{Roelfsema}, R., {Schmid}, H.~M., {Pragt}, J., {et~al.} 2010, in \procspie,
  Vol. 7735, Ground-based and Airborne Instrumentation for Astronomy III, 4

\bibitem[{{Rouan} {et~al.}(2000){Rouan}, {Riaud}, {Boccaletti}, {Cl{\'e}net},
  \& {Labeyrie}}]{Rouan00}
{Rouan}, D., {Riaud}, P., {Boccaletti}, A., {Cl{\'e}net}, Y., \& {Labeyrie}, A.
  2000, \pasp, 112, 1479

\bibitem[{{Sauvage} {et~al.}(2016{\natexlab{a}}){Sauvage}, {Fusco}, {Lamb},
  {Girard}, {Brinkmann}, {Guesalaga}, {Wizinowich}, {O'Neal}, {N'Diaye},
  {Vigan}, {Mouillet}, {Beuzit}, {Kasper}, {Le Louarn}, {Milli}, {Dohlen},
  {Neichel}, {Bourget}, {Haguenauer}, \& {Mawet}}]{Sauvage16b}
{Sauvage}, J.-F., {Fusco}, T., {Lamb}, M., {et~al.} 2016{\natexlab{a}}, in
  \procspie, Vol. 9909, Adaptive Optics Systems V, 990916

\bibitem[{{Sauvage} {et~al.}(2016{\natexlab{b}}){Sauvage}, {Fusco}, {Petit},
  {Costille}, {Mouillet}, {Beuzit}, {Dohlen}, {Kasper}, {Suarez}, {Soenke},
  {Baruffolo}, {Salasnich}, {Rochat}, {Fedrigo}, {Baudoz}, {Hugot}, {Sevin},
  {Perret}, {Wildi}, {Downing}, {Feautrier}, {Puget}, {Vigan}, {O'Neal},
  {Girard}, {Mawet}, {Schmid}, \& {Roelfsema}}]{Sauvage16a}
{Sauvage}, J.-F., {Fusco}, T., {Petit}, C., {et~al.} 2016{\natexlab{b}},
  Journal of Astronomical Telescopes, Instruments, and Systems, 2, 025003

\bibitem[{{Sauvage} {et~al.}(2007){Sauvage}, {Fusco}, {Rousset}, \&
  {Petit}}]{Sauvage07}
{Sauvage}, J.-F., {Fusco}, T., {Rousset}, G., \& {Petit}, C. 2007, Journal of
  the Optical Society of America A, 24, 2334

\bibitem[{{Scarrott} {et~al.}(1983){Scarrott}, {Warren-Smith}, {Pallister},
  {Axon}, \& {Bingham}}]{Scarrott83}
{Scarrott}, S.~M., {Warren-Smith}, R.~F., {Pallister}, W.~S., {Axon}, D.~J., \&
  {Bingham}, R.~G. 1983, \mnras, 204, 1163

\bibitem[{{Schiavon} {et~al.}(2012){Schiavon}, {Dalessandro}, {Sohn}, {Rood},
  {O'Connell}, {Ferraro}, {Lanzoni}, {Beccari}, {Rey}, {Rhee}, {Rich}, {Yoon},
  \& {Lee}}]{Schiavon12}
{Schiavon}, R.~P., {Dalessandro}, E., {Sohn}, S.~T., {et~al.} 2012, \aj, 143,
  121

\bibitem[{{Schmid} {et~al.}(2017){Schmid}, {Bazzon}, {Milli}, {Roelfsema},
  {Engler}, {Mouillet}, {Lagadec}, {Sissa}, {Sauvage}, {Ginski}, {Baruffolo},
  {Beuzit}, {Boccaletti}, {Bohn}, {Claudi}, {Costille}, {Desidera}, {Dohlen},
  {Dominik}, {Feldt}, {Fusco}, {Gisler}, {Girard}, {Gratton}, {Henning},
  {Hubin}, {Joos}, {Kasper}, {Langlois}, {Pavlov}, {Pragt}, {Puget}, {Quanz},
  {Salasnich}, {Siebenmorgen}, {Stute}, {Suarez}, {Szul{\'a}gyi}, {Thalmann},
  {Turatto}, {Udry}, {Vigan}, \& {Wildi}}]{Schmid17}
{Schmid}, H.~M., {Bazzon}, A., {Milli}, J., {et~al.} 2017, \aap, 602, A53

\bibitem[{{Schmid} {et~al.}(2006{\natexlab{a}}){Schmid}, {Beuzit}, {Feldt},
  {Gisler}, {Gratton}, {Henning}, {Joos}, {Kasper}, {Lenzen}, {Mouillet},
  {Moutou}, {Quirrenbach}, {Stam}, {Thalmann}, {Tinbergen}, {Verinaud},
  {Waters}, \& {Wolstencroft}}]{Schmid06a}
{Schmid}, H.~M., {Beuzit}, J.-L., {Feldt}, M., {et~al.} 2006{\natexlab{a}}, in
  IAU Colloq. 200: Direct Imaging of Exoplanets: Science \& Techniques, ed.
  C.~{Aime} \& F.~{Vakili}, 165--170

\bibitem[{{Schmid} {et~al.}(2012){Schmid}, {Downing}, {Roelfsema}, {Bazzon},
  {Gisler}, {Pragt}, {Cumani}, {Salasnich}, {Pavlov}, {Baruffolo}, {Beuzit},
  {Costille}, {Deiries}, {Dohlen}, {Dominik}, {Elswijk}, {Feldt}, {Kasper},
  {Mouillet}, {Thalmann}, \& {Wildi}}]{Schmid12}
{Schmid}, H.-M., {Downing}, M., {Roelfsema}, R., {et~al.} 2012, in \procspie,
  Vol. 8446, Ground-based and Airborne Instrumentation for Astronomy IV., 8

\bibitem[{{Schmid} {et~al.}(2006{\natexlab{b}}){Schmid}, {Joos}, \&
  {Tschan}}]{Schmid06b}
{Schmid}, H.~M., {Joos}, F., \& {Tschan}, D. 2006{\natexlab{b}}, \aap, 452, 657

\bibitem[{{Schneider} \& {Silverstone}(2003)}]{Schneider03}
{Schneider}, G. \& {Silverstone}, M.~D. 2003, in \procspie, Vol. 4860,
  High-Contrast Imaging for Exo-Planet Detection., ed. A.~B. {Schultz}, 1--9

\bibitem[{{Serkowski}(1974)}]{Serkowski74}
{Serkowski}, K. 1974, in IAU Colloq. 23: Planets, Stars, and Nebulae: Studied
  with Photopolarimetry, ed. T.~{Gehrels}, 135

\bibitem[{{Serkowski} {et~al.}(1975){Serkowski}, {Mathewson}, \&
  {Ford}}]{Serkowski75}
{Serkowski}, K., {Mathewson}, D.~S., \& {Ford}, V.~L. 1975, \apj, 196, 261

\bibitem[{{Serkowski} \& {Shawl}(2001)}]{Serkowski01}
{Serkowski}, K. \& {Shawl}, S.~J. 2001, \aj, 122, 2017

\bibitem[{{Sivaramakrishnan} {et~al.}(2001){Sivaramakrishnan}, {Koresko},
  {Makidon}, {Berkefeld}, \& {Kuchner}}]{Sivaramakrishnan01}
{Sivaramakrishnan}, A., {Koresko}, C.~D., {Makidon}, R.~B., {Berkefeld}, T., \&
  {Kuchner}, M.~J. 2001, \apj, 552, 397

\bibitem[{{Soummer} {et~al.}(2012){Soummer}, {Pueyo}, \& {Larkin}}]{Soummer12}
{Soummer}, R., {Pueyo}, L., \& {Larkin}, J. 2012, \apjl, 755, L28

\bibitem[{{Stenflo}(1996)}]{Stenflo96}
{Stenflo}, J.~O. 1996, \nat, 382, 588

\bibitem[{{Stenflo} \& {Keller}(1997)}]{Stenflo97}
{Stenflo}, J.~O. \& {Keller}, C.~U. 1997, \aap, 321, 927

\bibitem[{{Stolker} {et~al.}(2016){Stolker}, {Dominik}, {Avenhaus}, {Min}, {de
  Boer}, {Ginski}, {Schmid}, {Juhasz}, {Bazzon}, {Waters}, {Garufi},
  {Augereau}, {Benisty}, {Boccaletti}, {Henning}, {Langlois}, {Maire},
  {M{\'e}nard}, {Meyer}, {Pinte}, {Quanz}, {Thalmann}, {Beuzit}, {Carbillet},
  {Costille}, {Dohlen}, {Feldt}, {Gisler}, {Mouillet}, {Pavlov}, {Perret},
  {Petit}, {Pragt}, {Rochat}, {Roelfsema}, {Salasnich}, {Soenke}, \&
  {Wildi}}]{Stolker16}
{Stolker}, T., {Dominik}, C., {Avenhaus}, H., {et~al.} 2016, \aap, 595, A113

\bibitem[{{Thalmann} {et~al.}(2015){Thalmann}, {Mulders}, {Janson}, {Olofsson},
  {Benisty}, {Avenhaus}, {Quanz}, {Schmid}, {Henning}, {Buenzli}, {M{\'e}nard},
  {Carson}, {Garufi}, {Messina}, {Dominik}, {Leisenring}, {Chauvin}, \&
  {Meyer}}]{Thalmann15}
{Thalmann}, C., {Mulders}, G.~D., {Janson}, M., {et~al.} 2015, \apjl, 808, L41

\bibitem[{{Thalmann} {et~al.}(2008){Thalmann}, {Schmid}, {Boccaletti},
  {Mouillet}, {Dohlen}, {Roelfsema}, {Carbillet}, {Gisler}, {Beuzit}, {Feldt},
  {Gratton}, {Joos}, {Keller}, {Kragt}, {Pragt}, {Puget}, {Rigal}, {Snik},
  {Waters}, \& {Wildi}}]{Thalmann08}
{Thalmann}, C., {Schmid}, H.~M., {Boccaletti}, A., {et~al.} 2008, in \procspie,
  Vol. 7014, Ground-based and Airborne Instrumentation for Astronomy II., 3

\bibitem[{{Tinbergen}(1979)}]{Tinbergen79}
{Tinbergen}, J. 1979, \aaps, 35, 325

\bibitem[{{Tinbergen}(2007)}]{Tinbergen07}
{Tinbergen}, J. 2007, \pasp, 119, 1371

\bibitem[{{van Belle} {et~al.}(1996){van Belle}, {Dyck}, {Benson}, \&
  {Lacasse}}]{vanBelle96}
{van Belle}, G.~T., {Dyck}, H.~M., {Benson}, J.~A., \& {Lacasse}, M.~G. 1996,
  \aj, 112, 2147

\bibitem[{{van Boekel} {et~al.}(2017){van Boekel}, {Henning}, {Menu}, {de
  Boer}, {Langlois}, {M{\"u}ller}, {Avenhaus}, {Boccaletti}, {Schmid},
  {Thalmann}, {Benisty}, {Dominik}, {Ginski}, {Girard}, {Gisler}, {Lobo Gomes},
  {Menard}, {Min}, {Pavlov}, {Pohl}, {Quanz}, {Rabou}, {Roelfsema}, {Sauvage},
  {Teague}, {Wildi}, \& {Zurlo}}]{vanBoekel17}
{van Boekel}, R., {Henning}, T., {Menu}, J., {et~al.} 2017, \apj, 837, 132

\bibitem[{{van Holstein} {et~al.}(2017){van Holstein}, {Snik}, {Girard}, {de
  Boer}, {Ginski}, {Keller}, {Stam}, {Beuzit}, {Mouillet}, {Kasper},
  {Langlois}, {Zurlo}, {de Kok}, \& {Vigan}}]{vanHolstein17}
{van Holstein}, R.~G., {Snik}, F., {Girard}, J.~H., {et~al.} 2017, in
  \procspie, Vol. 10400, Techniques and Instrumentation for Detection of
  Exoplanets VIII, 1040015

\bibitem[{{Vernazza} {et~al.}(2017){Vernazza}, {Marchis}, {Carry}, {Marsset},
  \& {Hanus}}]{Vernazza17}
{Vernazza}, P., {Marchis}, F., {Carry}, B., {Marsset}, M., \& {Hanus}, J. 2017,
  European Planetary Science Congress, 11, EPSC2017

\bibitem[{{Vigan} {et~al.}(2016){Vigan}, {Bonnefoy}, {Ginski}, {Beust},
  {Galicher}, {Janson}, {Baudino}, {Buenzli}, {Hagelberg}, {D'Orazi},
  {Desidera}, {Maire}, {Gratton}, {Sauvage}, {Chauvin}, {Thalmann}, {Malo},
  {Salter}, {Zurlo}, {Antichi}, {Baruffolo}, {Baudoz}, {Blanchard},
  {Boccaletti}, {Beuzit}, {Carle}, {Claudi}, {Costille}, {Delboulb{\'e}},
  {Dohlen}, {Dominik}, {Feldt}, {Fusco}, {Gluck}, {Girard}, {Giro}, {Gry},
  {Henning}, {Hubin}, {Hugot}, {Jaquet}, {Kasper}, {Lagrange}, {Langlois}, {Le
  Mignant}, {Llored}, {Madec}, {Martinez}, {Mawet}, {Mesa}, {Milli},
  {Mouillet}, {Moulin}, {Moutou}, {Orign{\'e}}, {Pavlov}, {Perret}, {Petit},
  {Pragt}, {Puget}, {Rabou}, {Rochat}, {Roelfsema}, {Salasnich}, {Schmid},
  {Sevin}, {Siebenmorgen}, {Smette}, {Stadler}, {Suarez}, {Turatto}, {Udry},
  {Vakili}, {Wahhaj}, {Weber}, \& {Wildi}}]{Vigan16}
{Vigan}, A., {Bonnefoy}, M., {Ginski}, C., {et~al.} 2016, \aap, 587, A55

\bibitem[{{Vigan} {et~al.}(2014){Vigan}, {Langlois}, {Dohlen}, {Zurlo},
  {Moutou}, {Costille}, {Gry}, {Madec}, {Le Mignant}, {Gluck}, \&
  {Sauvage}}]{Vigan14}
{Vigan}, A., {Langlois}, M., {Dohlen}, K., {et~al.} 2014, in \procspie, Vol.
  9147, Ground-based and Airborne Instrumentation for Astronomy V, 91474T

\bibitem[{{Whittet} {et~al.}(1992){Whittet}, {Martin}, {Hough}, {Rouse},
  {Bailey}, \& {Axon}}]{Whittet92}
{Whittet}, D.~C.~B., {Martin}, P.~G., {Hough}, J.~H., {et~al.} 1992, \apj, 386,
  562

\bibitem[{{Wildi} {et~al.}(2009){Wildi}, {Mouillet}, {Beuzit}, {Feldt},
  {Dohlen}, {Fusco}, {Petit}, {Desidera}, {Gratton}, {Schmid}, {Langlois},
  {Vigan}, {Charton}, {Claudi}, {Roelfsema}, {Baruffolo}, \& {Puget}}]{Wildi09}
{Wildi}, F., {Mouillet}, D., {Beuzit}, J.-L., {et~al.} 2009, in \procspie, Vol.
  7440, Techniques and Instrumentation for Detection of Exoplanets IV, 74400Q

\bibitem[{{Wildi} {et~al.}(2010){Wildi}, {Michaud}, {Crausaz}, {Dubosson},
  {Mouillet}, {Dohlen}, {Schmid}, \& {Beuzit}}]{Wildi10}
{Wildi}, F.~P., {Michaud}, B., {Crausaz}, M., {et~al.} 2010, in \procspie, Vol.
  7735, Ground-based and Airborne Instrumentation for Astronomy III, 77352V

\bibitem[{{Zurlo} {et~al.}(2016){Zurlo}, {Vigan}, {Galicher}, {Maire}, {Mesa},
  {Gratton}, {Chauvin}, {Kasper}, {Moutou}, {Bonnefoy}, {Desidera}, {Abe},
  {Apai}, {Baruffolo}, {Baudoz}, {Baudrand}, {Beuzit}, {Blancard},
  {Boccaletti}, {Cantalloube}, {Carle}, {Cascone}, {Charton}, {Claudi},
  {Costille}, {de Caprio}, {Dohlen}, {Dominik}, {Fantinel}, {Feautrier},
  {Feldt}, {Fusco}, {Gigan}, {Girard}, {Gisler}, {Gluck}, {Gry}, {Henning},
  {Hugot}, {Janson}, {Jaquet}, {Lagrange}, {Langlois}, {Llored}, {Madec},
  {Magnard}, {Martinez}, {Maurel}, {Mawet}, {Meyer}, {Milli},
  {Moeller-Nilsson}, {Mouillet}, {Orign{\'e}}, {Pavlov}, {Petit}, {Puget},
  {Quanz}, {Rabou}, {Ramos}, {Rousset}, {Roux}, {Salasnich}, {Salter},
  {Sauvage}, {Schmid}, {Soenke}, {Stadler}, {Suarez}, {Turatto}, {Udry},
  {Vakili}, {Wahhaj}, {Wildi}, \& {Antichi}}]{Zurlo16}
{Zurlo}, A., {Vigan}, A., {Galicher}, R., {et~al.} 2016, \aap, 587, A57

\end{thebibliography}

\end{document}